\documentclass[prc,twocolumn,showpacs,showkeys,preprintnumbers,floatfix,nofootinbib,superscriptaddress,longbibliography]{revtex4-1}

\usepackage{soul}
\usepackage{amsfonts} 
\usepackage{amssymb} 
\usepackage{amsmath} 
\usepackage{graphicx} 
\usepackage{subfigure} 
\usepackage{array} 
\usepackage{dcolumn} 
\usepackage{bm} 
\usepackage{latexsym} 
\usepackage{longtable} 
\usepackage{hyperref} 
\usepackage{bbold}
\usepackage{color}
\usepackage{multirow}
\usepackage{rotating}
\usepackage{comment}
\usepackage{mathtools}
\usepackage{algorithm}
\usepackage[noend]{algpseudocode}
\usepackage{ulem}
\usepackage{lipsum}

\usepackage[dvipsnames,usenames]{xcolor}
\input{Qcircuit}

\DeclareMathOperator{\Tr}{Tr}

\definecolor{green}{rgb}{0.1, 0.8, 0.1}



\begin{document}


\title{Preparation of excited states for nuclear dynamics on a quantum computer}

\preprint{INT-PUB-20-037} 
\preprint{LA-UR-20-27411} 

\author{Alessandro Roggero}
\affiliation{Institute for Nuclear Theory, University of Washington, Seattle, WA 98195, USA}

\author{Chenyi Gu}
\affiliation{Department of Physics and Astronomy, University of Tennessee, Knoxville, TN 37996, USA}

\author{Alessandro Baroni}
\affiliation{Theoretical Division, Los Alamos National Laboratory, Los Alamos, NM 87545, USA}

\author{Thomas Papenbrock}
\affiliation{Department of Physics and Astronomy, University of Tennessee, Knoxville, TN 37996, USA}
\affiliation{Physics Division, Oak Ridge National Laboratory, Oak Ridge, TN 37831, USA}

\date{\today}

\begin{abstract}
We study two different methods to prepare excited states on a quantum computer, a key initial step to study nuclear dynamics within linear response theory. The first method uses unitary evolution for a short time $T=\mathcal{O}(\sqrt{1-F})$ to approximate the action of an excitation operator $\hat{O}$ with fidelity $F$ and success probability $P\approx1-F$. The second method  probabilistically applies the excitation operator using the Linear Combination of Unitaries (LCU) algorithm. We benchmark these techniques on emulated and real quantum devices, using a toy model for thermal neutron-proton capture.
Despite its larger memory footprint,  the LCU-based method is efficient even on current generation noisy devices and can be implemented at a lower gate cost than a naive analysis would suggest. These findings show that quantum techniques designed to achieve good asymptotic scaling on fault tolerant quantum devices might also provide  practical benefits on devices with limited connectivity and gate fidelity.
\end{abstract}

\maketitle

\section{Introduction}
Quantum computing~\cite{steane1998,ladd2010} holds a lot of promise for low-energy nuclear theory~\cite{martinez2016,kaplan2017,Dumitrescu2018,klco2018,roggero2019,roggero2020A}. An error-corrected quantum computer with about 100 logical qubits, for instance, would make it possible to solve nuclear shell models~\cite{caurier2005} across the Segr{\`e} chart. Other examples of problems that are hard to solve with classical computers include time evolution of nuclear processes such as reactions, fission, and fusion. Present quantum devices consist of tens of noisy qubits, and decoherence is expected to remain a challenge for the foreseeable future. In this era of noisy intermediate-scale quantum (NISQ) devices~\cite{preskill2018}, it is interesting to explore how to prepare, evolve, and solve for quantum states on quantum computers, and how to mitigate or minimize the effects of decoherence~\cite{temme2017,kandala2019}.

In this paper we study how to prepare quantum states on existing superconducting devices. The nuclear physics problem we are motivated by is the $n(p,d)\gamma$ reaction, which is relevant for big-bang nucleosynthesis~\cite{adelberger2011}.
While this reaction is well understood and precise computations exist~\cite{park1998,chen1999,chen1999b}, it is a prototypical nuclear reaction and perhaps the simplest one to start with. How would one approach this problem on a quantum computer? 

We note that quantum computers consists of interconnected qubits, i.e. two-level systems that can be thought of as spin-$1/2$ states. The Jordan-Wigner or Bravyi-Kitaev~\cite{bravyi2002} transformations allow one to map fermionic creation and annihilation operators onto spin lowering and raising operators, respectively (see also~\cite{whitfield2016,setia2019,steudtner2019,derby2020} for other recent mappings). Having mapped a nuclear Hamiltonian onto qubits, quantum computing consists of two steps, namely (i) the preparation of states via unitary operations, and (ii) the readout via projection onto the canonical (spin-up/down) basis. As in classical computing, we speak of ``gates'' and ``cicuits'' when referring to individual unitary operations and a product of them.  Any unitary operation between a set of qubits can be decomposed efficiently into products of unitary operations that involve only one or two qubits (see, for instance, Sec.~4.5 of~Ref.\cite{nielsen2010}). On present superconducting devices single-qubit operations have a relatively high fidelity of more than 99\% while two-qubit gates are noisier with errors at the few percent level. Thus, it is an important task to reduce the number of two-qubit gates in quantum circuits, for instance by keeping Hamiltonians and operators of interest as local as possible~\cite{lloyd1996}. Finally, the readout is a quantum mechanical measurement process and irreversible. Its outcome is stochastic, and expectation values result from many repeated experiments of state preparations followed by readouts. This measurement process is also plagued by noise, with readout fidelities larger than 95\% on present superconducting devices. For an introduction to quantum computing, we refer the reader to Ref.~\cite{nielsen2010}.

The general problem we are interested in can be stated as follows: given a Hermitian excitation operator $O$ and an initial state $\ket{\Psi_0}$ (for instance the ground state of a many-body Hamiltonian) we want to find a protocol to prepare the normalized state
\begin{equation}
\label{eq:excit_state}
\ket{\Phi_E} = \frac{1}{\eta} O\rvert\Psi_0\rangle\quad\text{with}\quad \eta = \| O\rvert\Psi_0\rangle\|\;,
\end{equation}
on the same quantum register containing the original state. We note that the norm $\eta$ in Eq.~\eqref{eq:excit_state} above is the vector 2-norm, and $\ket{\Phi_E}$ is thus obtained from $\rvert\Psi_0\rangle$ by an unitary operation. An efficient implementation of this procedure is important for the performance of quantum algorithms developed to study exclusive scattering in linear response \cite{roggero2019}, could speedup the inclusion of pair-correlations with Jastrow factors~\cite{mazzola2019}, and more generally serve as a proxy to perform quantum computations with non-unitary gates \cite{ueda2003}.

This paper is organized as follows. In Sect.~\ref{sec:stateprep} we introduce the time evolution method and the LCU~\cite{childs2012} to generate the excited state of Eq.~\eqref{eq:excit_state}. Section~\ref{sec:res} presents the results for both methods, obtained from simulations and from running on quantum hardware. In Sect.~\ref{sec:npdg} we apply the methods to the $n(p,d)\gamma$ reaction. We present our conclusions in Sect.~\ref{sec:concl}. Several technical details can be found in the Appendices.

\section{State preparation}
\label{sec:stateprep}
In this Section we present two procedures to perform the state preparation in Eq.~\eqref{eq:excit_state}. In both cases we start with a quantum register already initialised in the target quantum state $\ket{\Psi_0}$. For applications such as computing scattering cross-sections, this state is the ground state of a many-body Hamiltonian that describes the nuclear target (but could also be a Gibbs state for applications at finite temperature). We note that the procedures we propose in this paper do not make use of any particular property of this state and can therefore be applied in general.

As with most numerical algorithms, both classical and quantum, the state-preparation techniques we propose 
are designed to be correct up to some finite error tolerance. Our goal is to find quantum procedures that produce an approximation $\ket{\Phi_A}$ to the target state $\ket{\Phi_E}$ of Eq.~\eqref{eq:excit_state} with a guaranteed fidelity $F=\lvert\langle\Phi_E\vert\Phi_A\rangle\rvert^2$ and a success probability $P_s$.
The efficiency of different algorithms is then related to the number of quantum operations needed to guarantee the desired ``precision'' metrics $F$ and $P_s$.

When such algorithms are executed on NISQ devices without any form of error correction, the average fidelity of operations implemented on the quantum device provide effectively an upper bound $f_U$ on the achievable state fidelity $F$. To give an example, let us assume that $f_q$ is the average fidelity of a $q$-qubit gate operation. For a circuit consisting of $n_1$ 1-qubit and $n_2$ 2-qubit gates we thus estimate the maximum expected circuit fidelity as $f_U\equiv f_1^{n_1}f_2^{n_2}$. For practical purposes the target state fidelity fulfills then $F\approx f_U$. However,  it is still desirable to devise schemes that are able to reach fidelities $F$ arbitrarily close to one in an efficient way as they would be useful on a fully error-corrected machine.

To quantify fidelities and success probabilities, and to discuss the generation of excited states, we decompose the excitation operator $O$ in terms of a linear combination of unitaries as
\begin{equation}
\label{eq:ex_op_lcu}
O = \sum_{k=0}^L \lambda_k U_k\quad\lambda_k>0\;.
\end{equation}
It is also convenient for later use to introduce the 1-norm of the coefficient vector as
\begin{equation}
\label{eq:lambda_norm}
\Lambda = \sum_{k=0}^L \lambda_k \geq \|O\|\;,
\end{equation}
with $\|\cdot\|$ the operator spectral norm.
As we will see below, one can use the LCU algorithm~\cite{childs2012} to prepare the excited state $\ket{\Phi_E}$ with success probability $P_s\propto1/\Lambda^2$.

\subsection{Time-dependent method}
\label{subsec:tdm}

\begin{figure}[tbh]
 \centering
 \includegraphics[width=0.49\textwidth]{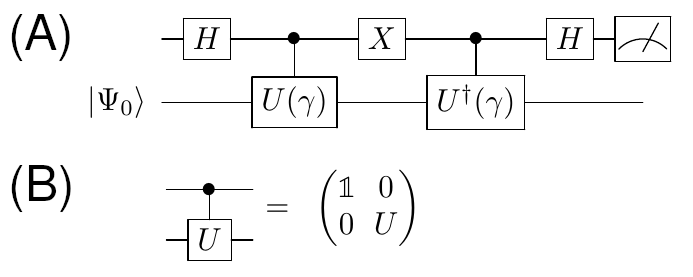}
 \caption{Quantum circuits defined in the text: (A) state preparation circuit and (B) generic controlled unitary.}
\label{fig:circuitA}
\end{figure}

The first state preparation algorithm we consider was proposed in Ref.~\cite{roggero2019}. Here, we will also analyze its efficiency. The main idea is to exploit the time-evolution operator 
\begin{equation}
\label{eq:timevop}
U(\gamma) = \exp\left(-i \gamma O\right) = \cos\left(\gamma O\right)-i\sin\left(\gamma O\right)\;,
\end{equation}
associated with the excitation operator $O$ 
to produce the approximate state
\begin{equation}
\ket{\Psi_A(\gamma)}\propto \sin\left(\gamma O\right)\ket{\Psi_0} = \ket{\Phi_E}+\mathcal{O}(\gamma^2)\;.
\end{equation}
Here $\gamma>0$ is a ``time'' parameter. Using the unitary $U(\gamma)$, controlled by an auxiliary qubit or 'ancilla', we perform this operation with the circuit shown in Fig.~\ref{fig:circuitA}(A).
Here and in what follows we use the convention that if no initial state is specified on the left hand side, then this defines the unitary operator corresponding to the circuit independently on the initial state.
We also denote the Pauli matrices $\sigma_x$, $\sigma_y$ and $\sigma_z$ compactly as $X$, $Y$ and $Z$. The gate denoted by $H$ is the Hadamard gate, defined as
\begin{equation}
\label{eq:Hadamard}
H = \frac{1}{\sqrt{2}}\begin{pmatrix}
1&1\\
1&-1\\
\end{pmatrix}\;,
\end{equation}
while the meter at the end of the ancilla line represents a projective measurement of the ancilla qubit polarization. Finally, the controlled-$U$ operations are defined on the total Hilbert space as shown in Fig.~\ref{fig:circuitA}(B).
Thus, they apply the unitary $U$ to the qubit register containing the state $\ket{\Psi_0}$ when the ancilla qubit is in $\ket{1}$ and the identity operator $\mathbb{1}$ otherwise.
It is convenient also to express the unitary operation depicted in the circuit from Fig.~\ref{fig:circuitA}(A) explicitly in the relevant two-dimensional space spanned by the states $\ket{0}\otimes \ket{\Psi_0}$ and $\ket{1}\otimes\ket{\Psi_0}$ as 
\begin{equation}
\label{eq:tdm_unitary}
V(\gamma)=\begin{pmatrix}
\cos\left(\gamma O\right)& i\sin\left(\gamma O\right)\\
-i\sin\left(\gamma O\right)& -\cos\left(\gamma O\right)
\end{pmatrix}\; .
\end{equation}
This is useful because the full evolution is contained in this subspace. Note that this operation differs slightly from the one proposed in Ref.~\cite{roggero2019} and it is simpler to implement.

Preparing the ancilla qubit in the state $\ket{0}$ and applying the circuit above [or equivalently the unitary matrix~\eqref{eq:tdm_unitary}]  produces the final state
\begin{equation}
\ket{\Omega(\gamma)}\!=\!\ket{0}\!\otimes\!\cos\left(\gamma O\right)\ket{\Psi_0}\!-\!i\ket{1}\!\otimes \!\sin\left(\gamma O\right)\ket{\Psi_0}.
\label{state1}
\end{equation}
We can now use the projective measurement on the ancilla qubit displayed in the circuit in Fig.~\ref{fig:circuitA}(A) to select only the second component of the state in Eq.~\eqref{state1}. Thus, if the measurement finds the ancilla in the state $\ket{1}$ the system register will be left in the state
\begin{equation}
\label{eq:tdm_state}
\ket{\Psi_A(\gamma)} = \frac{-i}{\sqrt{\langle\Psi_0\lvert \sin^2(\gamma O) \rvert\Psi_0\rangle}}\sin(\gamma O) \rvert\Psi_0\rangle\;.
\end{equation}
In case that we measure the ancilla in $\ket{0}$ we restart the entire process from scratch and try again. Due to the presence of a measurement, this procedure is inherently stochastic and the success probability is
\begin{equation}
\label{eq:psucc_tdm}
P_s=\bra{\Psi_0}\sin^2(\gamma O)\ket{\Psi_0}=\mathcal{O}\left(\gamma^2\right)\;.
\end{equation}

It follows from Eqs.~\eqref{eq:tdm_state} and \eqref{eq:psucc_tdm} that  a small error in the state preparation requires a small time extent $\gamma$;  however, small values of $\gamma$ will also lead to a small success probability. This tension between our goals suggests that the time-evolution technique might be inefficient when a small infidelity $\Delta_f=1-F$ and a large success probability are sought. In what follows we further explore this tension and present an appropriate compromise.

Let us constrain the parameter $\gamma$ to small values, as measured by the relevant operator norm $\Lambda$ [see Eq.~\eqref{eq:lambda_norm}], and use  $\gamma\in[0,\pi/2\Lambda]$. With this constraint we derive the bounds for the success probability
\begin{equation}
\label{eq:bound_prob}
\min\left[\sin^2\left(\gamma\Lambda\right),\gamma^2\eta^2\right]\geq P_s \geq \gamma^2\eta^2 \left(1-\frac{\gamma^2\Lambda^2}{3}\right)\;,
\end{equation}
and for the state fidelity
\begin{equation}
\begin{split}
F(\gamma) &= \left|\langle\Psi_E\vert\Psi_A(\gamma)\rangle\right|^2\\ &=\frac{1}{\eta^2}\frac{\left|\langle\Psi_0\lvert O \sin(\gamma O) \rvert\Psi_0\rangle\right|^2}{\langle\Psi_0\lvert \sin^2(\gamma O) \rvert\Psi_0\rangle}\\
&\geq \left(1-\frac{\gamma^2\Lambda^2}{6}\right)\;.
\end{split}
\end{equation}
Details of the derivation are presented in Appendix~\ref{app:method0_bounds}. 
Based on these bounds a minimum fidelity $F_{\rm min}$ for the state preparation results when the time extent $\gamma$ fulfills
\begin{equation}
\label{eq:gamma_bound}
\gamma \leq \frac{\sqrt{6\left(1-F_{\rm min}\right)}}{\Lambda} \ .
\end{equation}
This can be easily found by noticing that if we want to ensure $F(\gamma)\geq F_{\rm min}$ it is sufficient to choose $\gamma$ so that $\left(1-\frac{\gamma^2\Lambda^2}{6}\right)\geq F_{\rm min}$.
This yields a success probability bounded from above by
\begin{equation}
\label{eq:opt_ps}
P_s \leq \sin^2\left(\sqrt{6\left(1-F_{\rm min}\right)}\right)\approx 6\left(1-F_{\rm min}\right)\;,
\end{equation}
in the limit of small infidelities $1-F_{\rm min}\ll 1$.
Thus, the success probability $P_s$ is proportional to the infidelity $\Delta_f=1-F_{\rm min}$. We note that this result does not depend on the norm $\eta$ from Eq.~\eqref{eq:excit_state}. Similar bounds with the full $\eta$ dependence can  also be obtained directly from Eq.~\eqref{eq:bound_prob}. 

Especially important is the lower bound which guarantees a success probability of at least
\begin{equation}
P_s \geq \frac{\eta^2}{\Lambda^2}\left(2F_{\rm min}-1\right)\;,
\end{equation}
for any target fidelity $F_{\rm min}\geq(1-\pi^2/24)\approx 0.59$. This lower bound on $F_{\rm min}$ is based on $\gamma\leq\pi/2\Lambda$. 

So far the estimates for the efficiency rely on the availability of an {\it exact} implementation of the time-evolution unitary $U(\gamma)$ of Eq.~\eqref{eq:timevop}. In practice this is not a realistic assumption, and instead one has to make use of an approximate unitary operation $\widetilde{U}(\gamma)$ accurate only up to some additive error $\delta_U$. Given the importance of such approximations we need to consider these next.

\subsubsection{Imperfect time evolution}
\label{subsub:imperfect}

Let us assume that we only have an approximation to the time evolution operator $U(\gamma)$ and that the error is $\delta_U$. Using this approximate operation in the full circuit from Fig.~\ref{fig:circuitA}(A) generates a unitary operation $\widetilde{V}(\gamma)$. We have 
\begin{equation}
\label{eq:errV}
\|\widetilde{V}(\gamma)-V(\gamma)\|\leq \delta_V\;,
\end{equation}
and a total error $\delta_V\leq2\delta_U$. In Eq.~\eqref{eq:errV} we denoted the spectral norm of the operator by the double bars. The final state before the ancilla measurement is given by
\begin{equation}
\begin{split}
\label{eq:noisy_exp}
\ket{\widetilde{\Omega}(\gamma)}&=\widetilde{V}(\gamma)\ket{0}\otimes\ket{\Psi_0}\\
&=\cos(\alpha)\ket{\Omega(\gamma)} + \sin(\alpha)\ket{\xi(\gamma)}\;.
\end{split}
\end{equation}
Here we introduced in the second line the normalized state $\ket{\xi(\gamma)}$ which is orthogonal to the error-free vector $\ket{\Omega(\gamma)}=V(\gamma)\ket{0}\otimes\ket{\Psi_0}$. Using the inequality~\eqref{eq:errV} we find that the overlap with the wanted state is at least
\begin{equation}
\label{eq:alphabound}
\cos(\alpha)\geq1-\frac{\delta_V^2}{2}\;.
\end{equation}
A more detailed derivation of this result is presented in the Appendix~\ref{app:faultytimeev}. The new approximate state can now be obtained by postselection on the ancilla being in $\ket{1}$.  It reads
\begin{equation}
\ket{1}\otimes\ket{\widetilde{\Psi_A}(\gamma)}=\frac{1}{\sqrt{\widetilde{P_s}}}\Pi_1\ket{\widetilde{\Omega}(\gamma)}\;.
\end{equation}
Here we have introduced the projector $\Pi_1=\rvert1\rangle\langle1\lvert\otimes\mathbb{1}$, and the new success probability $\widetilde{P_s}$ satisfies 
\begin{equation}
\left|\widetilde{P_{s}}-P_{s}\right|\leq\left|\sin(\alpha)\right|\;.
\end{equation}
The angle $\alpha$ is defined in Eq.~\eqref{eq:noisy_exp}, and we again refer to the  Appendix~\ref{app:faultytimeev} for the derivation. Using Eq.~\eqref{eq:alphabound} we now find that the approximate success probability $\widetilde{P_s}$ is close to $P_s$, because
\begin{equation}
\label{eq:boundpstilde}
P_s-\delta_V\sqrt{1-\frac{\delta_V^2}{4}}\leq \widetilde{P_s}\leq P_s+\delta_V\sqrt{1-\frac{\delta_V^2}{4}}\;.
\end{equation}
One can use the bounds on the success probability from Eq.~\eqref{eq:bound_prob} to find the corresponding bounds for $\widetilde{P_s}$. For the fidelity $\widetilde{F}$ we follow a similar approach and find
\begin{equation}
\begin{split}
\widetilde{F} &\geq \frac{P_s F -\delta_V\sqrt{1-\delta_V^{2}/4}}{\widetilde{P_s}}\\
&\geq\left(1-\frac{\gamma^2\Lambda^2}{2}\right)-\left(2-\frac{\gamma^2\Lambda^2}{3}\right)\frac{\delta_V}{\gamma^2\eta^2}\;.
\end{split}
\end{equation}
In the last line we used both Eqs.~\eqref{eq:bound_prob} and \eqref{eq:boundpstilde} to eliminate the explicit dependence on the success probabilities. 

To guarantee a fidelity of at least $\widetilde{F}_{\rm min}$ we can now take
\begin{equation}
\label{eq:gammabound2}
\gamma\leq\frac{\sqrt{1-F_{\rm min}}}{\Lambda}\;,
\end{equation}
for the time step. This ensures that the approximation error induced by the approximate time-evolution is bounded by
\begin{equation}
\label{eq:deltabound}
\delta_V\leq\frac{\eta^2}{4}\gamma^4\eta^2\Lambda^2\leq\frac{\eta^2}{2\Lambda^2}\left(1-F_{\rm min}\right)^2\;.
\end{equation}
The error $\delta_U$ for the time-evolution unitary is
\begin{equation}
\label{eq:deltaboundU}
\delta_U(\gamma)\leq\frac{\eta^2}{8}\gamma^4\eta^2\Lambda^2\leq\frac{\eta^2}{4\Lambda^2}\left(1-F_{\rm min}\right)^2\;.
\end{equation}
This last inequality suggests that even for small time intervals $\gamma$ higher-order Trotter-Suzuki decompositions are preferred thanks to their better error scaling. Alternatively, one should use error-efficient approximation schemes such as the Taylor expansion technique from~\cite{berry2015} or quantum signal processing~\cite{low2017} that allow the computational cost to scale as $\mathcal{O}(\log(1/\delta))$.

\subsection{LCU-based method}
\label{sec:lcu-method}

The operation in Eq.~\eqref{eq:excit_state} can be performed in an exact way with the LCU technique~\cite{childs2012}. Here it is important that the decomposition in Eq.~\eqref{eq:ex_op_lcu} of the excitation operator employs an efficient implementation of unitaries $U_k$ and that the total number of terms, $L+1$, grows as a low-order polynomial with increasing system size.

\begin{figure}[tbh]
 \centering
 \includegraphics[width=0.49\textwidth]{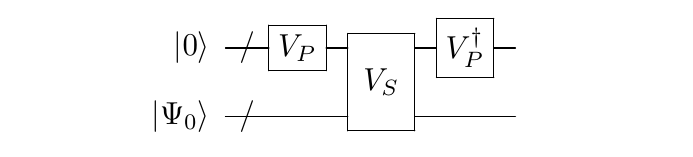}
 \caption{Linear combination of unitaries (LCU) circuit diagram.}
\label{fig:circuitB}
\end{figure}

The procedure requires the implementation of two unitaries. The first one is the {\it prepare} unitary
\begin{equation}
\label{eq:prepare}
V_P \rvert0\rangle = \sum_{k=0}^L\sqrt{\frac{\lambda_k}{\Lambda}}\ket{k}\;,
\end{equation}
acting on an ancilla register of size $M=\left\lceil \log_2(L+1)\right\rceil$ initialized in the reference state $\ket{0}$.
The second unitary operation is the {\it select} unitary
\begin{equation}
\label{eq:select}
V_S = \sum_{k=0}^L \rvert k\rangle\langle k\lvert\otimes U_k \ .
\end{equation}
Here the projectors $\ket{k}\bra{k}$ act on the ancilla register. We follow convention and assume that $U_0=\mathbb{1}$. This simplifies the execution of this unitary controlled by an ancilla qubit. We then execute the circuit displayed in Fig.~\ref{fig:circuitB}
whose corresponding unitary we denote by $W$. The final state of this operation can be decomposed as (see, e.g., Refs.~\cite{childs2012,childs2017})
\begin{equation}
\label{eq:fstate_LCU}
\ket{\Omega} = W \rvert0\rangle\otimes\ket{\Psi_0} = \frac{1}{\Lambda}\rvert0\rangle\otimes O\ket{\Psi_0} + \ket{\Phi^\perp} \ .
\end{equation}
Here $\ket{\Phi^\perp}$ is orthogonal to the $\rvert0\rangle$ state of the ancilla register
\begin{equation}
\left(\rvert0\rangle\langle0\rvert\otimes\mathbb{1}\right)\ket{\Phi^\perp} = 0\;.
\end{equation}
Thus, we can isolate the first component of $\ket{\Omega}$ in Eq.~\eqref{eq:fstate_LCU} by measuring the ancilla register and discard the experiment if we do not measure $\ket{0}$. This post-selection allows us to obtain the (not normalized) state
\begin{equation}
\label{eq:lcu_state_prep}
\left(\rvert0\rangle\langle0\rvert\otimes\mathbb{1}\right)\ket{\Omega} = \frac{1}{\Lambda} \ket{0}\otimes\ket{\Psi_0} = \frac{\eta}{\Lambda}\ket{0}\otimes\ket{\Phi_E}\;,
\end{equation}
with a finite success probability
\begin{equation}
\label{eq:lcu_prob}
P_s^{\rm LCU} = \frac{\eta^2}{\Lambda^2}\;.
\end{equation}

From an operational standpoint we can use this strategy to compute expectation values over the excited state $\ket{\Phi_E}$ by first estimating empirically the success probability $P_s^{\rm LCU}$ and then consider the estimator
\begin{equation}
\langle \Phi_E\lvert A\rvert\Phi_E\rangle = \frac{\langle \Omega\lvert\mathbb{1}\otimes A\rvert\Omega\rangle}{P_s^{\rm LCU}}\;.
\end{equation}

Comparing the result presented in Eq.~\eqref{eq:lcu_prob} with the bounds obtained for the time-dependent method in Eqs.~\eqref{eq:bound_prob} and \eqref{eq:gamma_bound} shows that the LCU-based method has a higher success probability (but not necessarily shorter depth)  whenever the target infidelity satisfies $\Delta_f\leq1/6$.

\section{Results}
\label{sec:res}

In this Section, we analyze the performance of the two approaches by using a simple excitation operator. 
To quantify the quality of the results of the quantum computation and to assess the efficacy of our error mitigation procedure, we introduce two different metrics. Let us consider a set of $N_O$ observables (e.g. the success probabilities for different excitation operators). We use $v^{(t)}_k$ to denote the expected theoretical value for the $k$th observable, and  $v^{(e)}_k$ and $\varepsilon^{(e)}_k$, respectively, its experimental value and estimated error. Throughout this paper we use two quality metrics. First, the chi squared
\begin{equation}
    \chi^2 = \sum_{k=1}^{N_O} \frac{\left(v^{(e)}_k-v^{(t)}_k\right)^2}{\left(\varepsilon^{(e)}_k\right)^2}\;,
\end{equation}
quantifies if the data is compatible with expected results within errors. We regard the results which are more close to a value of one as better.

Second, the normalized sum of squared deviations
    \begin{equation}
    {\rm nssd}(r) = \sqrt{\frac{\sum_{k=1}^{N_O}\left(v^{(e)}_k-v^{(t)}_k\right)^2}{\sum_{k=1}^{N_O}\left(r v^{(t)}_k\right)^2}}\;,
    \end{equation}
quantifies the accuracy of the calculation. The numerator is the 2-norm of the residuals while the denominator is the 2-norm of residuals corresponding to a relative error $r$. Unless stated otherwise, we will use $r=0.1$ and drop the argument, i.e. ${\rm nssd}\equiv {\rm nnsd}(0.1)$. We expect ${\rm nssd}$ to be close to zero. In other words, ${\rm nssd}\ll 1$ indicates that the relative error is much smaller than $10\%$.  To summarize, the $\chi^2$ is used to assess the degree of compatibility with the expected result, while the ${\rm nnsd}$ is used to quantify the accuracy.

\subsection{Time-dependent method}
\label{sec:res_m0A}
To explore the efficiency of the first method we use $\theta\in[0,\pi]$ for the simple excitation operator
\begin{equation}
\label{eq:m0_eo}
O(\theta) = \cos(\theta)X + \sin(\theta) \mathbb{1}\, ,
\end{equation}
which acts on a single qubit. In this basis we identify $\ket{\Psi_0}=\ket{0}$ with the initial state of the reaction and $\ket{1}$ as the final state. 
The time evolution operator for this excitation operator is a rotation around the qubit's $X$ axis and can be implemented exactly as
\begin{equation}
\label{eq:op1}
e^{-i\gamma O}=e^{-i\gamma\sin(\theta)}R_X\left[2\gamma\cos(\theta)\right]\;,
\end{equation}
with $R_X$ being a rotation around the $X$ direction (see Appendix~\ref{app:gate} for the explicit definition).

Based on this result the full state preparation circuit from Fig.~\ref{fig:circuitA}(A) can be implemented using four CNOT gates, four $Z$-rotations and five additional  single-qubit gates (see Appendix~\ref{Cir: method 0 simple} for a full derivation). We note that this CNOT count is suboptimal, because a generic 2-qubit unitary only requires three CNOT gates~\cite{vidal2004,vatan2004}. However, such a circuit comes at the price of needing $15$ single-qubit rotations.

\begin{figure}[tbh]
 \centering
 \includegraphics[width=0.49\textwidth]{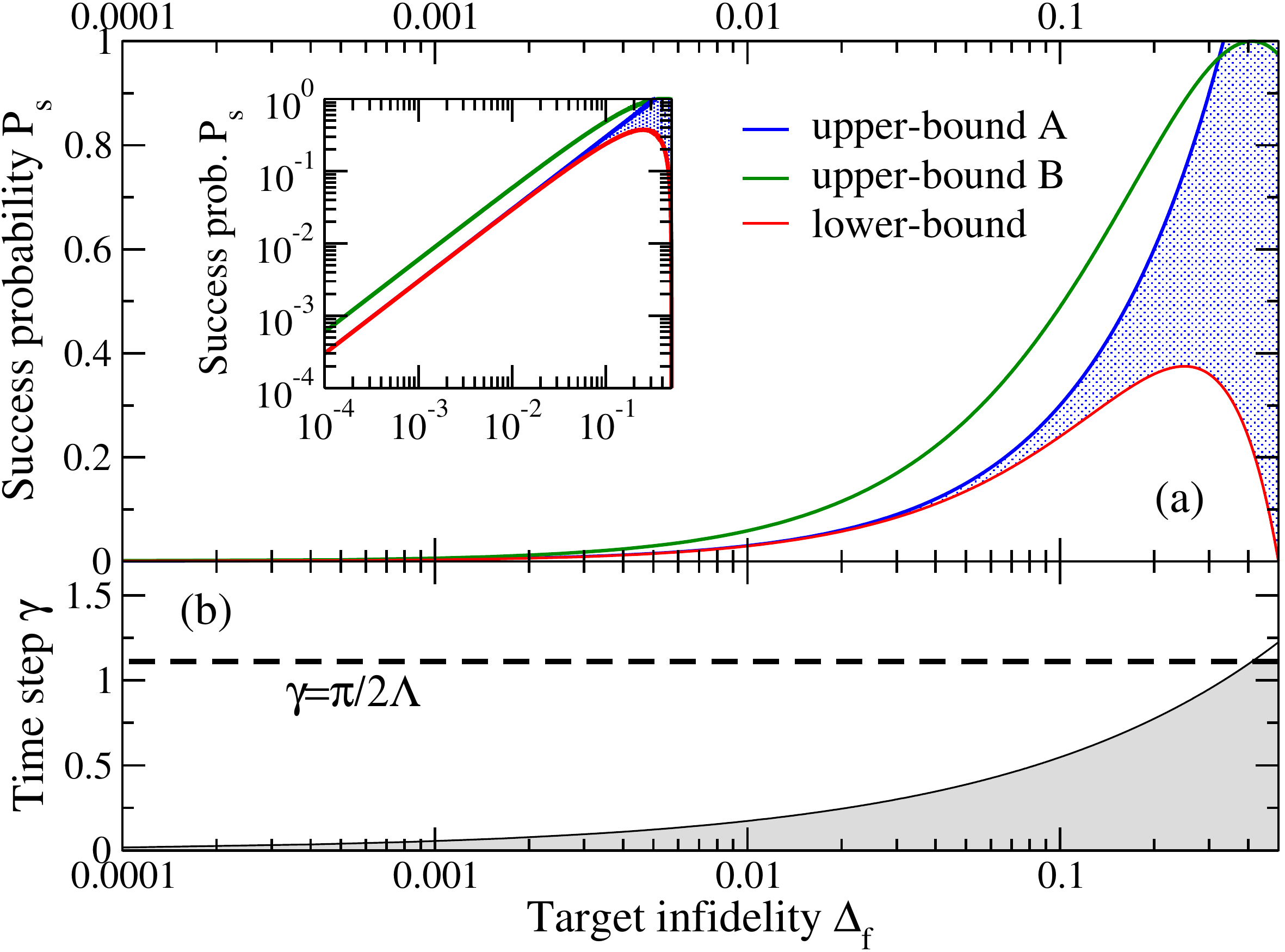}
 \caption{(Color online) Estimated bounds for both the success probability $P_s$ (a) and the maximum time-step parameter $\gamma$ (b) as a function of the target infidelity $\Delta_f$. The inset in the top panel shows more details for small values of $\Delta_f$.}
\label{fig:expected_fidelity}
\end{figure}

For the operator Eq.~\eqref{eq:m0_eo} the excited state norm from Eq.~\eqref{eq:excit_state} simply is $\eta=1$ for every angle, and this will facilitate the analysis of the results. In addition we also have a simple bound for this operator's  norm, as
\begin{equation}
\label{eq:lambda_theta}
\Lambda(\theta) = |\sin(\theta)| + |\cos(\theta)| \leq \sqrt{2} \equiv \Lambda_{\rm max}\;.
\end{equation}
In what follows we will use the simple bound $\Lambda_{\rm max}$ instead of $\Lambda(\theta)$. This allows us to simplify the presentation, because we can choose a $\theta$-independent value for $\gamma$ in Eq.~(\ref{eq:op1}). Using the bounds derived in the previous Section we can choose a sensible value for the ``time'' parameter $\gamma$ as a function of the target maximum infidelity $\Delta_f$. From Eq.~\eqref{eq:gamma_bound} we find $\gamma\leq\sqrt{3\Delta_f}$, and the corresponding allowed values of $\gamma$ are depicted in the lower panel of Fig.~\ref{fig:expected_fidelity} as the gray region.

In the top part of Fig.~\ref{fig:expected_fidelity} we show the corresponding bounds on the success probability $P_s$, see Eq.~\eqref{eq:bound_prob}. The two upper bounds correspond to the arguments on the left-hand side of Eq.~\eqref{eq:bound_prob}, i.e. the upper-bound A is $\sin^2(\gamma\Lambda)$ and the upper-bound B is $\gamma^2\eta^2$. The lower bound is from the right-hand side of Eq.~\eqref{eq:bound_prob}. For our excitation operator Eq.~\eqref{eq:m0_eo} we have $\eta=1$ and $\Lambda>1$, and the upper bound A is tighter for most values of $\Delta_f$. Based on these results we set $\gamma=0.3$ because this choice will guarantee a high fidelity $F(0.3)\geq97\%$ while keeping the success probability (and therefore the effective sample size) sufficiently large. We have  
\begin{equation}
\label{eq:ps_bound_method0}
9\%\geq P_s(0.3) \geq 8.46\%\;,
\end{equation}
and better bounds could have been obtained using the tighter bound $\Lambda(\theta)$ instead of $\Lambda_{\rm max}$.

\begin{figure}
    \centering
    \includegraphics[width=0.25\textwidth]{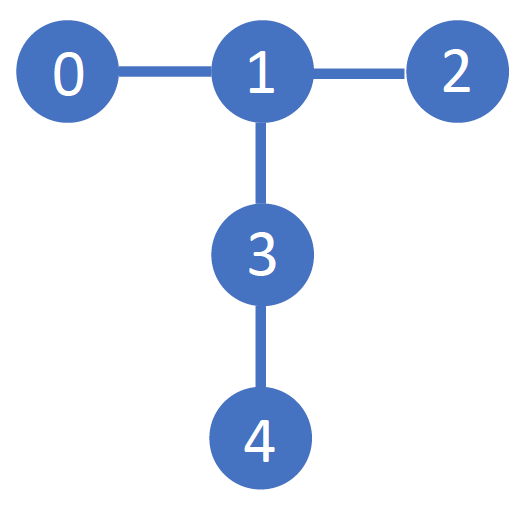}
    \caption{(Color online) Layout of the IBM quantum device Vigo~\cite{IBMQ_Vigo}. Shown are the qubits, labeled from 0 to 4, and their connectivity.}
    \label{fig:vigo}
\end{figure}

We used Qiskit~\cite{qiskit} and implemented the relevant circuits on the IBM quantum device Vigo~\cite{IBMQ_Vigo}.
The layout of this quantum chip is shown in Fig.~\ref{fig:vigo}. Our quantum computation used qubits 1 and 2 of Vigo, and we employed both, the virtual machine (VM) and the quantum processor unit (QPU).

\begin{figure}
 \centering
 \includegraphics[width=0.49\textwidth]{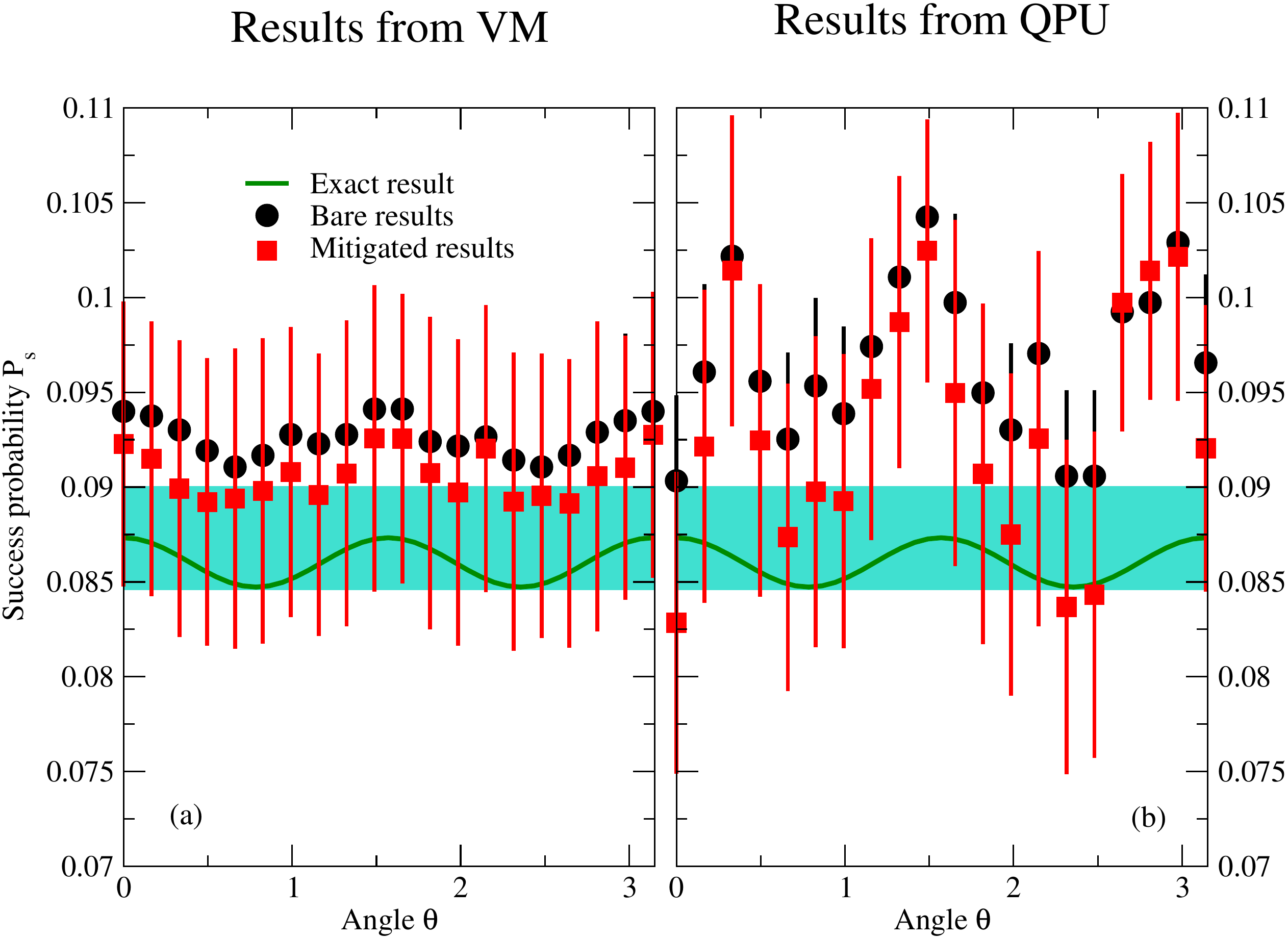}
 \caption{(Color online) Results on the success probability $P_s$ for time dependent method with the VM run (a) and QPU run (b) and exact analysis (green line), as function of angle $\theta$ in excitation operator Eq.~\eqref{eq:m0_eo}. Results are given with (red squares) and without (black squares) full mitigation. This convention apply to all the plots in this paper. The turquoise band is the bound Eq.~\eqref{eq:ps_bound_method0}.}
\label{fig:method0_ps_vm_vs_qpu}
\end{figure}

We present in Fig.~\ref{fig:method0_ps_vm_vs_qpu} the results obtained with the VM and the QPU, and compare them to the exact results from Eq.~\eqref{eq:psucc_tdm}. Note that the VM simulations use a noise model that is initialized from calibration data coming directly from the device.

The bare results are directly collected from runs on the VM or QPU. The mitigated results are obtained after applying an error mitigation procedure described below. We perform two types of error mitigation: a read-out correction on the measured distributions and a noise extrapolation assuming that the dominant noise channel is the one associated with the execution of entangling two-qubit gates like CNOT (see Refs.~\cite{Dumitrescu2018,li2017,Endo2018}).  The detailed error mitigation scheme we use is explained in Appendix~\ref{app:err_mitigation}. We see that the errors are strongly correlated, i.e. not random on the VM, and that error mitigation moves the bare results toward the benchmark. On the QPU, noise is significant, and error mitigation only somewhat improves the results. 

We present in Tab.~\ref{tab:ps_tdep} the quality metrics for the results on the success probability shown in Fig.~\ref{fig:method0_ps_vm_vs_qpu}. Here, we listed the metrics for the bare result, read-out (RO) error mitigated results, and fully mitigated results (using readout correction and CNOT extrapolation). On average the results for running on the QPU differ from the exact results by about 10\% (as ${\rm nssd}\approx 1$), and apparently the error extrapolation technique is not successful in reducing this discrepancy. In contrast, we can see from the results in Tab.~\ref{tab:ps_tdep} that error mitigation improves the accuracy on the VM. The reason for this apparent discrepancy can be understood by noticing that the value of the $\chi^2$ for the QPU calculation is reduced by a factor of two as a result of the error extrapolation. This suggests that, even though the central values do not improve considerably, the dispersion of results is reduced by the error extrapolation step while keeping the results compatible with the reference values.

\begin{table}
\subtable[VM]{
\begin{tabular}{l|r|r}
 & $\chi^2$ & ${\rm  nssd}$ \\ \hline
bare & 2.08 & 0.765 \\
RO mit. & 1.46 & 0.656 \\
full mit. & 0.37 & 0.539 
\end{tabular}
}
\subtable[QPU]{
\begin{tabular}{l|r|r}
 & $\chi^2$ & ${\rm nssd}$ \\ \hline
bare & 5.64 & 1.299  \\
RO mit. & 3.00 & 1.004 \\
full mit. & 1.47 & 1.052 
\end{tabular}
}
\caption{Quality metrics for the success probability $P_s$ obtained (a) on the VM run and (b) from the QPU run. Shown are bare results, and with error correction for read out, and full error mitigation. \label{tab:ps_tdep}}
\end{table}

\begin{figure}
 \centering
 \includegraphics[width=0.49\textwidth]{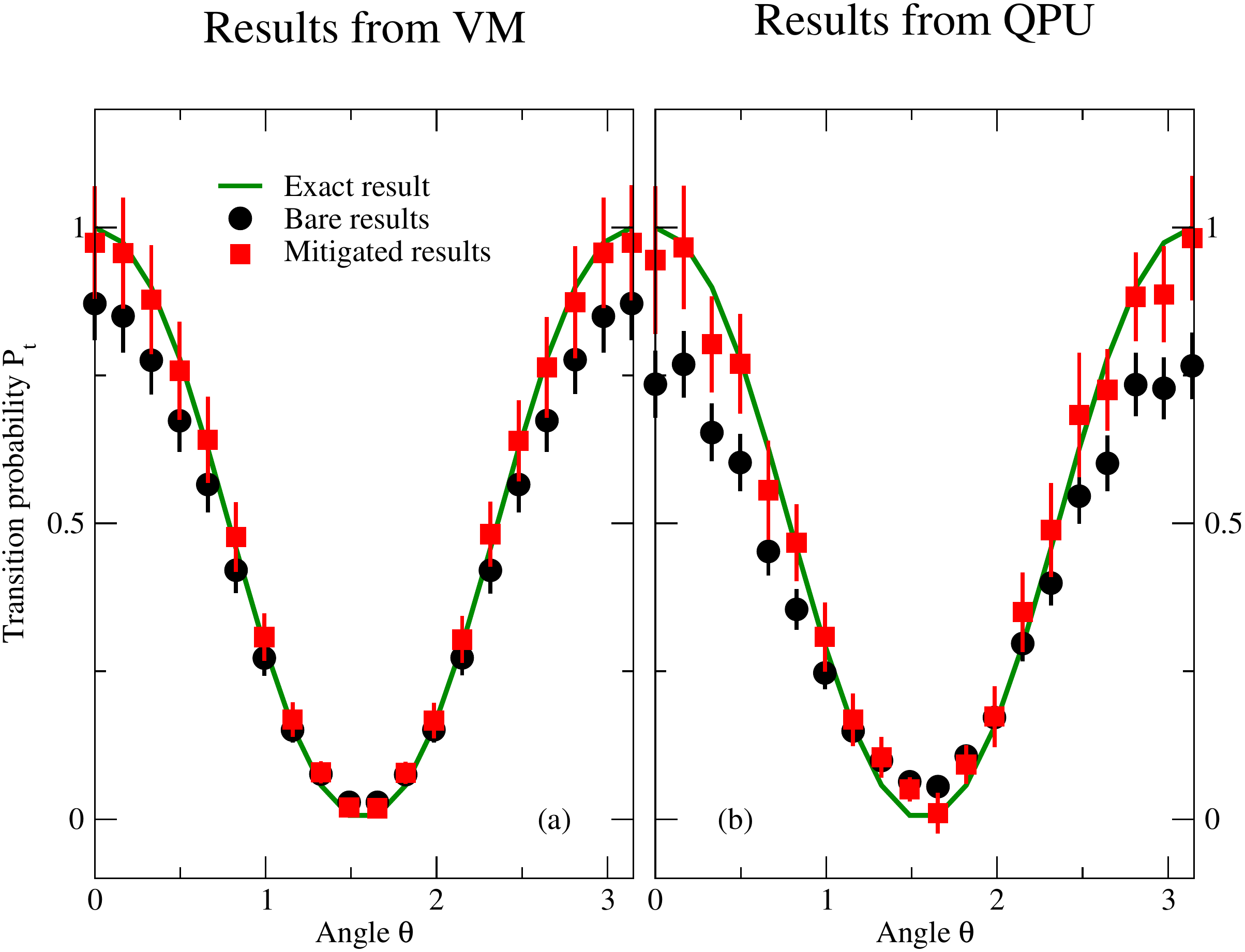}
 \caption{(Color online) Results on the transition probability $P_t$ for time dependent method with the VM run (a) and QPU run (b) and exact analysis (green line). Results are given with (red squares) and without (black squares) full mitigation.}
\label{fig:method0_matel_vm_vs_qpu}
\end{figure}

To assess the quality of the prepared state $\ket{\Psi_A}$ from Eq.~\eqref{eq:tdm_state} we directly measure its projection $P_t$ onto the final state $\ket{1}$ and compare it with the expected result. We have for the transition probability
\begin{equation}
\label{eq:p1_tdm}
\begin{split}
P_t &\equiv \left|\langle\Psi_A\vert1\rangle\right|^2\\
&= \frac{\left|\langle 00 \lvert U(\gamma)\rvert11\rangle\right|^2 }{\left|\langle 00 \lvert U(\gamma)\rvert11\rangle\right|^2+\left|\langle 00 \lvert U(\gamma)\rvert01\rangle\right|^2}\\
&=\frac{\left|\langle 00 \lvert U(\gamma)\rvert11\rangle\right|^2 }{P_s(\gamma)} \ .
\end{split}
\end{equation}
In the second line we re-introduced the ancilla state used in the implementation of the unitary $U(\gamma)$ from Eq.~\eqref{eq:tdm_unitary}.

We show the results for $P_t$ in Fig.~\ref{fig:method0_matel_vm_vs_qpu}.
The left and right panels correspond to the results from the VM and QPU, respectively. The exact results are calculated from Eq.~\eqref{eq:p1_tdm}. The corresponding quality metrics are shown in  Tab.~\ref{tab:matel_tdep}. On average the transition probability for running on real QPU differs from the exact results only by $7\%$, and $\chi^2 = 0.65$ shows that $P_t$ it is compatible with the exact results within the uncertainties. Even though not as complete as a full state tomography, we can use these results as strong indicators that the time dependent method runs with reasonable efficiency on a real quantum computer. We also note that the error mitigation work reasonably well as the $\chi^2$ and ${\rm nssd}$ are decreased.

\begin{table}[t]
\subtable[VM]{
\begin{tabular}{l|r|r}
 & $\chi^2$ & ${\rm nssd}$ \\ \hline
bare & 2.80 & 1.259 \\
RO mit. & 0.87 & 0.757 \\
full mit. & 0.39 & 0.386
\end{tabular}
}
\subtable[QPU]{
\begin{tabular}{l|r|r}
 & $\chi^2$ & ${\rm nssd}$ \\ \hline
bare & 11.27 & 2.324  \\
RO mit. & 3.19 & 1.059 \\
full mit. & 0.65 & 0.718 
\end{tabular}
}
\caption{Quality metrics for the transition probability obtained (a) on the VM run and (b) from the QPU run.\label{tab:matel_tdep}}
\end{table}

\subsection{LCU-based method}
\label{sec:lcuresults}
We now discuss the results based on the state preparation using the LCU technique described in Sec.~\ref{sec:lcu-method}. The single-qubit excitation operator Eq.~\eqref{eq:m0_eo} can easily be implemented with this method. The excitation operator is a sum of two unitaries, and we only need $M=1$ ancilla in addition to the qubit to represent the system. For every angle $\theta\in[0,\pi]$ the {\it prepare} unitary from Eq.~\eqref{eq:prepare} can be implemented using a single rotation $R_y$ around the $Y$ axis (cf. Appendix~\ref{app:gate}), and the resulting state is
\begin{equation}
\ket{\Phi_1} = R_y(\phi_1) \ket{0} = \cos{\phi_1 \over 2}\ket{0} +\sin{\phi_1 \over 2}\ket{1}\;. 
\end{equation}
\begin{figure}[tbh]
 \centering
 \includegraphics[width=0.49\textwidth]{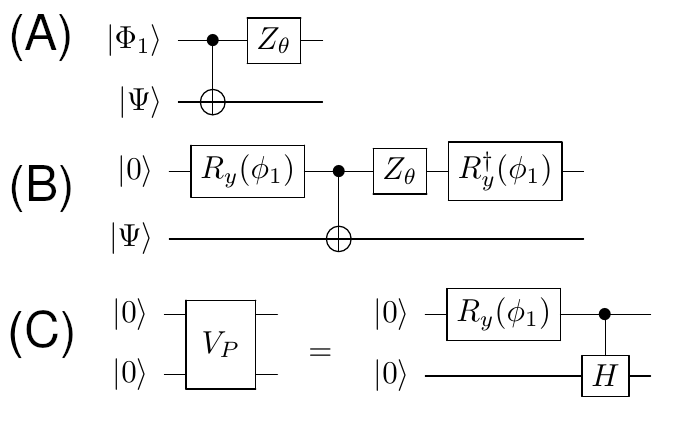}
 \caption{Explicit circuit construction for the unitaries defined in the text: (A) is the {\it select} unitary, (B) the full LCU circuit for the excitation in Eq.~\eqref{eq:m0_eo}, and (C) the {\it prepare} unitary in second quantization, Eq.~\eqref{eq:phi2_state}.
}
\label{fig:circuitC}
\end{figure}

Here, the angle is 
\begin{equation}
\label{eq:phi1_angle}
\phi_1 = 2\arcsin\left(\sqrt{\frac{\lvert\cos(\theta)\rvert}{\lvert\cos(\theta)\rvert+\sin(\theta)}}\right)\;.
\end{equation}

Given the state $\ket{\Phi_1}$, the {\it select} unitary from Eq.~\eqref{eq:select} can be implemented with the circuit displayed in Fig.~\ref{fig:circuitC}(A), 
where $Z_\theta$ is the unitary whose matrix representation in the computational basis is
\begin{equation}
\label{eq:ztheta}
\begin{split}
Z_\theta &\equiv \bigg\{\begin{matrix} \mathbb{1}&\text{for}\quad\theta\in[0,\pi/2]\\
Z&\text{for}\quad\theta\in[\pi/2,\pi]\end{matrix}\\
&= \begin{pmatrix} 1&0\\
0&1-2H(\theta-\pi/2)
\end{pmatrix} \ .\\
\end{split}
\end{equation}
Here $H(x)$ is the Heaviside step function. This additional operation is necessary to account for the sign of the cosine for $\theta>\pi/2$.
The complete circuit is presented in Fig.~\ref{fig:circuitC}(B).

This circuit is too simple for a realistic test of the LCU method. For a more serious challenge, we  implemented the excitation operator Eq.~\eqref{eq:m0_eo} in second quantization using two qubits for the system register. In this formulation, the initial state will be $\ket{10}$ while the final state is mapped to $\ket{01}$. The excitation operator defined in Eq.~\eqref{eq:m0_eo} becomes
\begin{equation}
\begin{split}
\widetilde{O}(\theta) &= \sin(\theta) (c_1^\dagger c_1 + c_0^\dagger c_0)+\cos(\theta)(c_0^\dagger c_1+c_1^\dagger c_0)\;.
\end{split}
\label{otilde}
\end{equation}
We use the Jordan-Wigner transformation to represent the creation/annihilation operators in terms of Pauli operators and define
\begin{equation}
\label{eq:mapping}
c_k = \frac{X_k - i Y_k}{2}\quad,\quad c^\dagger_k = \frac{X_k + i Y_k}{2}\;,
\end{equation}
which act on qubit $k$. Thus, the excitation operator Eq.~\eqref{otilde} becomes
\begin{equation}
\label{eq:m1_eo}
\widetilde{O}(\theta) = \frac{\cos(\theta)}{2}\left(X_0X_1 + Y_0Y_1\right) + \sin(\theta) \mathbb{1}\;.\\
\end{equation}
\begin{figure}[tbh]
 \centering
 \includegraphics[width=0.49\textwidth]{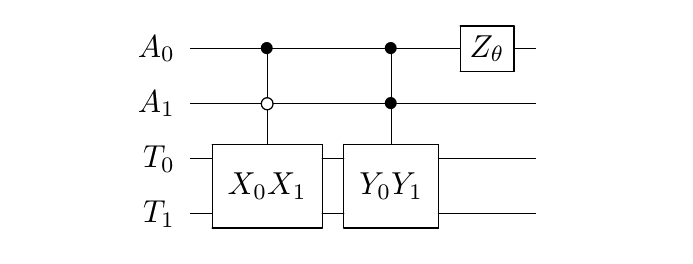}
 \caption{Implementation of the {\it select} unitary for the excitation operator in Eq.~\eqref{eq:m1_eo}.}
\label{fig:circuitD}
\end{figure}

In this case we can use a {\it prepare} unitary very similar to the one above and producing the state
\begin{equation}
\label{eq:phi2_state}
\ket{\Phi_2} = \cos(\phi_1)\ket{00} +\sin(\phi_1)\ket{1}\otimes\frac{\ket{0}+\ket{1}}{\sqrt{2}}\;,
\end{equation}
using the implementation reported in Fig.~\ref{fig:circuitC}(C),
and an explicit decomposition of the needed two qubit gate can be found in App.~\ref{app:full_lcuA}. Note that the angle $\phi_1$ entering Eq.~\eqref{eq:phi2_state} coincides with Eq.~\eqref{eq:phi1_angle} thanks to the factor $1/2$ in front of the first two terms in Eq.~\eqref{otilde}.

Let us consider the possible schematic implementation of the {\it select} unitary $V_s$ as presented in Fig.~\ref{fig:circuitD},
where the labels $(A_0,A_1)$ indicate the ancilla and labels $(T_0,T_1)$ the target qubits respectively. 

 \begin{figure*}[bth]
 \centering
 \includegraphics[width=0.8\textwidth]{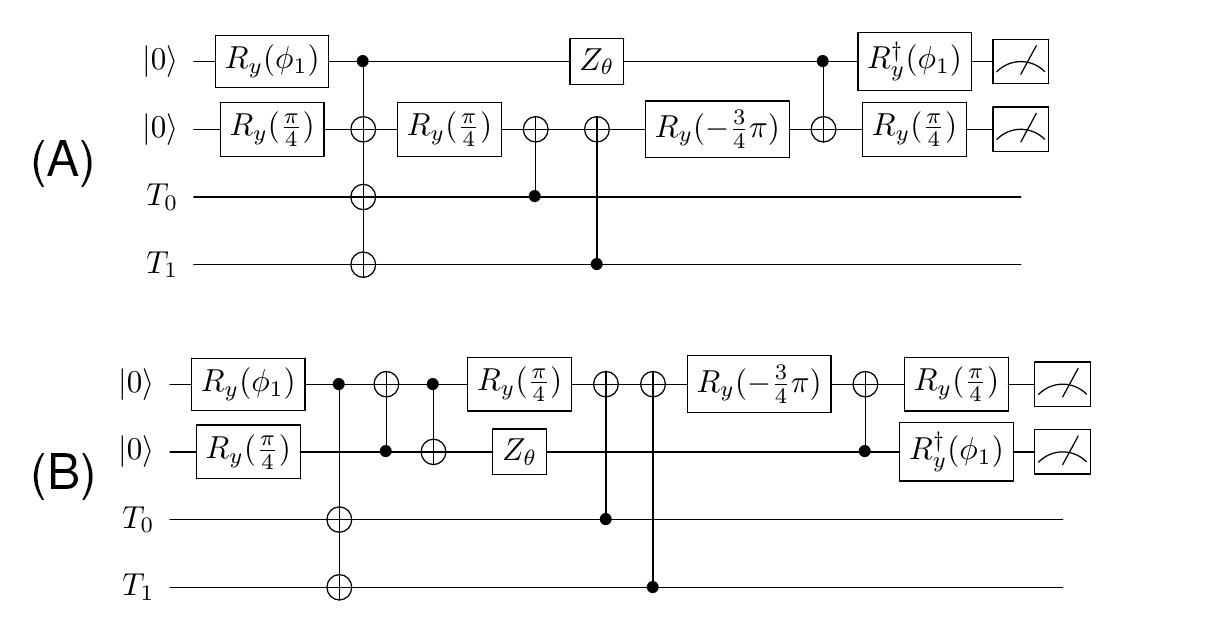}
 \caption{Complete LCU circuits in second quantization: (A) assumes all-to-all connectivity while (B) is designed for implementation on the device displayed in Fig.~\ref{fig:vigo}.}
\label{fig:circuitE}
\end{figure*}

A direct implementation of this circuit requires four Toffoli gates (see App.~\ref{app:gate} for definition) gates and at most seven Clifford gates (among $X,Z,S,S^\dagger$). Using the optimal decomposition presented in \cite{shende2009} of six CNOT gates per Toffoli we arrive at a total of 24 CNOT gates. Constraints from the qubit connectivity on QPUs might further increase this count. 

It is important to note however that for the operator in Eq.~\eqref{eq:m1_eo} we are only using three of the four possible unitaries we could implement using two qubits. This implies that, when used in conjunction with the {\it prepare} operation, we would obtain the same final outcome if we used a more general unitary of the form
\begin{equation}
\widetilde{V_S} = V_S + \sum_{k=L+1}^{2^M-1}\rvert k\rangle\langle k\lvert\otimes F_k\;,
\end{equation}
with arbitrary operations $F_k$. In our case, for instance, we could perform any unitary operation controlled on $\ket{01}$ as the state $\ket{\Phi_2}$ is orthogonal to it. If done properly, a circuit implementation of $\widetilde{V_S}$ could be cheaper in CNOT cost than the original $V_S$. Furthermore one can try to cancel operations needed for $\widetilde{V_S}$ with those used for the prepare steps before and after.

Based on these insights we were able to implement the entire four-qubit circuit needed for the LCU with only six CNOT and up to seven single qubit gates as displayed in Fig.~\ref{fig:circuitE}(A).
A complete derivation of this implementation can be found in the Appendix~\ref{app:full_lcuA}.
We note that this construction is general as it does not rely on a particular input state for the target qubits.

Unfortunately the circuit in Fig.~\ref{fig:circuitE}(A) requires the four qubits to be almost fully connected and cannot be implemented natively on the QPUs with the topology shown in Fig.~\ref{fig:vigo}. A simple modification that solves the problem with minimal additional cost is presented in Fig.~\ref{fig:circuitE}(B), 
obtained by applying a SWAP gate (see App.~\ref{app:gate} for definition) between the $A_0$ ancilla qubit and the $A_1$ ancilla qubit. This circuit requires just one CNOT more than the one in Fig.~\ref{fig:circuitE}(A) and needs the $A_0$ qubit to have connections with the other three qubits. For the connectivity map shown in Fig.~\ref{fig:vigo} this corresponds to qubit 1.

We define two estimators for the transition probability. First, since in this case we have $\eta=1$ the success probability for the 
LCU method Eq.~\eqref{eq:lcu_prob} is completely determined  by the operator norm $\Lambda(\theta)$ which for this operator takes the same form as Eq.~\eqref{eq:lambda_theta}. This gives the estimator
\begin{equation}
\label{eq:pa_estim}
P_t^A(\theta) = \Lambda^2(\theta) \, {\rm Tr}\left[\rvert\Omega\rangle\langle\Omega\lvert \Pi_0\otimes\rvert\psi_f\rangle\langle\psi_f\lvert\right]\;,
\end{equation}
where the state $\ket{\Omega}\equiv\ket{\Omega(\theta)}$ is from Eq.~\eqref{eq:fstate_LCU}, $\Pi_0$ is the projector to the state of the ancilla register with all qubits in $\ket{0}$, and $\ket{\psi_f}=\ket{01}$ is the final state. The trace operation is denoted as ${\rm Tr}$. 

An alternative estimator can be obtained by rescaling with the empirical success probability instead of the expected one~\footnote{Note that this corresponds to postselection on successful runs}, leading to the ratio estimator
\begin{equation}
\label{eq:estimator_2}
P_t^B(\theta) \equiv \frac{{\rm Tr}\left[\rvert\Omega\rangle\langle\Omega\lvert \Pi_0\otimes\rvert\psi_f\rangle\langle\psi_f\lvert\right]}{{\rm Tr}\left[\rvert\Omega\rangle\langle\Omega\lvert \Pi_0\otimes\mathbb{1}\right]}\;.
\end{equation}
In absence of systematic errors the two estimators $P_t^A$ and $P_t^B$ should coincide.

As for the time-dependent method above, we performed the quantum computations using Qiskit~\cite{qiskit} and the IBM quantum device Vigo~\cite{IBMQ_Vigo} using both a VM and the real QPU.   In this case however, we mapped  Vigo's qubits 1, 0, 3, and 2 onto the ancilla qubits $A_0$, $A_1$, and target qubits $T_0$, $T_1$ in Fig.~\ref{fig:circuitE}(B), respectively.

\begin{figure}
 \centering
 \includegraphics[width=0.49\textwidth]{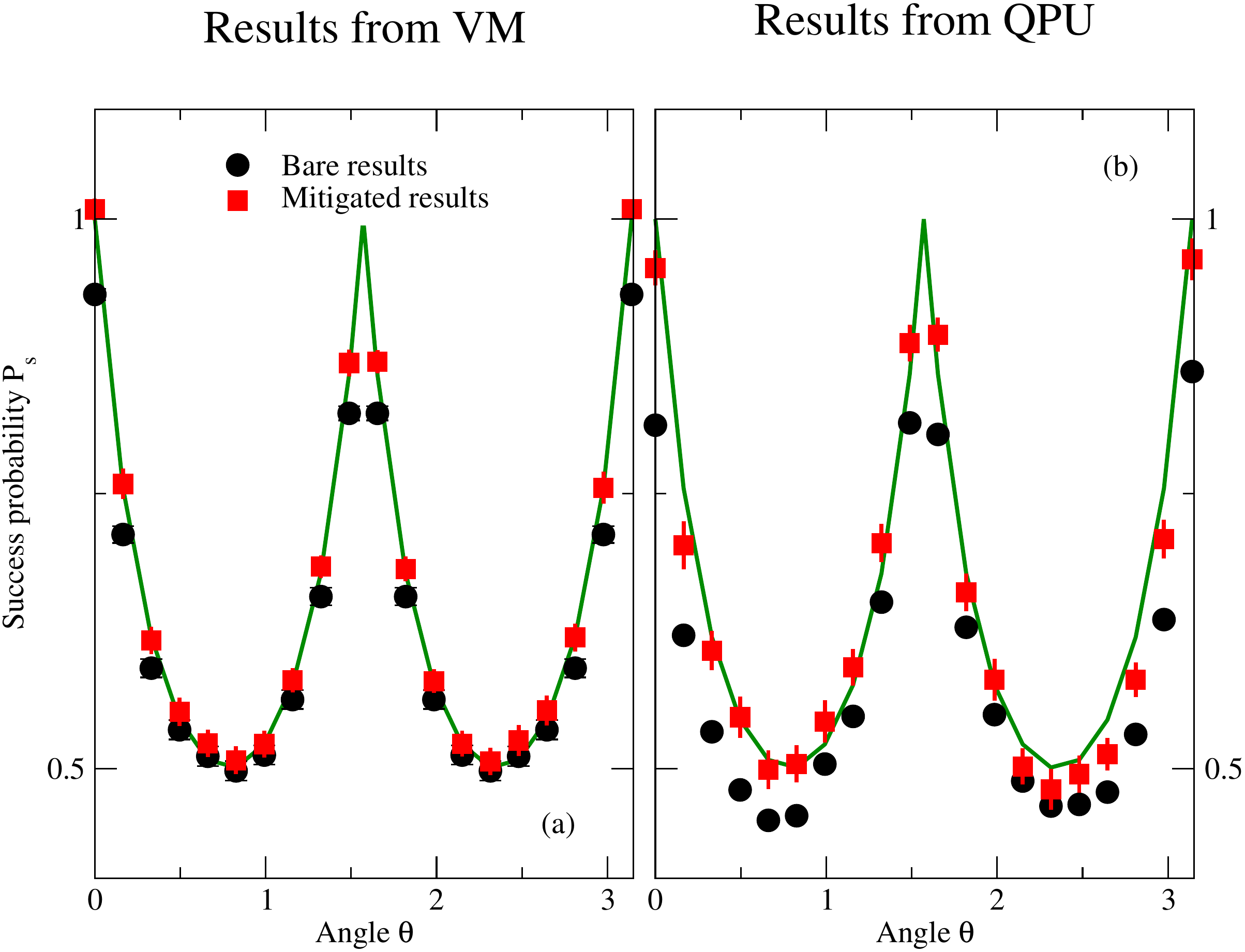}
 \caption{(Color online) Results on the success probability $P_s$ for LCU method, as function of angle $\theta$ in excitation operator Eq.~\eqref{eq:m1_eo}. The symbols follow the same convention as Fig.~\ref{fig:method0_ps_vm_vs_qpu}.}
\label{fig:method1_ps_vm_vs_qpu}
\end{figure}

In Fig.~\ref{fig:method1_ps_vm_vs_qpu} we show the results for the success probability, and in Tab.~\ref{tab:ps_2q2a} report the values for the quality metrics in both VM and QPU runs. The comparison with Fig.~\ref{fig:method0_ps_vm_vs_qpu} indicates that, as anticipated in Sec.~\ref{sec:stateprep}, the LCU method has a much larger success probability than the time evolution technique. Furthermore, the error mitigation works remarkably well and is able to bring the results close to their error-free value. This effect is especially visible in the evolution of the $\chi^2$ measure which in the real QPU case is reduced by almost two orders of magnitude using error mitigation.

\begin{table}[ht]
\subtable[VM]{
\begin{tabular}{l|r|r}
 & $\chi^2$ & ${\rm nssd}$ \\ \hline
bare & 26.52 & 0.453 \\
RO mit. & 3.63 & 0.211 \\
full mit. & 0.44 & 0.118 
\end{tabular}
}
\subtable[QPU]{
\begin{tabular}{l|r|r}
 & $\chi^2$ & ${\rm nssd}$ \\ \hline
bare & 112.56 & 1.190  \\
RO mit. & 70.17 & 1.085 \\
full mit. & 2.61 & 0.415 
\end{tabular}
}
\caption{Quality metrics for the success probability from (a) the VM and (b) the QPU using the LCU method. \label{tab:ps_2q2a}}
\end{table}

\begin{figure}
 \centering
 \includegraphics[width=0.49\textwidth]{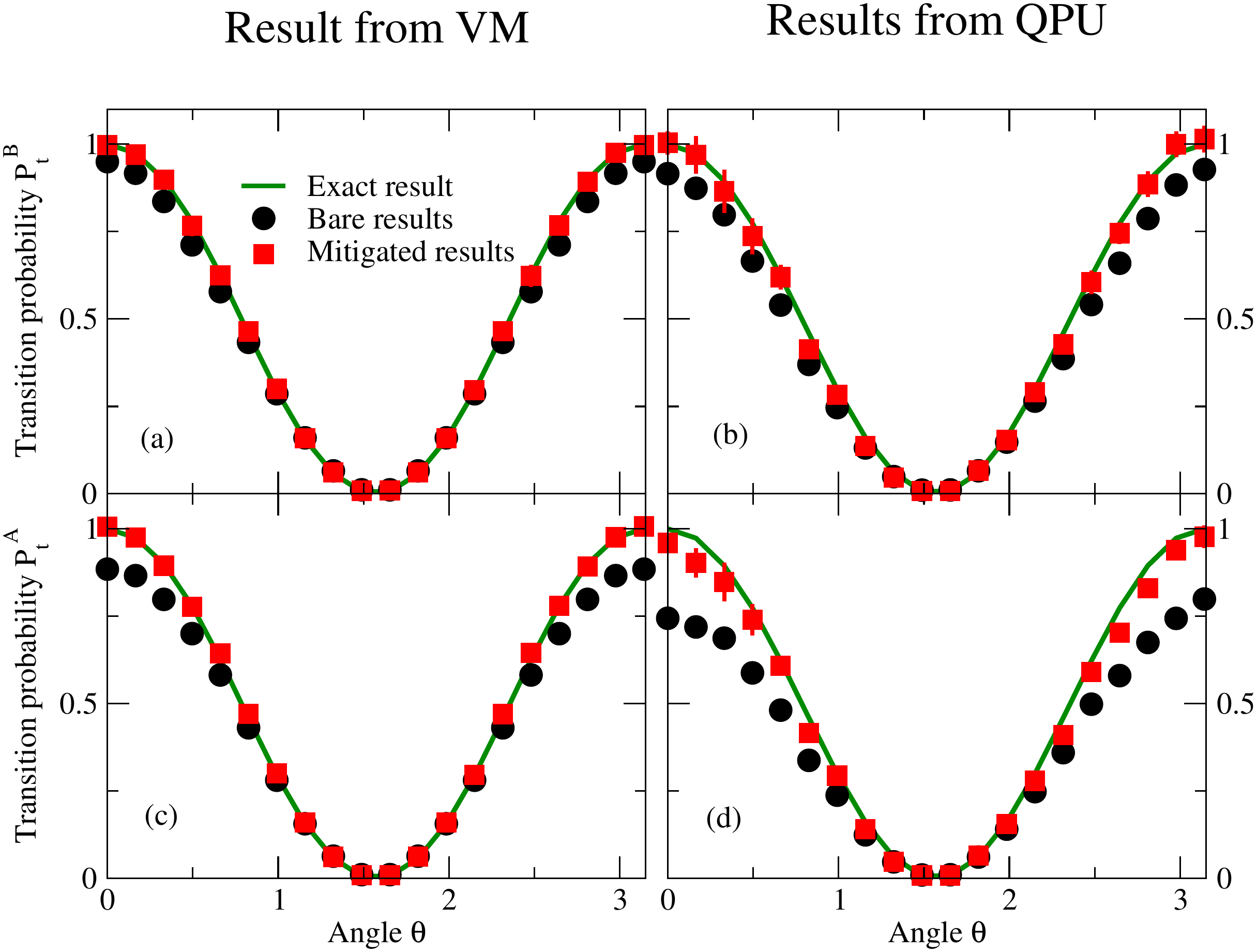}
 \caption{(Color online) Results for the  transition probabilities $P_t^A$ (c,d) and $P_t^B$ (a,b) for LCU method, as function of angle $\theta$ in excitation operator Eq.~\eqref{eq:m1_eo}. The symbols follow the same convention as Fig.~\ref{fig:method0_ps_vm_vs_qpu}.}
\label{fig:method1_matel_vm_vs_qpu}
\end{figure}

The corresponding results for the transition probability are presented in Fig.~\ref{fig:method1_matel_vm_vs_qpu} using both estimator $P_t^A$ from Eq.~\eqref{eq:pa_estim} and $P_t^B$ from Eq.~\eqref{eq:estimator_2}. The first striking difference between these estimators is the quality of the bare results as quantified by the $\chi^2$ measure: for both the simple noise model implemented in the VM and the real noise present in the hardware we see an order of magnitude difference between $P_t^A$ and $P_t^B$, with the latter being the better. This effect can be understood qualitatively assuming the noise model to be a simple depolarizing channel (see eg. Sec.8.3.4 of~\cite{nielsen2010}) and we discuss the idea in detail in Appendix~\ref{app:ratio}.

Error mitigation works very well in this case too, and this is particularly evident for the lower two panels of Fig.~\ref{fig:method1_matel_vm_vs_qpu} corresponding to the $P_t^A$ estimator. The bare values for the QPU runs of the transition probability at both $\theta\approx0$ and $\theta\approx\pi$ are off by about $25\%$ but the fully mitigated results bring this down by a factor of 5. As we can see from the results in Tab.~\ref{tab:matel_2q2a} , this behavior is also tracked by the average precision metric $nssd$ and by the value of $\chi^2$ which drops by two orders of magnitude for the $P_t^A$ observable and down to values around unity for the second estimator.

\begin{table}[ht]
\subtable[VM]{
\begin{tabular}{l|rr|rr}
 & $\chi^2$($P_t^A$)& $\chi^2$($P_t^B$) & ${\rm  nssd}$($P_t^A$)& ${\rm nssd}$($P_t^B$)\\  \hline
bare & 76.20 & 10.09 & 1.022 & 0.623 \\
RO mit. & 6.57 & 0.29 & 0.321 & 0.106 \\
full mit. & 0.29 & 0.12 & 0.129 & 0.053
\end{tabular}
}
\subtable[QPU]{
\begin{tabular}{l|rr|rr}
 & $\chi^2$($P_t^A$)& $\chi^2$($P_t^B$) & ${\rm nssd}$($P_t^A$)& ${\rm nssd}$($P_t^B$)\\ \hline
bare & 251.70 & 23.00 & 2.367 & 1.151 \\
RO mit. & 109.78 & 4.51 & 1.661 & 0.487 \\
full mit. & 2.48 & 1.13 & 0.580 & 0.326
\end{tabular}
}
\caption{Quality metrics for the results on the transition probabilities $P_t^A$ and $P_t^B$ with (a)the VM run and (b)the QPU run for LCU method. \label{tab:matel_2q2a}}
\end{table}

To summarize this Section, we used both the time-dependent and LCU methods to implement the excitation operator Eq.~\eqref{eq:m0_eo} and its generalization to second quantization Eq.~\eqref{eq:m1_eo}. While both methods are clearly useful 
on existing quantum devices, the LCU-based method is more efficient and error tolerant, and also more accurate if we use estimators like Eq.~\eqref{eq:estimator_2}. It is important to note at this point that, even though the success probability could be estimated off-line on a classical computer since it corresponds to a static structure factor, the more common situation is when we do not have access to this information (as in the next section) and we are instead forced to choose Eq.~\eqref{eq:estimator_2}. It is encouraging to see that this is also the preferred way in general.

\section{Application to the $n(p,d)\gamma$ reaction}
\label{sec:npdg}
In this Section we apply the methods described in Sec.~\ref{sec:stateprep} and benchmarked in Sec.~\ref{sec:res} to a simple model for the excitation operator that induces the $M1$ transition in the $n(p,d)\gamma$ reaction. In the two-dimensional space spanned by the $^1S_0$ and $^3S_1$ states the excitation operator takes the form
\begin{equation}
\label{eq:em_op_1st}
O(\theta) = \alpha\mathbb{1}+\beta X-\alpha Z = \begin{pmatrix}
0 & \beta\\
\beta&2\alpha
\end{pmatrix}\;,
\end{equation}
where the two real constants are given by
\begin{align}
\alpha&=\sin(\theta^\prime)\frac{g_p+g_n}{4}\mu_N\;,\\
\beta&=\frac{\mu_N}{2\sqrt{2}}(g_p-g_n)\cos(\theta^\prime)\;.
\end{align}
Here, $\mu_N$ is the nuclear magneton and $g_{p,n}$ denotes the proton and neutron $g$ factor, respectively. The derivation of this expression is presented in Appendix~\ref{app:m1_transition}. In what  follows, we set $\mu_N = 1$ as we can always scale the operator by multiplying by a constant. 

In the following we initialize the system in the $^1S_0$ state (represented by $\ket{0}$ in the computational basis) and compute the transition probability to the $^3S_1$ ground-state (represented by $\ket{1}$ in the computational basis).

\subsection{First quantization version}
As we discussed in Sec.~\ref{subsec:tdm}, the main obstacle to an efficient implementation of the time-dependent method is the low success probability for a high target fidelity. In order to find a reasonable value for the time interval $\gamma$ we use the bounds derived in Eq~\eqref{eq:bound_prob} and carry out a similar analysis to what was done in Sec.~\ref{sec:res_m0A}. The result of this analysis is that the value $\gamma=0.3$ used in Sec.~\ref{sec:res_m0A} is also a good choice in this case, with a guaranteed high fidelity $F(0.3)\geq98.8\%$ together with relatively high success probability
\begin{equation}
\label{eq:prob_td_nucex}
73.8\% \geq P_s(0.3) \geq 4.5\% \;.
\end{equation}
The exact lower bound for $P_s$ in this case is $6.8\%$ and our lower bound (obtained from Eq.~\eqref{eq:bound_prob}) comes very close.

In order to implement the circuit shown in Fig.~\ref{fig:circuitA}(A) we use a construction similar to the simpler case studied in Sec.~\ref{sec:res_m0A} which needs 4 CNOT gates and additional single qubit rotations. We present a detailed derivation of the construction in Sec.~\ref{app:circ_td_npdg}.

For the LCU-based method instead there is no hyperparameter to tune and the success probability is guaranteed to be the value computed from Eq.~\eqref{eq:lcu_prob}
\begin{equation}
\begin{split}
\label{eq:prob_lcu_nucex}
P_s &= \frac{\langle 1 \lvert O^2(\theta) \rvert 1 \rangle}{\Lambda^2}= \frac{\beta^2+4\alpha^2}{4\alpha^2+\beta^2+4\alpha|\beta|}\geq 0.5\;.
\end{split}
\end{equation}
The three operators appearing in the definition Eq.~\eqref{eq:em_op_1st} of the excitation operator require an auxiliary register made of 2 ancillas. As described in detail in Appendix~\ref{sec:lcu_nuc_op} the LCU circuit can be implemented with only 3 CNOT gates and additional one-qubit rotations, even with the restricted connectivity shown in Fig.~\ref{fig:vigo}.

Both algorithms where executed on the Vigo QPU~\cite{IBMQ_Vigo} using physical qubits $(1,2)$ (ancilla in $1$ and system in $2$) for the time-dependent method and qubits $(2,3,1)$ for the LCU-based method. In the latter case, qubits $(2,3)$ represented ancilla qubits and qubit $1$ was the system.

As we have seen from the results of the previous section, simulations using a Virtual Machine are not realistic enough to predict accurately the behavior of the real QPU and for this reason we choose to only show results obtained from QPU runs in this section.
\begin{figure}
 \centering
 \includegraphics[width=0.49\textwidth]{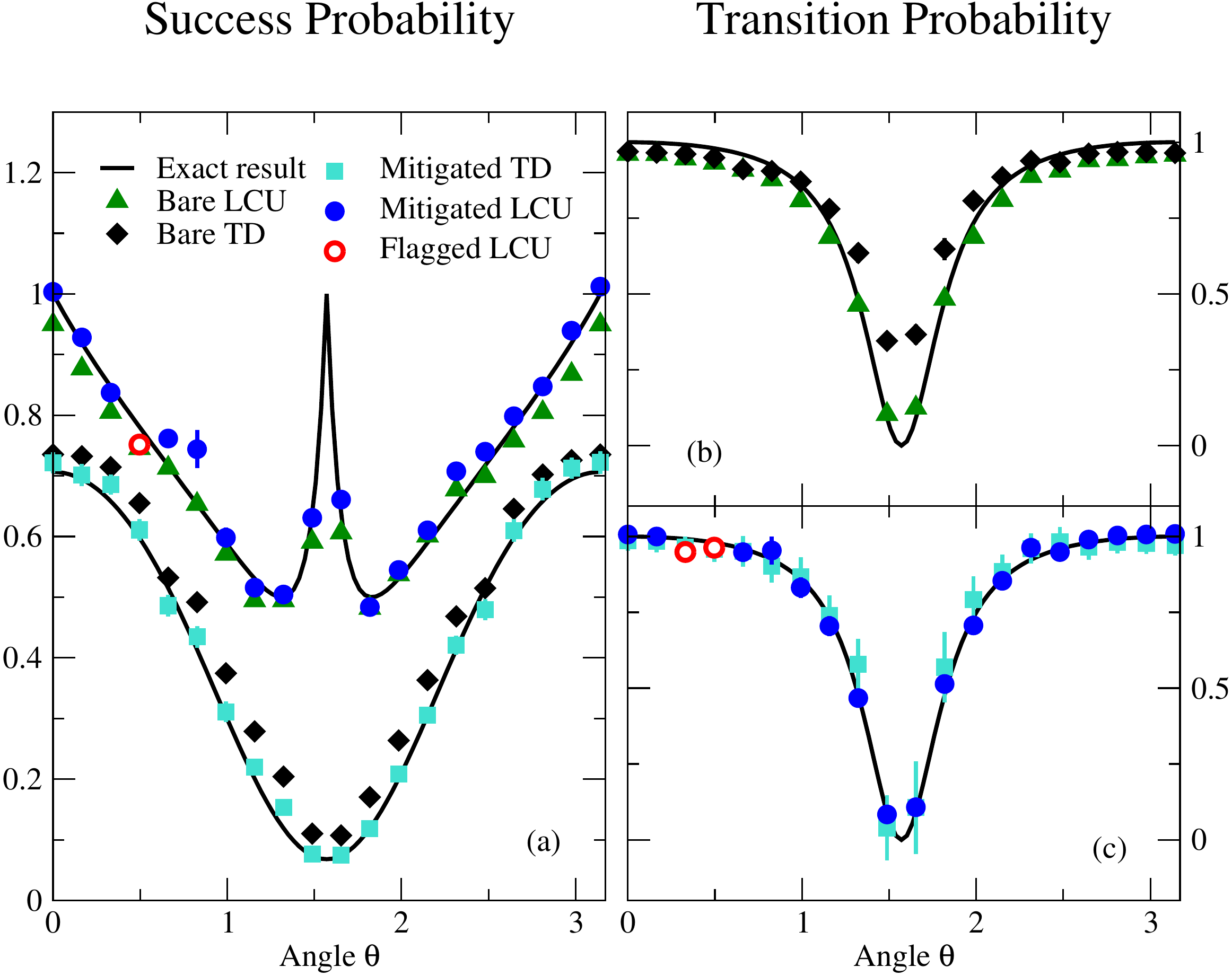}
 \caption{(Color online) Results on the success probability [panel (a) on the left] and transition probability (right panels) for the excitation operator in Eq.~\eqref{eq:em_op_1st} using both the time-dependent method (TD) and the LCU based one. Panel (b) shows the bare results from the QPU and panel (c) the results obtained after error mitigation.}
\label{fig:nptodg_1stq}
\end{figure}

In Fig.~\ref{fig:nptodg_1stq} we report the results on both the success probability (left panel) and the ${^1S_0}\to {^3S_1}$ transition probability (right panels). The red empty circles are results that were flagged by our error mitigation scheme because no self-consistent extrapolation was found. The reported values are only mitigated with respect to the read-out error (see App.~\ref{app:err_mitigation} for more details on our framework). Note that this is the only situation in this work where this problem was present, this could be either due to difficulties in executing our specific implementation or could simply be an effect coming from calibration of the device (the data for the time-dependent method was collected on July 20 2020 while the LCU data was generated on August 27 2020). As we can see from the plot, this flagged point does not show any discernible systematic error and it could also be possible that our flagging procedure is too conservative. 

As we have anticipated with Eq.~\eqref{eq:prob_td_nucex} and Eq.~\eqref{eq:prob_lcu_nucex}, the success probability for the LCU method [top curve on panel (a)] is larger than the one of the time dependent method (indicated with TD in the plot). A direct effect of this is the reduced size of the statistical sample of experimental results that can be used to compute observables of the excited state like the transition probability. This is especially evident for the bare TD results for the transition probability [black diamonds in panel (b)] which are off by a factor of about $3$ in the region of lowest success probability around $\theta=\pi/2$. This effect is also present in the mitigated results shown in panel (c) (the turquoise squares) which show large fluctuations in the same region. The results for the transition probability with the LCU technique on the other hand are more stable, with bare results already twice as accurate than with the TD method on average (see values of the nssd in Tab.~\ref{tab:npdg_tp_m0}). Interestingly we can see from Tab.~\ref{tab:npdg_tp_m0} that the read-out correction for this observable is the dominant source of improvement over the bare results with the additional noise extrapolation step only changing the results slightly.

\begin{table}[ht]
\subtable[TD]{
\begin{tabular}{l|r|r}
 & $\chi^2$ & ${\rm nssd}$ \\ \hline
bare & 11.77     & 1.198    \\
RO mit. & 6.04  & 1.013 \\
full mit. & 0.31 & 0.436
\end{tabular}
}
\subtable[LCU]{
\begin{tabular}{l|r|r}
 & $\chi^2$ & ${\rm  nssd}$ \\ \hline
bare & 13.62     & 0.494   \\
RO mit. & 1.41  & 0.173 \\
full mit. & 2.19  & 0.253
\end{tabular}
}
\caption{Quality metrics for the results on the ${^1S_0}\to {^3S_1}$ transition probability in the $n(p,d)\gamma$ reaction with (a) the time dependent (TD) method and (b) the LCU-based method. Both sets of results were obtained using the Vigo QPU~\cite{IBMQ_Vigo}.\label{tab:npdg_tp_m0}}
\end{table}

\subsection{Second quantization}

The second quantization formalism is very useful in order to scale up simulations of nuclear dynamics since it guarantees a polynomial increase of the computational cost as a function of the basis size. In this Section we implement the transition operator Eq.~\eqref{eq:em_op_1st} in the Fock space spanned by the two angular momentum states using a register of $n=2$ qubits. This excitation operator has only real matrix elements and can be expressed in second quantization as
\begin{equation}
\begin{split}
\widetilde{O} &= 2\alpha c_1^\dagger c_1+\beta(c_0^\dagger c_1+c_1^\dagger c_0)\;.
\end{split}
\end{equation}
Again we use the Jordan-Wigner transformation Eq.~\eqref{eq:mapping} to represent the creation/annihilation operators in terms of Pauli operators. This yields the excitation operator
\begin{equation}
\label{eq:obar3qA}
\begin{split}
\widetilde{O} =&\alpha \mathbb{1} +\frac{\beta}{2}(X_0X_1+Y_0Y_1) + \alpha Z_1\\
&=\begin{pmatrix}
2\alpha&0&0&0\\
0&0&\beta&0\\
0&\beta&2\alpha&0\\
0&0&0&0\\
\end{pmatrix}\;,
\end{split}
\end{equation}
and we recover the original operator in the wanted subspace of one-particle states (the innermost $2\times 2$ matrix). 

Note that both operators in Eq.~\eqref{eq:obar3qA} and Eq.~\eqref{eq:obar3q} contain non-commuting operators and therefore the implementation of the time-dependent method will require a suitable approximation of the evolution operator $\exp\left(-i\gamma O\right)$ and, as shown in the previous section, will give a relatively low success probability. On the other hand the LCU method can be applied directly to either operator without making any approximation while maintaining a large success probability. For these reasons we study here only the extension of the LCU to the second quantized case. 

We could also obtain the same operator in this subspace by using an alternative definition
\begin{equation}
\label{eq:obar3q}
\begin{split}
\overline{O}&=\alpha \mathbb{1} +\frac{\beta}{2}(X_0X_1+Y_0Y_1)-\frac{\alpha}{2}(Z_0-Z_1)\\
&=\begin{pmatrix}
\alpha&0&0&0\\
0&0&\beta&0\\
0&\beta&2\alpha&0\\
0&0&0&\alpha\\
\end{pmatrix}\;,
\end{split}
\end{equation}
which can be implemented with LCU with less depth. As we show in Sec.~\ref{sec:lcu_nuc_op}, the {\it select} unitary for the operator in Eq.~\eqref{eq:obar3q} can be implemented with only seven CNOT gates together with three for the {\it prepare} unitary, while the former requires 11 CNOT and two CNOT, respectively (see~\cite{vatan2004} for the prepare).

In order to increase the density of points in the region around $\pi/2$ where the cross section drops to zero, we chose to implement instead the simpler operator
\begin{equation}
\begin{split}
\label{eq:omtheta}
\overline{O}_M (\theta) &= \sin(\theta)\mathbb{1} +\frac{\cos(\theta)}{2}(X_0X_1+Y_0Y_1)\\
&-\frac{\sin(\theta)}{2}(Z_0-Z_1)\;,
\end{split}
\end{equation}
for uniformly spaced values of the angle $\theta$. The original excitation operator is then obtained as
\begin{equation}
\label{eq:theta_change}
\overline{O} = \sqrt{\alpha^2+\beta^2}\; \overline{O}_M (\phi_{\alpha\beta})\quad\phi_{\alpha\beta}=\arctan\left(\frac{\alpha}{\beta}\right)\;.
\end{equation}
Note that, since every observable we compute is independent on a global scale factor like $\sqrt{\alpha^2+\beta^2}$, we only need to perform the change of variables in the angle.

This simulation was executed on the Vigo QPU~\cite{IBMQ_Vigo} using the circuit from Fig.~\ref{fig:circuitM}(D). This requires 15 CNOT gates for the connectivity graph in Fig.~\ref{fig:vigo}. For this simulation, we used qubits 1,4,3 for the ancilla qubits $A_0$, $A_1$ and $A_2$ and qubits 0 and 2 for the target qubits $T_0$, $T_1$ respectively.

\begin{figure}
 \centering
 \includegraphics[width=0.49\textwidth]{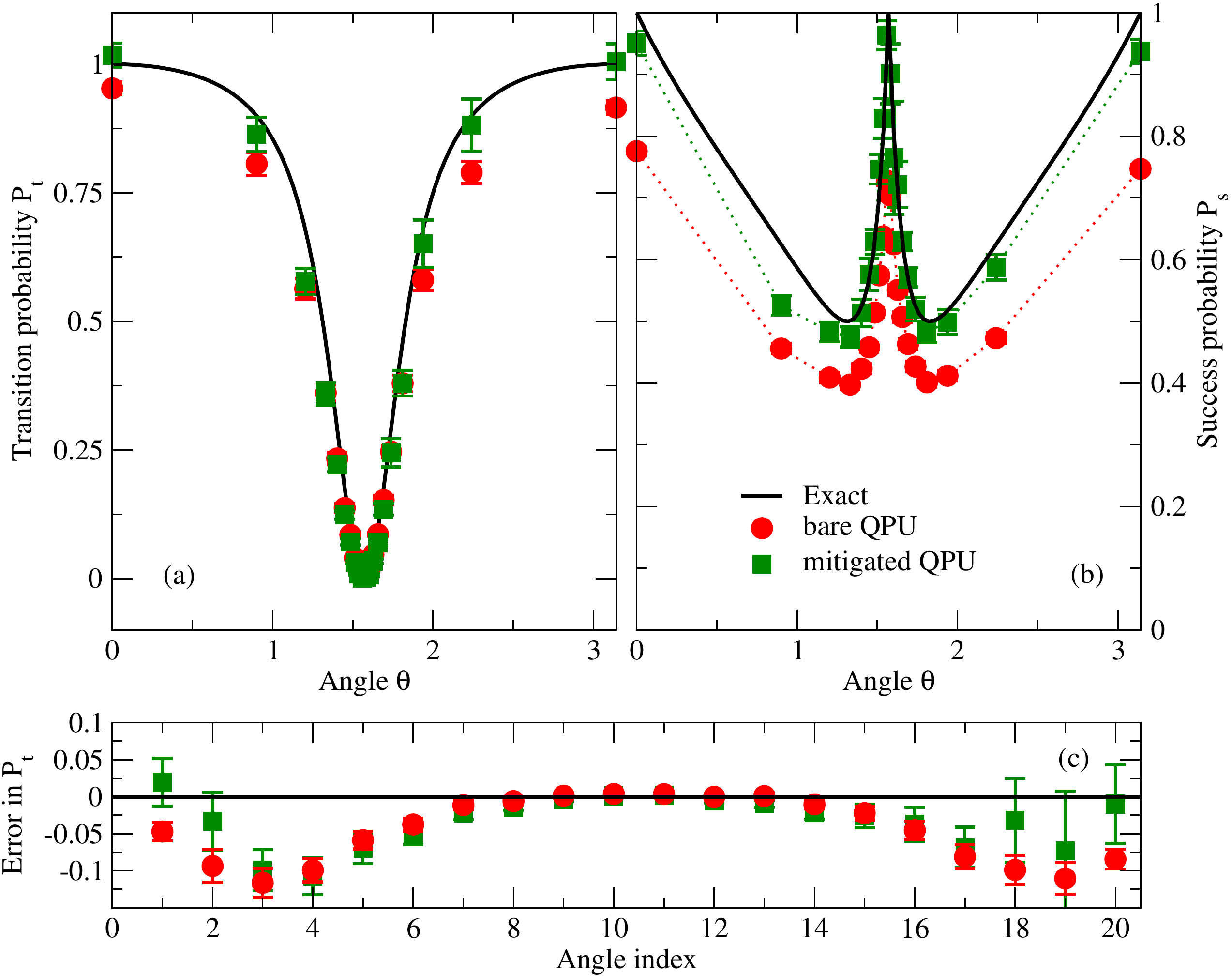}
 \caption{(Color online) Results on the transition probability $P_T$ (a) and success probability $P_s$ (b) for the $np\to d\gamma$ reaction obtained using the LCU method on the Vigo QPU~\cite{IBMQ_Vigo}. In panel (c) we present, for every angle considered here, the deviation from the expected result. The dotted lines in panel (b) are just a guide to the eye.}
\label{fig:nucex_lcu_qpu}
\end{figure}

We present the results obtained from this simulation in Fig.~\ref{fig:nucex_lcu_qpu}. In panel (a) we report the result for the transition probability and in panel (b) the corresponding success probability for the LCU step. In all panels, black curves represented the expected exact values, red circles are bare result from the QPU and green squares are the fully mitigated results (with read-our correction and error extrapolation). Finally, in panel (c) at the bottom, we show the error in the transition probability for all the 20 angles we considered here (a uniform grid in $[0,\pi]$). 

Note that, as mentioned above, the change of variables in Eq.~\eqref{eq:theta_change} produces a larger density of results in the interesting central region. Once again we see that for the LCU method the quality of the results for the transition probability is higher than for the success probability thanks to cancellation of errors. For instance we can see that, even though the bare result for the success probability at $\theta=\pi$ is off by more than $25\%$, the corresponding (bare) result for the transition probability has only an error of about $8\%$. This explains why the difference between bare and mitigated results [very apparent in panel (b)] is not seen directly in the errors shown in panel (c) at the bottom. Interestingly, even though error mitigation is able to bring some results to be compatible with the exact case, for some of the angles the mitigation procedure fails to improve the agreement. We interpret this as further evidence that the calculation of ratio estimators, like the transition probability, enjoys a much reduced sensitivity to hardware noise.

This behavior is in fact not directly observed in the results on the success probability presented in Tab.~\ref{tab:lcu_nucex} where error extrapolation is responsible for the majority of the 2 orders of magnitude improvement in the value of the $\chi^2$.

\begin{table}[ht]
\subtable[Success Probability]{
\begin{tabular}{l|r|r}
 & $\chi^2$ & ${\rm nssd}$ \\ \hline
bare & 372.01     & 2.18    \\
RO mit. & 250.72  & 2.04 \\
full mit. & 3.81  & 0.52 
\end{tabular}
}
\subtable[Transition Probability]{
\begin{tabular}{l|r|r}
 & $\chi^2$ & ${\rm  nssd}$ \\ \hline
bare & 15.4      & 1.219        \\
RO mit.& 8.0     & 0.835     \\
full mit. & 5.8 & 0.928  
\end{tabular}
}
\caption{Quality metrics for the results on the success probability [panel (a)] and transition probability [panel (b)] for the excitation operator in Eq.~\eqref{eq:obar3q} using the LCU method on the Vigo QPU~\cite{IBMQ_Vigo}. \label{tab:lcu_nucex}}
\end{table}

\section{Conclusions}
\label{sec:concl}
We applied the time evolution method and the LCU to generate excited states on a quantum computer and studied their accuracy and efficiency. 
Both methods are feasible and efficient for the state preparation algorithm.
The time-evolution method is accurate in the quantum computation of transition matrix elements. However, keeping a small target infidelity goes together with a relatively low success probability. In contrast, the LCU method has a much larger success probability while also being accurate in the quantum computation of the transition matrix element.
It is also  more resilient to depolarizing noise with the help of ratio estimators. This is true for the examples studied in this paper and also for most cases where it is easy to add ancilla qubits and postselect on their measured state. However, there might still be some cases where the time-dependent method could be superior. It is an advantage that this method  only needs one ancilla qubit.

Note that these considerations would change when considering fault-tolerant quantum devices that allow us to perform calculation of arbitrary length: the success probability $P_s$ of a quantum algorithm can be increased to 1 using $O(1/\sqrt{P_s})$ additional unitary gates using standard amplitude amplification techniques. Alternatively, we could implement the excitation operator deterministically also using quantum signal processing~\cite{low2019} or its generalization, the Quantum Singular Value Transformation~\cite{qsvt2019}. In the context of nuclear physics, similar ideas were used in past work for the evolution operator itself~\cite{roggero2020A} and have the potential to substantially reduce the overall cost to carry out ab-initio calculations of both inclusive and exclusive nuclear cross sections~\cite{roggero2020B}.

Our quantum computations employed both the quantum virtual machine and a quantum processor, and we used noise mitigation to reduce readout noise and non-perfect fidelities of the single-qubit and two-qubit operations. We also derived bounds for success probabilities of the excitation operators and developed metrics that quantify the accuracy of the studied methods.      

Based on a very simple model, we also performed a quantum computation of the $n(p,d)\gamma$ reaction, using the time-dependent method and the LCU-based method. The applications of this work are a step towards computing nuclear reactions on quantum devices. 

\begin{acknowledgments}

We thank J. Carlson and L. Cincio for useful discussions.
The work was supported by the U.S. Department of Energy, Office of Science, Office of Advanced Scientific Computing Research (ASCR) quantum algorithm teams program, under field work proposal number ERKJ333, by the U.S. Department of Energy, Office of
Science, Office of Nuclear Physics, under Award Nos.~DE-FG02-00ER41132, DE-FG02-96ER40963, DE-SC0019478, DE-SC0018223 and DE-AC52-06NA25396 and by the U.S. Department of Energy HEP QuantISED Grant No. KA2401032. Oak Ridge National Laboratory is
supported by the Office of Science of the Department of Energy under contract No. DE-AC05-00OR22725.
We  acknowledge  use  of the  IBM  Q  for  this  work. The  views  expressed  are those of the authors and do not reflect the official policy or position of IBM or the IBM Q team. 
\end{acknowledgments}

\appendix
\label{sec:app}

\section{Proof of bounds for time-dependent method}
\label{app:method0_bounds}
Here we derive the bounds that were presented in Subsect.~\ref{subsec:tdm} of the main text. Using Taylor's theorem we know that for some $y\in[0,x]$ the following holds
\begin{equation}
\label{eq:sintapp}
\sin(x) = x-\frac{x^3}{6}\cos(y)\;.
\end{equation}

Using this relation we arrive at the following bounds on the expectations values
\begin{equation}
\langle X^2\rangle - \frac{\langle X^4\rangle}{6} \leq \langle X\sin(X)\rangle \leq \langle X^2\rangle
\end{equation}
and
\begin{equation}
\langle X^2\rangle - \frac{\langle X^4\rangle}{3} +\frac{\langle X^6\rangle}{36} \leq \langle \sin^2(X)\rangle \leq \langle X^2\rangle   \ . 
\end{equation}
Here $X$ is a bounded operator with $\|X\|\leq\pi/2$ and its averages $\langle X^k \rangle = Tr\left[X^k \rho\right]$ are computed on a generic input state $\rho$.
The lower bounds can be relaxed into a form more amenable for estimation by noticing that
\begin{equation}
\langle X^4\rangle = {\rm Var}[X^2]+\langle X^2\rangle^2\leq \Lambda^2\langle X^2\rangle \ , 
\end{equation}
where $\Lambda$ is a bound on the norm $\|X\|$.
The final inequalities used in the main text are then
\begin{equation}
\langle X^2\rangle \left(1- \frac{\Lambda^2}{6} \right)\leq \langle X\sin(X)\rangle \leq \langle X^2\rangle\;,
\end{equation}
and
\begin{equation}
\langle X^2\rangle \left(1- \frac{\Lambda^2}{3} \right)\leq \langle \sin(X)^2\rangle \leq \langle X\sin(X)\rangle\;,
\end{equation}
or the alternative upper bound
\begin{equation}
\langle X^2\rangle \left(1- \frac{\Lambda^2}{3} \right)\leq \langle \sin(X)^2\rangle \leq \sin(\Lambda)^2 \ ,
\end{equation}
which is obtained by noticing that $\sin^2(x)$ is monotonically increasing for $x\in[0,\pi/2]$.

\section{Details on bounds with imperfect time evolution}
\label{app:faultytimeev}
Here we derive the bounds that were used in Subsubsect.~\ref{subsub:imperfect} of the main  text.
As a starting point we take two normalized states
\begin{equation}
\ket{\phi}=A\ket{0}\quad\ket{\psi}=B\ket{0}\;,
\end{equation}
obtained using two unitaries $A$ and $B$ satisfying
\begin{equation}
\label{eq:opdiff}
\|A-B\|\leq\delta\;.
\end{equation}
Let us consider the (unnormalized) state obtained by taking their difference
\begin{equation}
\ket{\varepsilon} \equiv \ket{\phi} - \ket{\psi} = \left(A-B\right)\ket{0}\;.
\end{equation}
Its norm is bound as follows
\begin{equation}
\label{eq:errub}
\begin{split}
\|\ket{\varepsilon}\|^2 &=\bra{0}\left( A^\dagger - B^\dagger\right)\left( A - B\right)\ket{0}\\
&\leq \max_{\lambda\in\sigma[\left( A^\dagger - B^\dagger\right)\left( A - B\right)]} \lambda\\
&\equiv  \max_{s_i\in\{\text{singular values of} M=\left( A - B\right)\}} s_i^2\\
&\equiv \|A - B\|^2 \leq \delta^2\;,
\end{split}
\end{equation}
where in the second line  $\sigma[M]$ denotes the spectrum of the operator $M$, in the third line we used the definition of singular values, and in the last line we used Eq.~\eqref{eq:opdiff}. At this point we can decompose the second state as
\begin{equation}
\ket{\psi} = \cos(\theta)\ket{\phi} + \sin(\theta)\ket{\xi}\;,
\end{equation}
for some normalized state $\ket{\xi}$ orthogonal to $\ket{\phi}$. This yields
\begin{equation}
\ket{\varepsilon} = \left[1-\cos({\theta})\right]\ket{\phi}-\sin({\theta})\ket{\xi} \ ,
\end{equation}
so that we have
\begin{equation}
\begin{split}
\|\ket{\varepsilon}\| &= \left[1-\cos({\theta})\right]^2+\sin^2({\theta})\\
&=1+\cos^2(\theta)-2\cos(\theta)+1-\cos^2(\theta)\\
&=2-2\cos(\theta)\leq\delta^2\;.
\end{split}
\end{equation}
In the last line we used the result from Eq.~\eqref{eq:errub}.

Given a projection operator $\Pi$ with $\Pi^2=\Pi$, we have
\begin{equation}
\begin{split}
|\widetilde{P_{s}}-P_{s}| &\coloneqq|\langle\psi\lvert\Pi\rvert\psi\rangle-\langle\phi\lvert\Pi\lvert\phi\rangle| \\
& = \left|\Tr(\Pi\rho)-\Tr(\Pi\sigma)\right| \\
&= \left|\Tr\left(\Pi(\rho-\sigma)\right)\right| \\
&\leq D(\rho, \sigma)=\frac{1}{2}\Tr|\rho-\sigma|=\left|\sin(\theta)\right|\;.
\end{split}
\end{equation}
Here we have defined for convenience the density matrices $\rho=\rvert\psi\rangle\langle\psi\lvert$ and $\sigma=\rvert\phi\rangle\langle\phi\lvert$, and in the last line we used the definition of the trace distance $D(\rho,\sigma)$.

\section{$M1$ transition matrix elements}
\label{app:m1_transition}
Here we derive in more detail the form of the $M1$ transition operator used in Sect.~\ref{sec:npdg} of the main text. 
The dominant process for the $np\to d\gamma$ reaction is from the continuum $^1S_0$ state (with spin isospin $T=1,T_z=0$) to the deuteron bound state. For a calculation of this process in pion-less EFT, we refer the reader to Refs.~\cite{park1998,chen1999,chen1999b}. In leading order the bound-state is the $^3S_1$ state with isospin $T=T_z=0$. Thus, the electromagnetic transition will be of $M1$ multipole order, and the magnetic moment
\begin{equation}
\mathbf{m} = 2\mu_N \mathbf{l} + g_p\mu_N\mathbf{S}_p + g_n\mu_N\mathbf{S}_n\;,
\end{equation}
is the transition operator. Here, $\mu_N$ is the nuclear magneton, and $g_{p,n}$ are the proton and neutron $g$ factors, respectively. For the $^1S_0\to {^3S_1}$ transition, the orbital angular momentum $\mathbf{l}$ does not contribute, and the spatial wave functions only contribute through the overlap between a continuum state $u_1(r)\approx (a-r)/a$ and a bound state $u_3(r)=e^{-\gamma r}$, with $a$ and $\gamma$ real parameters. Focusing on the spin-isospin part of the initial $|i\rangle$ and final $|f\rangle$ states we write
\begin{equation}
\begin{split}
|i\rangle &\equiv {1\over 2}\left(|p\uparrow n\downarrow\rangle + |n\uparrow p\downarrow\rangle -|p\downarrow n\uparrow\rangle - |n\downarrow p\uparrow\rangle\right) \\
|f\rangle &\equiv {1\over \sqrt{2}}\left(|n\uparrow p\uparrow\rangle - |p\uparrow n\uparrow\rangle\right) .
\end{split}
\end{equation}
Thus,
\begin{equation}
\langle f|\mathbf{m}|i\rangle = {\mu_N\over\sqrt{2}} (g_p-g_n)\langle\uparrow|\mathbf{S}|\downarrow\rangle .
\end{equation}
Using $\mathbf{S} = {1\over 2}\mathbf{\sigma}$ finally yields
\begin{equation}
\langle f|\mathbf{m}|i\rangle = {\mu_N\over 2\sqrt{2}} (g_p-g_n)
\left(\begin{array}{c}
1\\
-i\\
0
\end{array}\right) .
\end{equation}
	
Thus, $|\langle f|\mathbf{m}|i\rangle|^2 = \mu_N^2(g_p-g_n)^2/4$, and we have $g_p=5.586$, and $g_n=-3.826$.
	
Two comments are in order. First, the spatial matrix element is a number that we cannot compute on a quantum chip using a minimum number of states. Our states are structureless, i.e. they contain only spin/isospin information. Second, we could also compute the $E2$ transition to the $^3D_1$ component of the deuteron wave function. This contribution is suppressed as the asymptotic $D$ to $S$ state amplitude is small.   
	
More generally we have also the following diagonal matrix elements
\begin{equation}
\langle f|\mathbf{m}|f\rangle = \frac{g_p+g_n}{2}\mu_N\begin{pmatrix}
0\\
0\\
1\\
\end{pmatrix}\;,
\end{equation}
and
\begin{equation}
\langle i|\mathbf{m}|i\rangle = 0\;.
\end{equation}
If we take the polarization vector $\mathbf{\xi}$ on the XZ-plane we find
\begin{equation}
\langle f|\mathbf{m}\cdot\mathbf{\xi}|f\rangle = \sin(\theta^\prime)\frac{g_p+g_n}{2}\mu_N\;,
\end{equation}
\begin{equation}
\langle i|\mathbf{m}\cdot\mathbf{\xi}|i\rangle = 0\;,
\end{equation}
and
\begin{equation}
\langle f|\mathbf{m}\cdot\mathbf{\xi}|i\rangle = \frac{\mu_N}{2\sqrt{2}}(g_p-g_n)\cos(\theta^\prime)\;.
\end{equation}
	
In this two-dimensional space we can then express the excitation operator as
\begin{equation}
\label{eq:nuclear_op}
O(\theta') = \alpha\mathbb{1}+\beta X+\gamma Z = \begin{pmatrix}
\alpha+\gamma & \beta\\
\beta&\alpha-\gamma
\end{pmatrix}\;,
\end{equation}
with
\begin{align}
\alpha&=\sin(\theta^\prime)\frac{g_p+g_n}{4}\mu_N\;,\\
\beta&=\frac{\mu_N}{2\sqrt{2}}(g_p-g_n)\cos(\theta^\prime)\;,\\
\gamma&=-\sin(\theta^\prime)\frac{g_p+g_n}{4}\mu_N\;.
\end{align}
	
In particular we note that:
\begin{equation}
\alpha = -\gamma \geq 0 \quad;\quad \beta\bigg\{\begin{matrix}\geq0&\text{if}\;\;\theta'\in[0,\pi/2]\\
<0&\text{if}\;\;\theta'\in(\pi/2,\pi]\\
\end{matrix}\, .
\end{equation}

\section{Quantum gates}
\label{app:gate}

\begin{figure}[tbh]
 \centering
 \includegraphics[width=0.49\textwidth]{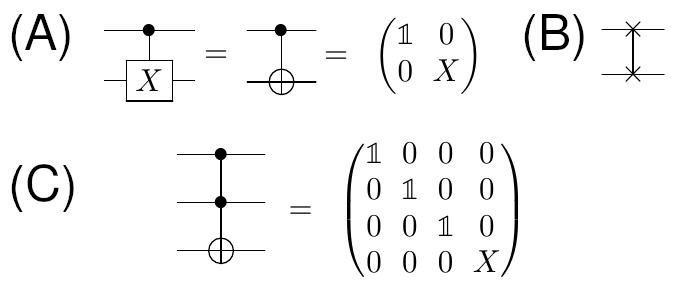}
 \caption{Circuit representation of common unitaries (see text).}
\label{fig:circuitF}
\end{figure}

In this section we provide more details on the conventions used in the definitions of the quantum gates used in the main text.
We indicate the single qubit Pauli matrices $\sigma_x$, $\sigma_y$ and $\sigma_z$ and the corresponding gates as $X$, $Y$ and $Z$. The Hadamard gate, denoted with $H$, was introduced in Eq.~\eqref{eq:Hadamard} in the main text.
The one qubit rotation gates $R_x$,  $R_{y}$ and $R_z$  used throughout this work are defined in the following way
\begin{eqnarray}
R_x(\phi)&=& \left(\begin{array}{cc}{\cos(\phi/2)} & {-i\sin(\phi/2)} \\ {-i\sin(\phi/2)} & {\cos(\phi/2)}\end{array}\right)\, ,\\
R_{y}(\phi) &=& \left(\begin{array}{cc}{\cos(\phi/2)} & {-\sin(\phi/2)} \\ {\sin(\phi/2)} & {\cos(\phi/2)}\end{array}\right)\, ,\\
R_z(\phi)&=& \left(\begin{array}{cc}{e^{-i\phi/2}} & 0\\ 0 & {e^{i\phi/2}}\end{array}\right)\label{eq:gateZ_def}\, ,
\end{eqnarray}
where we note that only the the rotation along the $z$-axis above differs from the implementation on IBM qiskit~\cite{qiskit} by an overall global phase $e^{i\phi/2}$.
The two-qubit gate used more frequently in the main text is the CNOT gate defined in Fig.~\ref{fig:circuitF}(A),
with $\mathbb{1}$ the $2\times2$ identity matrix.
We also use the ${\rm SWAP}$ gate which is represented by the symbol in Fig.~\ref{fig:circuitF}(B),
and described by the following unitary matrix
\begin{equation}
{\rm SWAP}=\begin{pmatrix}
1& 0& 0&0\\
0&0&1&0\\
0&1&0&0\\
0&0&0&1\\
\end{pmatrix}\;.
\end{equation}
Finally the 3-qubit Toffoli gate used in Fig.~\ref{fig:circuitD} of the main text and defined in Fig.~\ref{fig:circuitF}(C)
and compactly expressed as the following linear combination
\begin{eqnarray}
T&=&\left(\mathbb{1}-\ket{11}\bra{11}\right)\otimes \mathbb{1}+\ket{11}\bra{11}\otimes X\; .
\end{eqnarray}

\section{Circuit implementation details for the time dependent method}
Here we derive in detail the quantum circuits needed to implement the time-dependent state preparation described in Fig.~\ref{fig:circuitA}(A) of the main text.

We start by discussing the simplest case when the excitation operator $O$ is a single-qubit operator and express explicitly Fig.~\ref{fig:circuitA}(A) in terms of one qubit gates and CNOTs. Using the Euler decomposition for a unitary 2$\times$2 matrix, we can express the exact propagator $U(\gamma)=\exp(-i \gamma O)$ using 3 angles and a phase parameter
\begin{equation}
\label{Euler_rotation}
    e^{-i \gamma O}=e^{-i \delta }R_{z}(x_1) R_{y}(x_2) R_{z}(x_3)\;.
\end{equation}
The time-dependent method requires an implementation of this unitary controlled by an ancilla qubit and it is clear from this decomposition that this can be implemented using three controlled one qubit rotation gate and one controlled phase gate. Here we choose the decomposition $Z Y Z$ for the Euler angles since the rotation gate $R_z$ can usually be implemented with higher fidelity.

\begin{figure}[tbh]
 \centering
 \includegraphics[width=0.49\textwidth]{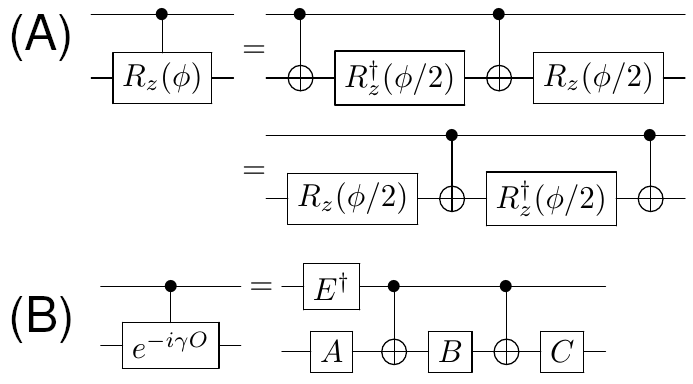}
 \caption{Controlled single qubit rotations: (A) single $z$ rotation, (B) general case.}
\label{fig:circuitG}
\end{figure}

A controlled $Z$ rotation can be written as in Fig.~\ref{fig:circuitG}(A).
A controlled $Y$ rotation can be written similarly to Fig.~\ref{fig:circuitG}(A) by replacing $R_z$ rotations with $R_y$ rotations. We can implement the controlled time-evolution operator by decomposing each controlled one qubit rotation separately. 

A more efficient approach is described in Ref.~\cite{barenco1995}(Lemma 5.1 and Lemma 5.2) and requires only two CNOTs. Using this strategy, the implementation of the controlled time-evolution operator is reported in Fig.~\ref{fig:circuitG}(B),
where we have defined the following gates
\begin{align}
    &A=\mathrm{R}_{\mathrm{z}}\left(\frac{x_3-x_1}{2}\right) ,\\
    &B=\mathrm{R}_{\mathrm{y}}\left(-\frac{x_2}{2}\right) \cdot \mathrm{R}_{\mathrm{z}}\left(-\frac{x_1+x_3}{2}\right) , \\
    &C=\mathrm{R}_{z}(x_1) \cdot \mathrm{R}_{\mathrm{y}}\left(\frac{x_2}{2}\right)  ,\\
    &E=\left(\begin{array}{cc}{1} & {0} \\ {0} & {e^{i \delta}}\end{array}\right).
\end{align}
Notice that the gate denoted by $E(\delta)$ is equivalent (up to an overall phase) to the $R_z(\delta)$ gate defined in Eq. ~(\ref{eq:gateZ_def}).

\subsection{Simple excitation operator}
\label{Cir: method 0 simple}
\begin{figure}[tbh]
 \centering
 \includegraphics[width=0.485\textwidth]{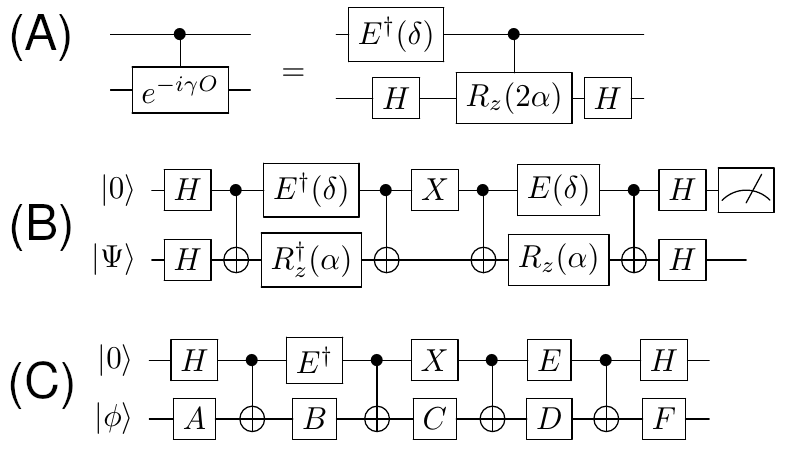}
 \caption{Quantum circuits described in the text.}
\label{fig:circuitH}
\end{figure}

Let us consider the simple model excitation
\begin{equation}
O (\theta)= \cos(\theta)X+\sin(\theta)\mathbb{1}\label{eq:ex_op_simple}\; .
\end{equation}
We first notice that the propagator $U(\gamma)$ is a rotation around the $X$ axis
\begin{equation}
e^{-i\gamma O}=e^{-i\gamma\sin(\theta)}R_x\left(2\gamma\cos(\theta)\right) \ , 
\end{equation}
because the constant term contributes only to a global phase.
We use the the identity $R_x(\theta) = H R_z(\theta) H$, and the controlled time-evolution operator can be expressed as in Fig.~\ref{fig:circuitH}(A)
with $\delta = \gamma \sin{\theta}$ and $\alpha = \gamma \cos{\theta}$.

The final implementation of the complete state-preparation circuit with the excitation operator Eq.~\eqref{eq:ex_op_simple} is given in  Fig.~\ref{fig:circuitH}(B).
Here we used the two different variants in Fig.~\ref{fig:circuitG}(A) to remove two rotations.

\subsection{Nuclear excitation operator}
\label{app:circ_td_npdg}
Here we present the quantum circuit for the nuclear excitation operator from Eq.~\eqref{eq:nuclear_op}
\begin{equation}
    O(\theta^{\prime}) = \alpha I+\beta X+\gamma Z \ .
\end{equation}
The rotation obtained by exponentiation of this operator can be again decomposed in Euler angles as above and the result for $U^{\dagger}(\gamma)$ reads
\begin{equation}
    \begin{aligned}
    e^{i \gamma O}&= e^{i \delta }R_{z}(-x_1) R_{y}(x_2) R_{z}(-x_3)\ .
\end{aligned}
\end{equation}
Here the rotation angles are those for the original $U^{\dagger}(\gamma)$ as in Eq.~\eqref{Euler_rotation} instead. We then use Fig.~\ref{fig:circuitG}(B) twice to get the final circuit as in Fig.~\ref{fig:circuitH}(C),
where
\begin{align}
C&=\mathrm{R}_{\mathrm{z}}\left(\frac{x_1-x_3}{2}\right) \cdot \mathrm{R}_{\mathrm{z}}(x_1) \cdot \mathrm{R}_{\mathrm{y}}\left(\frac{x_2}{2}\right) ,\\
D&=\mathrm{R}_{\mathrm{y}}\left(-\frac{x_2}{2}\right) \cdot \mathrm{R}_{\mathrm{z}}\left(\frac{x_1+x_3}{2}\right),\\
F&=\mathrm{R}_{\mathrm{z}}\left(-x_1\right) \cdot \mathrm{R}_{\mathrm{y}}\left(\frac{x_2}{2}\right)  \ ,
\end{align}
while gate $A$ and $B$ are the same as defined in Fig.~\ref{fig:circuitG}(B).

\section{Circuit implementation for the LCU method}
We derive and present the final expressions of the circuits used with the LCU method in the main text.
\subsection{Simple excitation operator}
\label{app:full_lcuA}
\begin{figure*}[bth]
 \centering
 \includegraphics[width=0.9\textwidth]{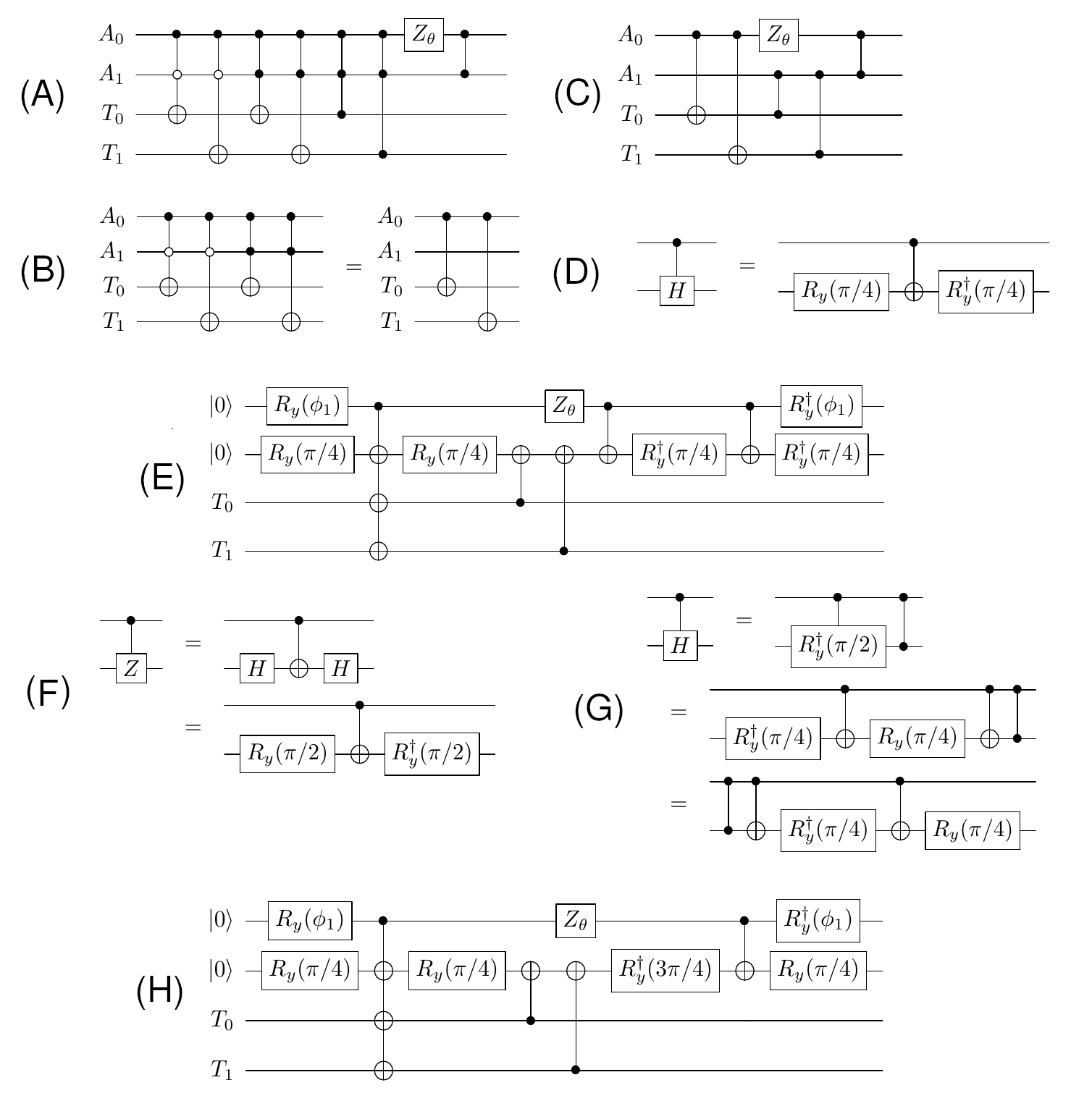}
 \caption{Quantum circuits described in the text.}
\label{fig:circuitI}
\end{figure*}

 The simple excitation operator of interest is
\begin{equation}
O_2(\theta) = \frac{\cos(\theta)}{2}\left(X_0X_1 + Y_0Y_1\right) + \sin(\theta) \mathbb{1}\; .    \label{eq:siple_op_LCU}
\end{equation}
Here $\mathbb{1}$ denotes the $4\times4$ identity.
We use the following mapping between ancillary qubit states and the operators on the left-hand side of Eq.~\eqref{eq:siple_op_LCU}
\begin{equation}
\label{eq:app_lcu_A_map}
\ket{00}\rightarrow\mathbb{1}\, ,\quad\ket{10}\rightarrow X_0 X_1\,,\quad\ket{11}\rightarrow Y_0Y_1\;.
\end{equation}
The circuit corresponding to the {\it select} unitary is displayed in Fig.~\ref{fig:circuitD} of the main text, 
where $A_0$ and $A_1$ refer, respectively, to the first and second ancillary qubit while $T_0$ and $T_1$ denote the target qubits.
We recall that the circuit above should be applied after the {\it prepare} unitary circuit from Fig.~\ref{fig:circuitC}(C).
In order to simplify the {\it select} circuit in Fig.~\ref{fig:circuitD}, 
we use the identity
\begin{equation}
\label{eq:app_xz_to_y}
X_0Z_0X_1Z_1 = -Y_0Y_1 \ .
\end{equation}
This leads to the equivalent {\it select} circuit displayed in Fig.~\ref{fig:circuitI}(A).
Here the last controlled-$Z$ operation has been added to correct the sign of the $Y_0Y_1$ term.

We note that the second and third gates in Fig.~\ref{fig:circuitI}(A) commute with each other. This implies the identity in Fig.~\ref{fig:circuitI}(B).
Furthermore, we are not using the state $\ket{01}$ of the ancilla register in the mapping Eq.~\eqref{eq:app_lcu_A_map}. Thus, we can directly remove the controls on the first ancillary qubit for the fifth and sixth gate in Fig.~\ref{fig:circuitI}(A). 
The final circuit for {\it select} becomes as in Fig.~\ref{fig:circuitI}(C).

We now turn to discuss the implementation of the {\it prepare} unitary and consider the circuit presented in Fig.~\ref{fig:circuitC}(C) of the main text. We start by using the decomposition 
\begin{equation}
\label{eq:app_H_dec}
\begin{split}
H &= X R_y(\pi/2)= R_y^\dagger(\pi/4) X R_y(\pi/4)\;,
\end{split}
\end{equation}
to implement the controlled-$H$ operation as described in Fig.~\ref{fig:circuitI}(D).
This only requires a single CNOT gate. Using these identities, the full circuit needed for LCU implementation of the operator in Eq.~\eqref{eq:siple_op_LCU} is reported in Fig.~\ref{fig:circuitI}(E).
In this circuit  we used for the first {\it prepare} unitary the implementation from the first line of Eq.~\eqref{eq:app_H_dec} while for the final {\it prepare} unitary we used the second line of Eq.~\eqref{eq:app_H_dec} and then took the inverse. In addition, we also used the identity from Fig.~\ref{fig:circuitI}(F)
which can be obtained using Eq.~\eqref{eq:app_H_dec}. It is also possible to eliminate one additional CNOT gate by using a different decomposition of the controlled-$H$ gate as described in Fig.~\ref{fig:circuitI}(G) and
obtained from $H=ZR^\dagger_y(\pi/2)$. We notice that if the bottom qubit is set to $\ket{0}$ before applying the circuit on the second line above one can drop the second CNOT gate as it only contributes a global phase. The same is true if instead we measure the projector in the bottom qubit state $\ket{0}$. We finally obtain the circuit in Fig.~\ref{fig:circuitI}(H).

\subsection{Nuclear excitation operator}
\label{sec:lcu_nuc_op}

\begin{figure*}[bth]
 \centering
 \includegraphics[width=0.9\textwidth]{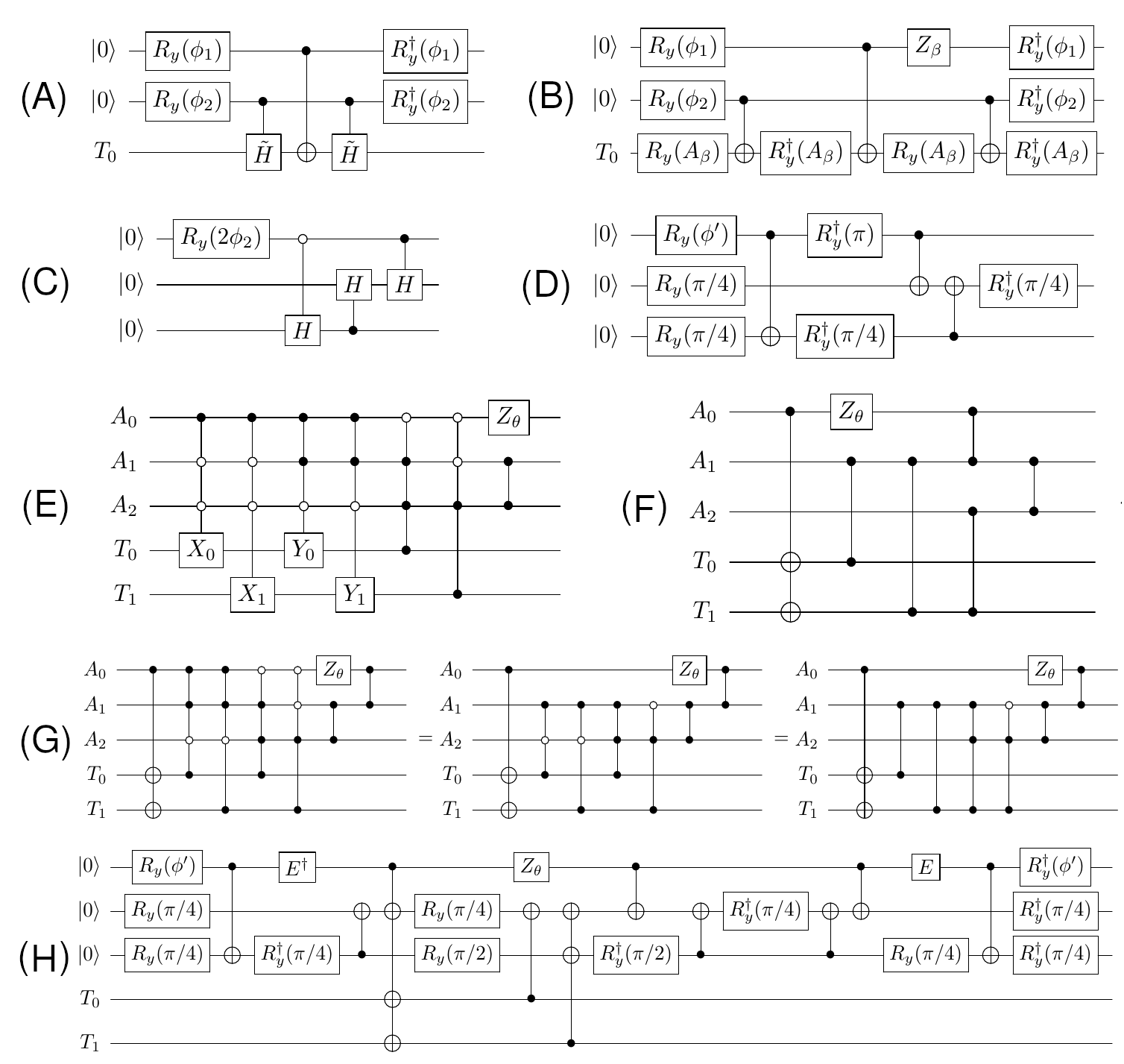}
 \caption{Quantum circuits described in the text.}
\label{fig:circuitL}
\end{figure*} 

We show how to implement the LCU oracle for the more complicated case of the nuclear excitation operator
\begin{equation}
\label{eq:lcu_nuc_op}
O(\theta) = \alpha\mathbb{1}+\beta X-\alpha Z \;,
\end{equation}
where the two real constants are given by Eq.~\eqref{eq:nuclear_op}. We choose the state-to-operator mapping as follows
\begin{equation}
\begin{split}
&\ket{00}\rightarrow\mathbb{1} \quad\ket{01}\rightarrow \mathbb{1} \quad\ket{10}\rightarrow 
X \quad\ket{11}\rightarrow -Z\;.
\end{split}
\end{equation}
For the case $\beta\geq 0$, the full circuit is given in Fig.~\ref{fig:circuitL}(A),
where the $\tilde{H}$ is defined as $\tilde{H}=R_{y}^{\dagger}(3 \pi / 4) X R_{y}(3 \pi / 4)$, we have the identity $\tilde{H} X \tilde{H} = -Z$. Also, the angles are
\begin{equation}
\begin{aligned}
\phi_1 &= 2\arcsin{\left(\sqrt{\frac{\alpha+|\beta|}{2\alpha+|\beta|}}\right)}\;,\\
\phi_2 &= 2\arcsin{\left(\sqrt{\frac{\alpha}{\alpha+|\beta|}}\right)}\;.
\end{aligned}
\end{equation}
If $\beta\leq 0$ instead, we express the Hadamard gate $H$ (modulo a phase) as $H=R_{y}(3 \pi / 4) X R_{y}^{\dagger}(3 \pi / 4)$ and use $H X H = Z$. We account for the negative sign in front of the last term in Eq.~\eqref{eq:lcu_nuc_op} by applying the $Z$ gate to the first ancilla qubit. The final circuit is shown in Fig.~\ref{fig:circuitL}(B),
where similarly to Eq.~\eqref{eq:ztheta} we use
\begin{equation}
\begin{split}
Z_\beta &\equiv \bigg\{\begin{matrix} \mathbb{1}&\text{for}\quad\beta\geq0\\
Z&\text{for}\quad\beta<0\end{matrix}\\
\end{split}
\end{equation}and the additional angle
\begin{equation}
\begin{split}
A_\beta &= \bigg\{\begin{matrix} \frac{3\pi}{4}&\text{for}\quad\beta\geq0\\
\frac{-3\pi}{4}&\text{for}\quad\beta<0\end{matrix} \;.\\
\end{split}
\end{equation}

As explained in the main text, to test the flexibility of the LCU method on a more scalable version of the problem we also consider the implementation of the nuclear excitation operator in second quantization. In the following, we show how to implement the LCU oracle for the operator $\overline{O}_M (\theta)$ introduced in Eq.~\eqref{eq:omtheta} as
\begin{equation}
\begin{split}
\label{eq:obar3q_M}
\overline{O}_M (\theta) &= \sin(\theta)\mathbb{1} +\frac{\cos(\theta)}{2}(X_0X_1+Y_0Y_1)\\
&-\frac{\sin(\theta)}{2}(Z_0-Z_1).
\end{split}
\end{equation}
As we note in the main text, this operator differs from the original second quantized operator in Eq.~\eqref{eq:obar3q} by a global scale factor $\sqrt{\alpha^2+\beta^2}$.

Using an ancilla register of three qubits, it is useful to consider the following state-to-operator map
\begin{equation}
\label{eq:app_3q_map}
\begin{split}
&\ket{000}\rightarrow\mathbb{1}\quad\ket{001}\rightarrow Z_1\quad\ket{011}\rightarrow -Z_0\\
&\quad\quad\ket{100}\rightarrow X_0X_1\quad\ket{110}\rightarrow Y_0Y_1\;,
\end{split}
\end{equation}
to construct the {\it prepare} and {\it select} unitaries. The former can be chosen as depicted in Fig.~\ref{fig:circuitL}(C),
with the angle
\begin{equation}
\label{eq:t2a3angle}
\phi_2 = \arcsin{\left(\sqrt{\frac{\lvert\cos(\theta)\rvert}{2\sin(\theta)+\lvert\cos(\theta)\rvert}}\right)}\;.
\end{equation}

\begin{figure*}[tbh]
 \centering
 \includegraphics[width=0.9\textwidth]{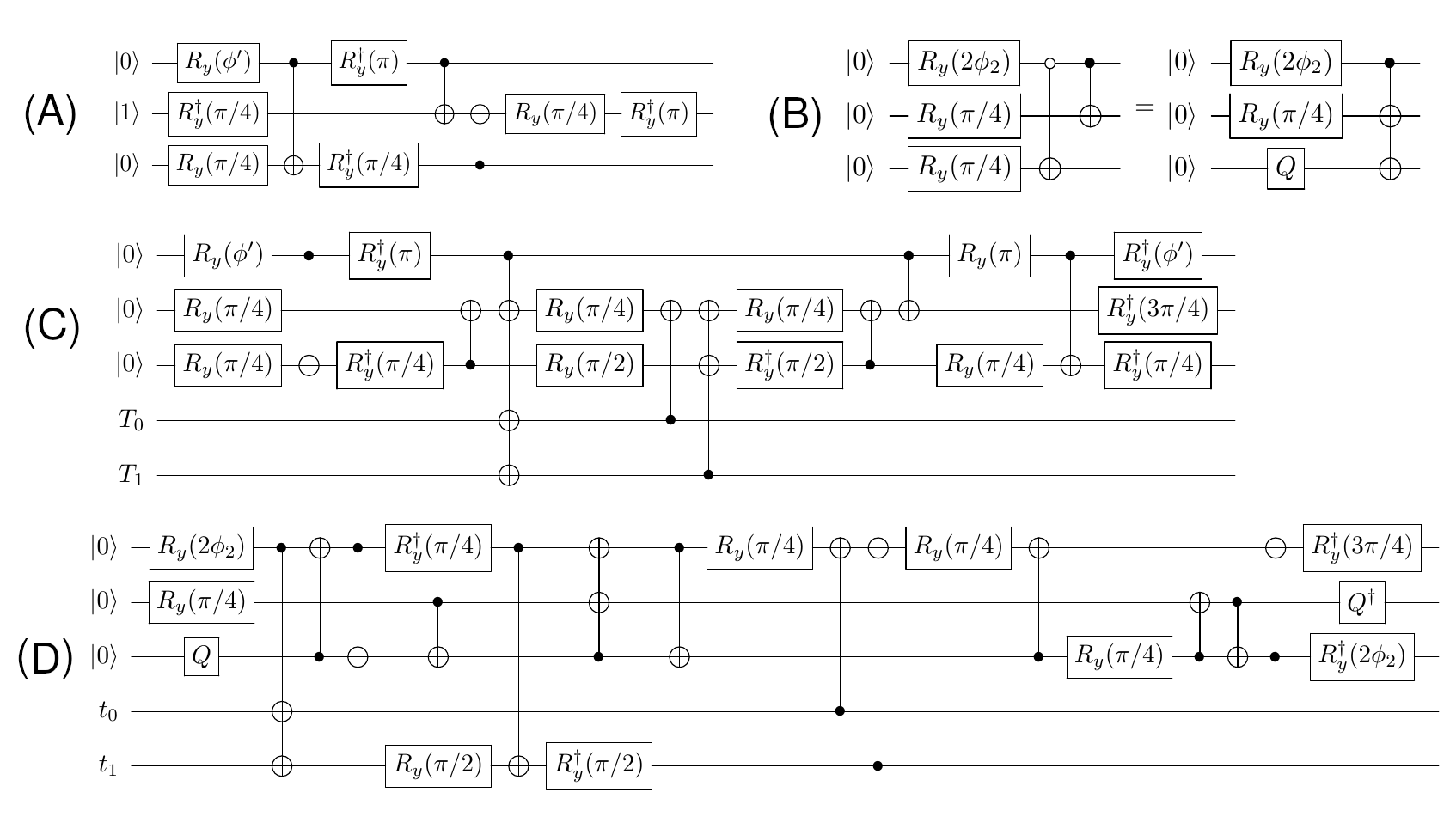}
 \caption{Quantum circuits described in the text.}
\label{fig:circuitM}
\end{figure*} 

The flag state prepared by this circuit reads
\begin{equation}
\begin{split}
\ket{\Phi_3} 
&=\frac{\cos(\phi_2)}{\sqrt{2}}\left(\ket{000}+\frac{\ket{001}+\ket{011}}{\sqrt{2}}\right)\\
&+\frac{\sin(\phi_2)}{\sqrt{2}}\left(\ket{100}+\ket{110}\right)\;.
\end{split}
\end{equation}

Using results and notations from the previous Section we can give an explicit implementation as depicted in Fig.~\ref{fig:circuitL}(D),
where $\phi^\prime=2\phi_2+\pi$.
This requires only three CNOT gates with fully connected qubits.
 
The {\it select} unitary corresponding to the mapping in Eq.~\eqref{eq:app_3q_map} can be represented as depicted in Fig.~\ref{fig:circuitL}(E),
with $Z_\theta$ defined in Eq.~\eqref{eq:ztheta} and used
to account for the sign associated with $\cos(\theta)$ and to recover the relative sign difference between $Z_0$ and $Z_1$ in Eq.~\eqref{eq:obar3q_M}. As done for the simpler four-qubit case in Sec.~\ref{sec:lcuresults}, we exploit the fact that three states in the ancilla register are left unused to construct the implementation of $V_S$ depicted in Fig.~\ref{fig:circuitL}(F).
This requires only seven entangling operations among pairs of qubits. In order to obtain this circuit we used the trick from Eq.~\eqref{eq:app_xz_to_y} and wrote the circuit in Fig.~\ref{fig:circuitL}(E) as reported in Fig.~\ref{fig:circuitL}(G).


With one additional step we obtain Fig.~\ref{fig:circuitL}(F). 
The total circuit from Fig.~\ref{fig:circuitB} of the main text can be expressed as in Fig.~\ref{fig:circuitL}(H).

This circuit has $13$ CNOT gates and can be further simplified as we show next.
We can absorb two of the controlled-$Z$ operation by unpreparing the following flag state (slightly different from the previously prepared state)
\begin{equation}
\begin{split}
\ket{\widetilde{\Phi}_3} 
&=\frac{\cos(\phi_2)}{\sqrt{2}}\left(\ket{000}+\frac{\ket{001}-\ket{011}}{\sqrt{2}}\right)\\
&+\frac{\sin(\phi_2)}{\sqrt{2}}\left(\ket{100}-\ket{110}\right)\;,
\end{split}
\end{equation}
which can be produced using a modified {\it prepare} unitary as depicted in  Fig.~\ref{fig:circuitM}(A).
Here, the second ancilla qubit is initialized in $\ket{1}$. This scheme is very similar in spirit to the asymmetric qubitization introduced in~\cite{babbush2019}.
The total circuit is now given by the expression in Fig.~\ref{fig:circuitM}(C)
and the number of CNOT gates is reduced to $11$.
The final hardware-efficient circuit takes into account the connectivity constraints. It consists of a total number of 14 CNOT gates, and we have explicitly the circuit in Fig.~\ref{fig:circuitM}(D).
Here, $Q=R_y(\pi/4)X$ is obtained as in Fig.~\ref{fig:circuitM}(B). 

\section{Depolarizing noise and ratio estimators}
\label{app:ratio}
In this section we show how the ratio estimators for the transition probabilities, like Eq.~\eqref{eq:p1_tdm} for the time-dependent case and Eq.~\eqref{eq:estimator_2} for the LCU method, exhibit error resilience for a simple error model. We consider the simple situation where the whole system is initially in a pure state $\rho=\rvert\Psi\rangle\langle\Psi\lvert$ 
and then becomes subject to a depolarizing noise channel
\begin{equation}
\label{eq:dep}
\Omega_\epsilon[\rho]=(1-\epsilon)\rho+\frac{\epsilon}{d}\mathbb{1}\;.
\end{equation}
Here $d=2^q$ for $q=n+k$ qubits, with $n$ qubits for the system register and $k$  ancillary qubits. We recall that for the time-dependent method $k=1$ and $k=\lceil\log(L+1)\rceil$ for the LCU-based method. The two observables we consider are the success probability $P_s$  and the transition probability $m$. They can be expressed  as
\begin{equation}
\label{eq:obs_errors}
P_s={\rm Tr}[\Pi_s \rho]\, ,\quad m=\frac{1}{P_s}{\rm Tr}[\Pi_s\Pi_{fin}\rho ]\, ,
\end{equation}
where $\Pi_s$ is a rank-$K$ projector with $K=q-k=n$ and $\Pi_{fin}$ is a rank-1 projector onto the final scattering state. For the LCU method we can find situations where the success probability is known beforehand to be equal to $\widetilde{P_s}$ (for instance in the problem discussed in Sec.~\ref{sec:res} where $\widetilde{P_s}=1/\Lambda^2$), in this case we want to consider the alternative estimator
\begin{equation}
\widetilde{m} = \frac{1}{\widetilde{P_s}}{\rm Tr}[\Pi_s\Pi_{fin}\rho ]\;,
\end{equation}
which is equivalent to $m$ in absence of errors.

We now want to compare the effect of the depolarizing channel Eq.~\eqref{eq:dep} on the two estimators $m$ and $\widetilde{m}$ and show that the former is more stable to such perturbations. We will denote observables evaluated on the modified density matrix $\rho_\epsilon=\Omega_\epsilon[\rho]$ with the superscript $E$. For the success probability we have
\begin{equation}
P_s^E = {\rm Tr}\left[\Pi_s\Omega_\epsilon[\rho]\right] = (1-\epsilon)P_s + \frac{\epsilon}{d}{\rm Tr}[\Pi_s]\, ,
\end{equation}
and from the definition of $\Pi_s$ we arrive at
\begin{equation}
P_s^E = (1-\epsilon)P_s + \frac{\epsilon}{2^{k}}\;.
\end{equation}
Similarly for the transition probabilities we have
\begin{equation}
\label{eq:mat_elem_trace}
\begin{split}
m^E &=\frac{1}{P_s^E} {\rm Tr}\left[\Pi_s\Pi_{fin}\Omega_\epsilon[\rho]\right] \\
&=\frac{1}{P_s^E}\left( (1-\epsilon)P_sm+\frac{\epsilon}{d} \right)\;,
\end{split}
\end{equation}
and for the alternative version instead
\begin{equation}
\label{eq:mat_elem_trace2}
\widetilde{m}^E =\frac{1}{\widetilde{P_s}} {\rm Tr}\left[\Pi_s\Pi_{fin}\Omega_\epsilon[\rho]\right] = (1-\epsilon)\widetilde{m}+\frac{\epsilon}{d\widetilde{P_s}} \;.
\end{equation}
In both of these derivations we used the definition of $\Pi_s\Pi_{fin}$ which implies ${\rm Tr}\left[\Pi_s\Pi_{fin}\right]=1$.

We start by considering the absolute error for the estimator in Eq.~\eqref{eq:mat_elem_trace2}, this is given by
\begin{equation}
 \Delta \widetilde{m} = \epsilon\left|\frac{1}{d\widetilde{P_s}}-\widetilde{m}\right|\approx\epsilon \widetilde{m}\; ,
\end{equation}
where the last approximation valid in the limit $d\widetilde{P_s}\gg1$.
We consider now the ratio estimator from Eq.~\eqref{eq:mat_elem_trace}, it is convenient to first express it as
\begin{equation}
\begin{split}
m^E &= \frac{(1-\epsilon)P_s m+\frac{\epsilon}{d}}{(1-\epsilon)P_s + \frac{\epsilon}{2^{k}}}\\
&= \frac{(1-\epsilon)P_s m+\frac{\epsilon}{d}}{(1-\epsilon)P_s}-\frac{\epsilon}{2^k}\frac{(1-\epsilon)P_sm+\frac{\epsilon}{d}}{\left((1-\epsilon)P_s + \epsilon\xi \right)^2}\\
&= m+\frac{\epsilon}{1-\epsilon} \frac{1}{dP_s} -\frac{\epsilon}{2^k}\frac{(1-\epsilon)P_sm+\frac{\epsilon}{d}}{\left((1-\epsilon)P_s + \epsilon\xi \right)^2}\;,\\
\end{split}
\end{equation}
for some $\xi\in[0,1/2^k]$. In the second line of the expression above we used directly Lagrange's remainder theorem as we did in Eq.~\eqref{eq:sintapp} above. This yields the error estimate
\begin{equation}
\label{eq:fin_ineq_noise}
\begin{split}
\Delta m &=\epsilon  \left|\frac{1}{1-\epsilon} \frac{1}{dP_s} -\frac{1}{2^k}\frac{(1-\epsilon)P_sm+\frac{\epsilon}{d}}{\left((1-\epsilon)P_s + \epsilon\xi \right)^2}\right|\\
&\leq\frac{\epsilon}{1-\epsilon}  \left( \frac{1}{dP_s} +\frac{(1-\epsilon)P_sm+\frac{\epsilon}{d}}{2^k(1-\epsilon)P_s^2 }\right)\\
&=\frac{\epsilon}{1-\epsilon}  \left( \frac{1}{dP_s} +\frac{m}{2^kP_s}+ \frac{\epsilon}{1-\epsilon} \frac{1}{d2^{k}P_s^2}\right)\\
&\approx\frac{\epsilon}{1-\epsilon} \frac{m}{2^kP_s}=\frac{\Delta \widetilde{m}}{(1-\epsilon)2^kP_s}\;,
\end{split}
\end{equation}
with the approximation in the last line valid for $d\widetilde{P_s}\gg1$ as above. Notice now that for the LCU method we also have $2^kP_s\gg 1$ whenever we use enough ancillas and therefore the relative error is exponentially suppressed.
For the time-dependent method instead the inequality in Eq.~\eqref{eq:fin_ineq_noise} does not lead to a suppression since $k=1$ in the algorithm described in the main text, note however that one can still add redundant qubits perform post-selection at the end. 

\section{Error Mitigation}
\label{app:err_mitigation}
 We want to describe now the error mitigation procedures employed in the main text. We used a modified version of the protocol of Ref.~\cite{roggero2020A} which is based on techniques from Refs.~\cite{li2017,temme2017,Endo2018}. The approach consists of two main steps. First we correct for the readout error noise (or measurement errors). This initial correction requires in principle $2^n$ measurements for $n$ qubits, however as we will show in the next subsection, we are able to reduce the total number of measurements to $2n$, assuming that the readout error is qubit independent. The second step of the error mitigation procedure  aims to correct gate noise. Here we assume, consistent with  the current error rates of IBMQ machine, that the dominant noise channel is the one associated with the execution of CNOT gates. Therefore given a circuit that implements the desired algorithm, we generate additional circuits, logically equivalent to the original one but with multiple CNOT gates. (We note that products of two CNOT gates are the identity.) We compute the wanted observables at different noise levels, and extrapolate them to the zero-noise limit. The zero-noise-extrapolation procedure is scalable as it adds only a linear overhead in the number of gates to the original circuit depth (this can be further improved using a stochastic variant, see eg.~\cite{he2020}).

 \subsection{Readout error mitigation}
We will review here the readout-error correction scheme used in this work (see also~\cite{kandala2017}). In the following we will assume the read-out errors are independent for different qubits and can be described in terms of two parameters:
\begin{itemize}
    \item $e_0$: probability to get $\ket{1}$ when we prepare $\ket{0}$
    \item $e_1$: probability to get $\ket{0}$ when we prepare $\ket{1}$
\end{itemize}
which are both zero in the limit of no noise. These parameters can be obtained by performing two independent calculations: the first one prepares $\ket{0}$ and measures, the second prepares $\ket{1}$ and measures. The result of this calibration experiment is a calibration matrix $P$ with entries
 \begin{equation}
 \label{eq:cab_mat}
 P = \begin{pmatrix}
 1-e_0 & e_1 \\
 e_0 & 1-e_1 \\ 
 \end{pmatrix} \;.
 \end{equation}

We can now use this information to mitigate this error source when computing observables, we now focus on the single qubit case and comment on generalizations to the more general multi-qubit scenario at the end. We will indicate with ${\bf p}=(p,1-p)$ the probability vector in the limit of no errors and with ${\bf p}_e=(p_e,1-p_e)$ the corresponding probabilities with read-out errors. The two are related through the calibration matrix and it's inverse
\begin{equation}
P \cdot {\bf p} = {\bf p}_e\;\Rightarrow\;P^{-1} \cdot {\bf p}_e = {\bf p}\;,
\end{equation}
and an explicit expression for the inverse transformation can be found
\begin{equation}
 P^{-1}=\frac{1}{d}\begin{pmatrix}
 1-e_1& -e_1\\
 -e_0& 1-e_0
 \end{pmatrix}\, ,
 \end{equation}
where we have introduced $d=1-e_0-e_1$. Note that the error parameters $(e_0,e_1)$ are known only with finite precision $\Delta e_0$ and $\Delta e_1$ and this can have an important effect when applying this inverse transformation directly. In particular, it is possible that the resulting mitigation probability vector ${\bf p}$ contains unphysical negative components. In the calculations presented in this paper, we first estimate the expected error on the mitigated probabilities as explained below, if we find negative probabilities compatible with zero within the uncertainty band we manually set it to zero, otherwise we perform an additional least square inversion enforcing the calibration matrix to be physical (as implemented in Qiskit-Ignis). We then use the resulting probabilities as corrected central values for the mitigated result while we keep the same error estimates coming from linear inversion. In practice we never had to use this fall-back procedure for the calculations reported in this work.

Here we will indicate with ${\rm Var}[M]$ the matrix containing the variances of the matrix elements of a matrix $M$. With this notation we have for the calibration matrix
\begin{equation}
{\rm Var}[P] = \begin{pmatrix}
\Delta e_0^2&\Delta e_1^2\\
\Delta e_0^2&\Delta e_1^2\\
\end{pmatrix}\;,
\end{equation}
while for the inverse matrix the variance to first order is
\begin{equation}
{\rm Var}[P^{-1}] =\frac{1}{d^2} \begin{pmatrix}
l_1&r_1\\
r_0&l_0\\
\end{pmatrix}\;,
\end{equation}
where we have introduced
\begin{equation}
\begin{split}
l_k & = \Delta e_k^2+(1-e_k)^2 \Delta_{01}^2\\
r_k & = \Delta e_k^2+e_k^2 \Delta_{01}^2\\
\end{split}
\end{equation}
together with
 \begin{equation}
\Delta_{01} = \frac{\Delta e_0^2+\Delta e_1^2}{d^2}\;.
 \end{equation}

It is now easy to propagate uncertainties to the final observable, here we describe in more detail the case of a single qubit observable but this can be easily generalized. Consider the following expectation value
\begin{equation}
\langle O\rangle = {\bf a}\cdot{\bf p} = a_0p_0+a_1p_1\;,
\end{equation}
we can estimate it's value using the error affected probabilities ${\bf p}_e$ using the inverse of the calibration matrix
\begin{equation}
\begin{split}
\langle \widetilde{O}\rangle &= {\bf a}\cdot P^{-1}{\bf p}_e = \sum_{ij} a_i \left[ P^{-1}\right]_{ij} \left[p_e\right]_j \;,
\end{split}
\end{equation}
together with the error estimate
\begin{equation}
\begin{split}
{\rm Var}[\langle \widetilde{O}\rangle] =& \sum_i a^2_i \left(\sum_j \left[ P^{-1}\right]_{ij}^2 \Delta [p_e]^2_j \right)\\
&+ \sum_i a^2_i \left(\sum_j {\rm Var}[P^{-1}]_{ij} \left[p_e\right]_j^2\right)\;.
\end{split}
\end{equation}

In the more general multi-qubit case the above equations carry out in a straightforward generalization, and for $n$ qubits we have in principle to perform $2^n$ measurements in order to build the calibration matrix given in Eq.~\eqref{eq:cab_mat}. This cost can be substantially reduced if it is reasonable to assume that errors happen independently on different qubits, in this case the full calibration matrix factors into a series of $n$ diagonal $2\times 2$ blocks and the number of measurement needed is only $2n$. For the results reported in this work we checked that the assumption of independence is realize to a good accuracy by performing explicitly the full $2^n$ calibration runs.

 \subsection{Zero-noise extrapolation}
 
We now describe in more detail the zero-noise extrapolation described briefly above. The idea is to collect data at different noise levels and then use a sensible parametrization of the noise dependence of an observable to extract a noise free estimator. Since, as discussed above, in the quantum devices used in this work the CNOT gate is the dominant source of noise, we will use the total number of CNOT $N_C$ in the circuit considered as a proxy to the noise level.  

Results at higher noise levels are obtained by generating a series of equivalent circuits adding even number of CNOT gates for each CNOT in the original circuit. We then compute the observables and the corresponding errors for each one of the circuits and denote them with $(O_{k},E_{k})$
where $k\geq0$ indicates the total number of CNOT gates, given by $N_C(k) = (2k-1)N_C$, used in the noise-amplified circuit used to extract the expectation value. For example $k=1$ and $k=2$ denote two circuits having respectively $N_C$ and $3N_C$ CNOT gates. We will also use $O_F$ to indicate the error-free expectation value we are trying to estimate.
In absence of gate noise all the pairs $(O_{k},E_{k})$ should be identical, but when CNOT gates have a failure probability $p_{\varepsilon}$ the effective failure probability will grow linearly with $k$.

 The mitigation is performed attempting three different types of zero-noise-extrapolation. In particular for low noise levels we perform a Richardson extrapolation, proposed in Ref.~\cite{temme2017}, and a polynomial extrapolation, while for larger noise levels we use a two point exponential extrapolation~\cite{Endo2018,endo2019}. These techniques are briefly described below.
 \begin{itemize}
 	\item Richardson extrapolation: for small error probabilities $p_\varepsilon$ we can expand the observable $O$ as a power series in the error $\epsilon = N_C p_\varepsilon$
 	\begin{equation}
 	O(\epsilon) = O_F + \sum_{j=1}^{M} c_j \epsilon^j + \mathcal{O}\left(\epsilon^{M+1}\right)\;,
 	\end{equation}
 	where $O_F$ is the error-free result and $c_j\in\mathbb{R}$ are expansion coefficients. The idea is to determine $O_F$ by inverting the polynomial expression above using $M+1$ evaluations of $O(\epsilon)$ for different values of $\epsilon$. We will use the error-amplified results $O_k$ defined above for this purpose. 
 	For instance, in the simple case $M=1$ we have the following estimator for $O_F$
 	\begin{equation}
 	O_F = \frac{1}{2}\left(3O_1-O_2\right) + \mathcal{O}\left(\epsilon^2\right)
 	\end{equation}
 	which is correct up to first order in the error $\epsilon$. This construction can be easily generalized to higher orders, in the calculations presented in the main text we attempt extrapolations up to order $k=4$ (corresponding to a 9-fold increase in CNOT count).
 	\item Polynomial: in this case we perform a polynomial fit to $O(\epsilon)$ using all the available results as it has been done in Refs.~\cite{li2017,Dumitrescu2018}. In this work we use 4 results at progressively higher noise levels ($O_1$ through $O_4$) and attempt to perform polynomial fits up to third order. The lowest order polynomial that achieves a result with $\chi^2\leq1$ is the preferred extrapolation (see also Sec.~\ref{sec:extr_valid} for additional details on how we combine different extrapolations)
 	\item Exponential: in this case we use a simple two point exponential fit to the results by assuming the following functional dependence of the observable from the noise strength $\epsilon$
  	\begin{equation}
 	\label{eq:Ohn_exp}
 	O(\epsilon) = O_F\exp(-A\epsilon) \quad\text{with}\quad A>0\;.
 	\end{equation}
 	We can now find the parameter $A$ and an estimate for the error free result $O_F$ by combining two values of $O(\epsilon)$ at different noise levels. For instance if we take $O_1=O(\epsilon)$ and $O_k=O((2k-1)\epsilon)$ we find
 	\begin{equation}
 	O_F = O_1 \exp(A\epsilon) = O_1 \left(\frac{O_1}{O_k}\right)^{\frac{1}{2(k-1)}}\;.
 	\end{equation}
 \end{itemize}

In the remaining subsections we describe how we test for consistency in these extrapolations and how we combine the extrapolated results into a single value. 

 \subsubsection{Validating extrapolation results}
 \label{sec:extr_valid}
 It is important to check a posteriori that the results of the zero noise extrapolations described above are consistent with the data. This can help in both detecting hardware problems in specific runs (cf. filtered results in~\cite{roggero2020A}) but also in identifying the best extrapolation procedure for a given dataset and target observable.
 
 In this work we use the same procedure devised in~\cite{roggero2020A} and explained in more details in this section. In our work we always attempt to use 3 additional data points at higher error levels (corresponding to the choices $k=2,3,4$ for the CNOT repetitions), and for convenience we denote with $N^{max}_{ex}$ the maximum order in this expansion (i.e. $N^{max}_{ex}=4$ in our case). We then use the procedure detailed below to check for internal consistency of the $N^{max}_{ex}$ extrapolated results $\{O^e_k\}_{k=1,\dots,N^{max}_{ex}}$ and their associated error estimates $\{E^e_k\}_{k=1,\dots,N^{max}_{ex}}$. In this notation, the result obtained after read-out error mitigation, denoted with a superscript $r$, corresponds to the pair $(O^e_1,E^e_1)=(O^r_1,E^r_1)$.
 
 Before describing the algorithm for consistency check, we need to introduce one more definition: we will say that two data points $(O_A,E_A)$ and $(O_B,E_B)$ are {\it compatible} with each other at the $m\sigma$ level if
 \begin{equation}
\left|O_A-O_B\right|\leq m \sqrt{E_A^2+E_B^2}\quad m\geq1\;.
 \end{equation}
 We will indicate this relation compactly as 
 \begin{equation}
(O_A,E_A) \stackrel{m}{==} (O_B,E_B)\;.
 \end{equation}
 
 We can now describe the consistency check procedure we used. We will consider the extrapolation at some order $k$ to have succeeded if the result $(O^e_k,E^e_k)$ is {\it compatible} at the $1\sigma$ level~\footnote{the effect of making a different choice for $m$ are explored in previous work~\cite{roggero2020A}} with all the higher ones $(O^e_l,E^e_l)$ for $l>k$, if this check fails for $k>1$ we increase by one an internal error count. The reason behind this is that if a results changes significantly by adding one more order, the lower order extrapolation is unlikely to have converged. At the same time we want to keep the order as small as possible since higher order extrapolations will rely on possibly much more noisy data.
 
 We use the algorithm described in pseudo code in Algorithm~\ref{alg:1} to determine the lowest order result that is compatible with the available higher order ones and also collect the number of failures using the error count. If the check returns this corresponds to $k=1$ then we flag the result as not needing extrapolation. If instead the check returns $k>1$ but an error count of zero, we flag the result as {\it error free}. Finally, if no order passes the consistency check we flag the result as {\it failed}.

\begin{algorithm}[H]
\caption{Consistency check\label{alg:1}}
    \begin{algorithmic}[1]
    \State define BOOL array {\it ctest} of size [$N^{max}_{ex}$]
    \State $error\_count \leftarrow 0$
    \For{$i = 1$ to $N^{max}_{ex}$}
        \State $ctest[i] \leftarrow True$
        \For{$j = i+1$ to $N^{max}_{ex}$}
            \If{{\bf not}  $\left[(O_A,E_A) \stackrel{m}{==} (O_B,E_B)\right]$ }
                \State $ctest[i] \leftarrow False$
                \State $error\_count \leftarrow error\_count + 1$
                \State {\bf break}
            \EndIf
        \EndFor
    \EndFor
    \State return array {\it ctest} and integer $error\_count$
\end{algorithmic}
\end{algorithm}
 
 Since the data obtained at high order in noise amplification could be too noisy to allow extrapolations that pass this test, in the case of a {\it failed} extrapolation we attempt again by removing the highest order result at $k=N^{max}_{ex}$ and repeat the procedure. Note that, as the new consistency checks are performed, the error count is not reinitialized and keeps collecting the number of failures. If the consistency check fails the second time we definitively flag the result as {\it failed} and avoid using it in the next manipulations.
 
 The procedure we just described attempts to use all the information collected in the experiment to select extrapolations in which we have a higher confidence and at the same time alerting us if the confidence level is too low. For this to work well it is important to check that the result at the highest noise levels do contain non-trivial information in the first place, here we use the same strategy used to find decohered runs in~\cite{roggero2020A} to avoid performing uncontrolled extrapolations. Contrary to what we found in~\cite{roggero2020A}, the results presented in this manuscript were never flagged by this procedure.

 \subsubsection{Combining different extrapolation results}
 At the end of the consistency check procedure described in the previous section, we collect the non {\it failed} results into a list of triples $(O^e_k,E^e_k,C_k)$ containing the best value $O^e_k$ and it's error estimate $E^e_k$, together with the total error count $C_k$ obtained for the given extrapolation method. Note that, as mentioned above, both the polynomial and exponential extrapolation strategies can have higher error counts than the Richardson since we are also checking for $\chi^2$ compatibility (in the polynomial case) and sign changes (in the exponential case).
 
 A final unique result is then obtained as follows: if the global linear fit was successful the we use that as our best result, if not then we average together the results with the smallest error count $C_k$ together. This allows to avoid problems in regimes where Richardson is well behaved but the exponential is not and vice-versa.
 
 \subsection{Error propagation}
 The results reported in the main text contain also an error which we use to quantify the effect of statistical fluctuations~\footnote{Since we are combining together results from different approximations the error bar contains information about the systematic errors too.}. In this work we obtained these errors using a simple resampling technique: we first obtain the read-out mitigated results $\{O^r_k\}$ at all the $N^{max}_{ex}$ noise levels with their corresponding errors $\{O^r_k\}$, we then associate to the $k$-th result a Gaussian distribution of mean $O^r_k$ and variance $E^r_k$ and obtain $M$ samples from this distribution. For each of the $M$ sets of $N^{max}_{ex}$ resampled observables we perform a full extrapolation neglecting the resulting error estimates. We finally compute the mean and $68\%$ probability confidence interval from the collection of $M$ extrapolated results.
 
 Results obtained by explicitly propagating the errors throughout the calculation (as done in~\cite{roggero2020A}) produces results that are very similar but with a mildly increased dispersion. This is likely an effect generated by both our choice of parameters for the consistency check and of the procedure to combine results together: when two results are only barely failing a consistency check this might result in an exclusion of that extrapolation method in the final average, but a small fluctuation can change this and possibly considerably shift the final optimal result. The resampling procedure we adopt ensures that these effects are smoothed out.


\begin{thebibliography}{44}%
\makeatletter
\providecommand \@ifxundefined [1]{%
 \@ifx{#1\undefined}
}%
\providecommand \@ifnum [1]{%
 \ifnum #1\expandafter \@firstoftwo
 \else \expandafter \@secondoftwo
 \fi
}%
\providecommand \@ifx [1]{%
 \ifx #1\expandafter \@firstoftwo
 \else \expandafter \@secondoftwo
 \fi
}%
\providecommand \natexlab [1]{#1}%
\providecommand \enquote  [1]{``#1''}%
\providecommand \bibnamefont  [1]{#1}%
\providecommand \bibfnamefont [1]{#1}%
\providecommand \citenamefont [1]{#1}%
\providecommand \href@noop [0]{\@secondoftwo}%
\providecommand \href [0]{\begingroup \@sanitize@url \@href}%
\providecommand \@href[1]{\@@startlink{#1}\@@href}%
\providecommand \@@href[1]{\endgroup#1\@@endlink}%
\providecommand \@sanitize@url [0]{\catcode `\\12\catcode `\$12\catcode
  `\&12\catcode `\#12\catcode `\^12\catcode `\_12\catcode `\%12\relax}%
\providecommand \@@startlink[1]{}%
\providecommand \@@endlink[0]{}%
\providecommand \url  [0]{\begingroup\@sanitize@url \@url }%
\providecommand \@url [1]{\endgroup\@href {#1}{\urlprefix }}%
\providecommand \urlprefix  [0]{URL }%
\providecommand \Eprint [0]{\href }%
\providecommand \doibase [0]{http://dx.doi.org/}%
\providecommand \selectlanguage [0]{\@gobble}%
\providecommand \bibinfo  [0]{\@secondoftwo}%
\providecommand \bibfield  [0]{\@secondoftwo}%
\providecommand \translation [1]{[#1]}%
\providecommand \BibitemOpen [0]{}%
\providecommand \bibitemStop [0]{}%
\providecommand \bibitemNoStop [0]{.\EOS\space}%
\providecommand \EOS [0]{\spacefactor3000\relax}%
\providecommand \BibitemShut  [1]{\csname bibitem#1\endcsname}%
\let\auto@bib@innerbib\@empty
\bibitem [{\citenamefont {Steane}(1998)}]{steane1998}%
  \BibitemOpen
  \bibfield  {author} {\bibinfo {author} {\bibfnamefont {Andrew}\ \bibnamefont
  {Steane}},\ }\bibfield  {title} {\enquote {\bibinfo {title} {Quantum
  computing},}\ }\href {\doibase 10.1088/0034-4885/61/2/002} {\bibfield
  {journal} {\bibinfo  {journal} {Reports on Progress in Physics}\ }\textbf
  {\bibinfo {volume} {61}},\ \bibinfo {pages} {117} (\bibinfo {year}
  {1998})}\BibitemShut {NoStop}%
\bibitem [{\citenamefont {{Ladd}}\ \emph {et~al.}(2010)\citenamefont {{Ladd}},
  \citenamefont {{Jelezko}}, \citenamefont {{Laflamme}}, \citenamefont
  {{Nakamura}}, \citenamefont {{Monroe}},\ and\ \citenamefont
  {{O'Brien}}}]{ladd2010}%
  \BibitemOpen
  \bibfield  {author} {\bibinfo {author} {\bibfnamefont {T.~D.}\ \bibnamefont
  {{Ladd}}}, \bibinfo {author} {\bibfnamefont {F.}~\bibnamefont {{Jelezko}}},
  \bibinfo {author} {\bibfnamefont {R.}~\bibnamefont {{Laflamme}}}, \bibinfo
  {author} {\bibfnamefont {Y.}~\bibnamefont {{Nakamura}}}, \bibinfo {author}
  {\bibfnamefont {C.}~\bibnamefont {{Monroe}}}, \ and\ \bibinfo {author}
  {\bibfnamefont {J.~L.}\ \bibnamefont {{O'Brien}}},\ }\bibfield  {title}
  {\enquote {\bibinfo {title} {{Quantum computers}},}\ }\href {\doibase
  10.1038/nature08812} {\bibfield  {journal} {\bibinfo  {journal} {Nature}\
  }\textbf {\bibinfo {volume} {464}},\ \bibinfo {pages} {45--53} (\bibinfo
  {year} {2010})},\ \Eprint {http://arxiv.org/abs/1009.2267} {arXiv:1009.2267
  [quant-ph]} \BibitemShut {NoStop}%
\bibitem [{\citenamefont {{Martinez}}\ \emph {et~al.}(2016)\citenamefont
  {{Martinez}}, \citenamefont {{Muschik}}, \citenamefont {{Schindler}},
  \citenamefont {{Nigg}}, \citenamefont {{Erhard}}, \citenamefont {{Heyl}},
  \citenamefont {{Hauke}}, \citenamefont {{Dalmonte}}, \citenamefont {{Monz}},
  \citenamefont {{Zoller}},\ and\ \citenamefont {{Blatt}}}]{martinez2016}%
  \BibitemOpen
  \bibfield  {author} {\bibinfo {author} {\bibfnamefont {E.~A.}\ \bibnamefont
  {{Martinez}}}, \bibinfo {author} {\bibfnamefont {C.~A.}\ \bibnamefont
  {{Muschik}}}, \bibinfo {author} {\bibfnamefont {P.}~\bibnamefont
  {{Schindler}}}, \bibinfo {author} {\bibfnamefont {D.}~\bibnamefont {{Nigg}}},
  \bibinfo {author} {\bibfnamefont {A.}~\bibnamefont {{Erhard}}}, \bibinfo
  {author} {\bibfnamefont {M.}~\bibnamefont {{Heyl}}}, \bibinfo {author}
  {\bibfnamefont {P.}~\bibnamefont {{Hauke}}}, \bibinfo {author} {\bibfnamefont
  {M.}~\bibnamefont {{Dalmonte}}}, \bibinfo {author} {\bibfnamefont
  {T.}~\bibnamefont {{Monz}}}, \bibinfo {author} {\bibfnamefont
  {P.}~\bibnamefont {{Zoller}}}, \ and\ \bibinfo {author} {\bibfnamefont
  {R.}~\bibnamefont {{Blatt}}},\ }\bibfield  {title} {\enquote {\bibinfo
  {title} {{Real-time dynamics of lattice gauge theories with a few-qubit
  quantum computer}},}\ }\href {\doibase 10.1038/nature18318} {\bibfield
  {journal} {\bibinfo  {journal} {Nature}\ }\textbf {\bibinfo {volume} {534}},\
  \bibinfo {pages} {516--519} (\bibinfo {year} {2016})},\ \Eprint
  {http://arxiv.org/abs/1605.04570} {arXiv:1605.04570 [quant-ph]} \BibitemShut
  {NoStop}%
\bibitem [{\citenamefont {{Kaplan}}\ \emph {et~al.}(2017)\citenamefont
  {{Kaplan}}, \citenamefont {{Klco}},\ and\ \citenamefont
  {{Roggero}}}]{kaplan2017}%
  \BibitemOpen
  \bibfield  {author} {\bibinfo {author} {\bibfnamefont {D.~B.}\ \bibnamefont
  {{Kaplan}}}, \bibinfo {author} {\bibfnamefont {N.}~\bibnamefont {{Klco}}}, \
  and\ \bibinfo {author} {\bibfnamefont {A.}~\bibnamefont {{Roggero}}},\
  }\bibfield  {title} {\enquote {\bibinfo {title} {{Ground States via Spectral
  Combing on a Quantum Computer}},}\ }\href
  {http://adsabs.harvard.edu/abs/2017arXiv170908250K} {\bibfield  {journal}
  {\bibinfo  {journal} {ArXiv e-prints}\ } (\bibinfo {year} {2017})},\ \Eprint
  {http://arxiv.org/abs/1709.08250} {arXiv:1709.08250 [quant-ph]} \BibitemShut
  {NoStop}%
\bibitem [{\citenamefont {Dumitrescu}\ \emph {et~al.}(2018)\citenamefont
  {Dumitrescu}, \citenamefont {McCaskey}, \citenamefont {Hagen}, \citenamefont
  {Jansen}, \citenamefont {Morris}, \citenamefont {Papenbrock}, \citenamefont
  {Pooser}, \citenamefont {Dean},\ and\ \citenamefont
  {Lougovski}}]{Dumitrescu2018}%
  \BibitemOpen
  \bibfield  {author} {\bibinfo {author} {\bibfnamefont {E.~F.}\ \bibnamefont
  {Dumitrescu}}, \bibinfo {author} {\bibfnamefont {A.~J.}\ \bibnamefont
  {McCaskey}}, \bibinfo {author} {\bibfnamefont {G.}~\bibnamefont {Hagen}},
  \bibinfo {author} {\bibfnamefont {G.~R.}\ \bibnamefont {Jansen}}, \bibinfo
  {author} {\bibfnamefont {T.~D.}\ \bibnamefont {Morris}}, \bibinfo {author}
  {\bibfnamefont {T.}~\bibnamefont {Papenbrock}}, \bibinfo {author}
  {\bibfnamefont {R.~C.}\ \bibnamefont {Pooser}}, \bibinfo {author}
  {\bibfnamefont {D.~J.}\ \bibnamefont {Dean}}, \ and\ \bibinfo {author}
  {\bibfnamefont {P.}~\bibnamefont {Lougovski}},\ }\bibfield  {title} {\enquote
  {\bibinfo {title} {Cloud quantum computing of an atomic nucleus},}\ }\href
  {\doibase 10.1103/PhysRevLett.120.210501} {\bibfield  {journal} {\bibinfo
  {journal} {Phys. Rev. Lett.}\ }\textbf {\bibinfo {volume} {120}},\ \bibinfo
  {pages} {210501} (\bibinfo {year} {2018})}\BibitemShut {NoStop}%
\bibitem [{\citenamefont {Klco}\ \emph {et~al.}(2018)\citenamefont {Klco},
  \citenamefont {Dumitrescu}, \citenamefont {McCaskey}, \citenamefont {Morris},
  \citenamefont {Pooser}, \citenamefont {Sanz}, \citenamefont {Solano},
  \citenamefont {Lougovski},\ and\ \citenamefont {Savage}}]{klco2018}%
  \BibitemOpen
  \bibfield  {author} {\bibinfo {author} {\bibfnamefont {N.}~\bibnamefont
  {Klco}}, \bibinfo {author} {\bibfnamefont {E.~F.}\ \bibnamefont
  {Dumitrescu}}, \bibinfo {author} {\bibfnamefont {A.~J.}\ \bibnamefont
  {McCaskey}}, \bibinfo {author} {\bibfnamefont {T.~D.}\ \bibnamefont
  {Morris}}, \bibinfo {author} {\bibfnamefont {R.~C.}\ \bibnamefont {Pooser}},
  \bibinfo {author} {\bibfnamefont {M.}~\bibnamefont {Sanz}}, \bibinfo {author}
  {\bibfnamefont {E.}~\bibnamefont {Solano}}, \bibinfo {author} {\bibfnamefont
  {P.}~\bibnamefont {Lougovski}}, \ and\ \bibinfo {author} {\bibfnamefont
  {M.~J.}\ \bibnamefont {Savage}},\ }\bibfield  {title} {\enquote {\bibinfo
  {title} {Quantum-classical computation of schwinger model dynamics using
  quantum computers},}\ }\href {\doibase 10.1103/PhysRevA.98.032331} {\bibfield
   {journal} {\bibinfo  {journal} {Phys. Rev. A}\ }\textbf {\bibinfo {volume}
  {98}},\ \bibinfo {pages} {032331} (\bibinfo {year} {2018})}\BibitemShut
  {NoStop}%
\bibitem [{\citenamefont {Roggero}\ and\ \citenamefont
  {Carlson}(2019)}]{roggero2019}%
  \BibitemOpen
  \bibfield  {author} {\bibinfo {author} {\bibfnamefont {Alessandro}\
  \bibnamefont {Roggero}}\ and\ \bibinfo {author} {\bibfnamefont {Joseph}\
  \bibnamefont {Carlson}},\ }\bibfield  {title} {\enquote {\bibinfo {title}
  {Dynamic linear response quantum algorithm},}\ }\href {\doibase
  10.1103/PhysRevC.100.034610} {\bibfield  {journal} {\bibinfo  {journal}
  {Phys. Rev. C}\ }\textbf {\bibinfo {volume} {100}},\ \bibinfo {pages}
  {034610} (\bibinfo {year} {2019})}\BibitemShut {NoStop}%
\bibitem [{\citenamefont {Roggero}\ \emph {et~al.}(2020)\citenamefont
  {Roggero}, \citenamefont {Li}, \citenamefont {Carlson}, \citenamefont
  {Gupta},\ and\ \citenamefont {Perdue}}]{roggero2020A}%
  \BibitemOpen
  \bibfield  {author} {\bibinfo {author} {\bibfnamefont {Alessandro}\
  \bibnamefont {Roggero}}, \bibinfo {author} {\bibfnamefont {Andy C.~Y.}\
  \bibnamefont {Li}}, \bibinfo {author} {\bibfnamefont {Joseph}\ \bibnamefont
  {Carlson}}, \bibinfo {author} {\bibfnamefont {Rajan}\ \bibnamefont {Gupta}},
  \ and\ \bibinfo {author} {\bibfnamefont {Gabriel~N.}\ \bibnamefont
  {Perdue}},\ }\bibfield  {title} {\enquote {\bibinfo {title} {Quantum
  computing for neutrino-nucleus scattering},}\ }\href {\doibase
  10.1103/PhysRevD.101.074038} {\bibfield  {journal} {\bibinfo  {journal}
  {Phys. Rev. D}\ }\textbf {\bibinfo {volume} {101}},\ \bibinfo {pages}
  {074038} (\bibinfo {year} {2020})}\BibitemShut {NoStop}%
\bibitem [{\citenamefont {Caurier}\ \emph {et~al.}(2005)\citenamefont
  {Caurier}, \citenamefont {Mart{\'i}nez-Pinedo}, \citenamefont {Nowacki},
  \citenamefont {Poves},\ and\ \citenamefont {Zuker}}]{caurier2005}%
  \BibitemOpen
  \bibfield  {author} {\bibinfo {author} {\bibfnamefont {E.}~\bibnamefont
  {Caurier}}, \bibinfo {author} {\bibfnamefont {G.}~\bibnamefont
  {Mart{\'i}nez-Pinedo}}, \bibinfo {author} {\bibfnamefont {F.}~\bibnamefont
  {Nowacki}}, \bibinfo {author} {\bibfnamefont {A.}~\bibnamefont {Poves}}, \
  and\ \bibinfo {author} {\bibfnamefont {A.~P.}\ \bibnamefont {Zuker}},\
  }\bibfield  {title} {\enquote {\bibinfo {title} {The shell model as a unified
  view of nuclear structure},}\ }\href {\doibase 10.1103/RevModPhys.77.427}
  {\bibfield  {journal} {\bibinfo  {journal} {Rev. Mod. Phys.}\ }\textbf
  {\bibinfo {volume} {77}},\ \bibinfo {pages} {427--488} (\bibinfo {year}
  {2005})}\BibitemShut {NoStop}%
\bibitem [{\citenamefont {Preskill}(2018)}]{preskill2018}%
  \BibitemOpen
  \bibfield  {author} {\bibinfo {author} {\bibfnamefont {John}\ \bibnamefont
  {Preskill}},\ }\bibfield  {title} {\enquote {\bibinfo {title} {Quantum
  {C}omputing in the {NISQ} era and beyond},}\ }\href {\doibase
  10.22331/q-2018-08-06-79} {\bibfield  {journal} {\bibinfo  {journal}
  {{Quantum}}\ }\textbf {\bibinfo {volume} {2}},\ \bibinfo {pages} {79}
  (\bibinfo {year} {2018})}\BibitemShut {NoStop}%
\bibitem [{\citenamefont {Temme}\ \emph {et~al.}(2017)\citenamefont {Temme},
  \citenamefont {Bravyi},\ and\ \citenamefont {Gambetta}}]{temme2017}%
  \BibitemOpen
  \bibfield  {author} {\bibinfo {author} {\bibfnamefont {Kristan}\ \bibnamefont
  {Temme}}, \bibinfo {author} {\bibfnamefont {Sergey}\ \bibnamefont {Bravyi}},
  \ and\ \bibinfo {author} {\bibfnamefont {Jay~M.}\ \bibnamefont {Gambetta}},\
  }\bibfield  {title} {\enquote {\bibinfo {title} {Error mitigation for
  short-depth quantum circuits},}\ }\href {\doibase
  10.1103/PhysRevLett.119.180509} {\bibfield  {journal} {\bibinfo  {journal}
  {Phys. Rev. Lett.}\ }\textbf {\bibinfo {volume} {119}},\ \bibinfo {pages}
  {180509} (\bibinfo {year} {2017})}\BibitemShut {NoStop}%
\bibitem [{\citenamefont {Kandala}\ \emph {et~al.}(2019)\citenamefont
  {Kandala}, \citenamefont {Temme}, \citenamefont {Córcoles}, \citenamefont
  {Mezzacapo}, \citenamefont {Chow},\ and\ \citenamefont
  {Gambetta}}]{kandala2019}%
  \BibitemOpen
  \bibfield  {author} {\bibinfo {author} {\bibfnamefont {A.}~\bibnamefont
  {Kandala}}, \bibinfo {author} {\bibfnamefont {K.}~\bibnamefont {Temme}},
  \bibinfo {author} {\bibfnamefont {A.~D.}\ \bibnamefont {Córcoles}}, \bibinfo
  {author} {\bibfnamefont {A.}~\bibnamefont {Mezzacapo}}, \bibinfo {author}
  {\bibfnamefont {J.~M.}\ \bibnamefont {Chow}}, \ and\ \bibinfo {author}
  {\bibfnamefont {J.~M.}\ \bibnamefont {Gambetta}},\ }\bibfield  {title}
  {\enquote {\bibinfo {title} {Error mitigation extends the computational reach
  of a noisy quantum processor},}\ }\href {\doibase 10.1038/s41586-019-1040-7}
  {\bibfield  {journal} {\bibinfo  {journal} {Nature}\ }\textbf {\bibinfo
  {volume} {567}},\ \bibinfo {pages} {491} (\bibinfo {year}
  {2019})}\BibitemShut {NoStop}%
\bibitem [{\citenamefont {Adelberger}\ \emph {et~al.}(2011)\citenamefont
  {Adelberger}, \citenamefont {Garc\'{\i}a}, \citenamefont {Robertson},
  \citenamefont {Snover}, \citenamefont {Balantekin}, \citenamefont {Heeger},
  \citenamefont {Ramsey-Musolf}, \citenamefont {Bemmerer}, \citenamefont
  {Junghans}, \citenamefont {Bertulani}, \citenamefont {Chen}, \citenamefont
  {Costantini}, \citenamefont {Prati}, \citenamefont {Couder}, \citenamefont
  {Uberseder}, \citenamefont {Wiescher}, \citenamefont {Cyburt}, \citenamefont
  {Davids}, \citenamefont {Freedman}, \citenamefont {Gai}, \citenamefont
  {Gazit}, \citenamefont {Gialanella}, \citenamefont {Imbriani}, \citenamefont
  {Greife}, \citenamefont {Hass}, \citenamefont {Haxton}, \citenamefont
  {Itahashi}, \citenamefont {Kubodera}, \citenamefont {Langanke}, \citenamefont
  {Leitner}, \citenamefont {Leitner}, \citenamefont {Vetter}, \citenamefont
  {Winslow}, \citenamefont {Marcucci}, \citenamefont {Motobayashi},
  \citenamefont {Mukhamedzhanov}, \citenamefont {Tribble}, \citenamefont
  {Nollett}, \citenamefont {Nunes}, \citenamefont {Park}, \citenamefont
  {Parker}, \citenamefont {Schiavilla}, \citenamefont {Simpson}, \citenamefont
  {Spitaleri}, \citenamefont {Strieder}, \citenamefont {Trautvetter},
  \citenamefont {Suemmerer},\ and\ \citenamefont {Typel}}]{adelberger2011}%
  \BibitemOpen
  \bibfield  {author} {\bibinfo {author} {\bibfnamefont {E.~G.}\ \bibnamefont
  {Adelberger}}, \bibinfo {author} {\bibfnamefont {A.}~\bibnamefont
  {Garc\'{\i}a}}, \bibinfo {author} {\bibfnamefont {R.~G.~Hamish}\ \bibnamefont
  {Robertson}}, \bibinfo {author} {\bibfnamefont {K.~A.}\ \bibnamefont
  {Snover}}, \bibinfo {author} {\bibfnamefont {A.~B.}\ \bibnamefont
  {Balantekin}}, \bibinfo {author} {\bibfnamefont {K.}~\bibnamefont {Heeger}},
  \bibinfo {author} {\bibfnamefont {M.~J.}\ \bibnamefont {Ramsey-Musolf}},
  \bibinfo {author} {\bibfnamefont {D.}~\bibnamefont {Bemmerer}}, \bibinfo
  {author} {\bibfnamefont {A.}~\bibnamefont {Junghans}}, \bibinfo {author}
  {\bibfnamefont {C.~A.}\ \bibnamefont {Bertulani}}, \bibinfo {author}
  {\bibfnamefont {J.-W.}\ \bibnamefont {Chen}}, \bibinfo {author}
  {\bibfnamefont {H.}~\bibnamefont {Costantini}}, \bibinfo {author}
  {\bibfnamefont {P.}~\bibnamefont {Prati}}, \bibinfo {author} {\bibfnamefont
  {M.}~\bibnamefont {Couder}}, \bibinfo {author} {\bibfnamefont
  {E.}~\bibnamefont {Uberseder}}, \bibinfo {author} {\bibfnamefont
  {M.}~\bibnamefont {Wiescher}}, \bibinfo {author} {\bibfnamefont
  {R.}~\bibnamefont {Cyburt}}, \bibinfo {author} {\bibfnamefont
  {B.}~\bibnamefont {Davids}}, \bibinfo {author} {\bibfnamefont {S.~J.}\
  \bibnamefont {Freedman}}, \bibinfo {author} {\bibfnamefont {M.}~\bibnamefont
  {Gai}}, \bibinfo {author} {\bibfnamefont {D.}~\bibnamefont {Gazit}}, \bibinfo
  {author} {\bibfnamefont {L.}~\bibnamefont {Gialanella}}, \bibinfo {author}
  {\bibfnamefont {G.}~\bibnamefont {Imbriani}}, \bibinfo {author}
  {\bibfnamefont {U.}~\bibnamefont {Greife}}, \bibinfo {author} {\bibfnamefont
  {M.}~\bibnamefont {Hass}}, \bibinfo {author} {\bibfnamefont {W.~C.}\
  \bibnamefont {Haxton}}, \bibinfo {author} {\bibfnamefont {T.}~\bibnamefont
  {Itahashi}}, \bibinfo {author} {\bibfnamefont {K.}~\bibnamefont {Kubodera}},
  \bibinfo {author} {\bibfnamefont {K.}~\bibnamefont {Langanke}}, \bibinfo
  {author} {\bibfnamefont {D.}~\bibnamefont {Leitner}}, \bibinfo {author}
  {\bibfnamefont {M.}~\bibnamefont {Leitner}}, \bibinfo {author} {\bibfnamefont
  {P.}~\bibnamefont {Vetter}}, \bibinfo {author} {\bibfnamefont
  {L.}~\bibnamefont {Winslow}}, \bibinfo {author} {\bibfnamefont {L.~E.}\
  \bibnamefont {Marcucci}}, \bibinfo {author} {\bibfnamefont {T.}~\bibnamefont
  {Motobayashi}}, \bibinfo {author} {\bibfnamefont {A.}~\bibnamefont
  {Mukhamedzhanov}}, \bibinfo {author} {\bibfnamefont {R.~E.}\ \bibnamefont
  {Tribble}}, \bibinfo {author} {\bibfnamefont {Kenneth~M.}\ \bibnamefont
  {Nollett}}, \bibinfo {author} {\bibfnamefont {F.~M.}\ \bibnamefont {Nunes}},
  \bibinfo {author} {\bibfnamefont {T.-S.}\ \bibnamefont {Park}}, \bibinfo
  {author} {\bibfnamefont {P.~D.}\ \bibnamefont {Parker}}, \bibinfo {author}
  {\bibfnamefont {R.}~\bibnamefont {Schiavilla}}, \bibinfo {author}
  {\bibfnamefont {E.~C.}\ \bibnamefont {Simpson}}, \bibinfo {author}
  {\bibfnamefont {C.}~\bibnamefont {Spitaleri}}, \bibinfo {author}
  {\bibfnamefont {F.}~\bibnamefont {Strieder}}, \bibinfo {author}
  {\bibfnamefont {H.-P.}\ \bibnamefont {Trautvetter}}, \bibinfo {author}
  {\bibfnamefont {K.}~\bibnamefont {Suemmerer}}, \ and\ \bibinfo {author}
  {\bibfnamefont {S.}~\bibnamefont {Typel}},\ }\bibfield  {title} {\enquote
  {\bibinfo {title} {Solar fusion cross sections. ii. the $pp$ chain and cno
  cycles},}\ }\href {\doibase 10.1103/RevModPhys.83.195} {\bibfield  {journal}
  {\bibinfo  {journal} {Rev. Mod. Phys.}\ }\textbf {\bibinfo {volume} {83}},\
  \bibinfo {pages} {195--245} (\bibinfo {year} {2011})}\BibitemShut {NoStop}%
\bibitem [{\citenamefont {Park}\ \emph {et~al.}(1998)\citenamefont {Park},
  \citenamefont {Kubodera}, \citenamefont {Min},\ and\ \citenamefont
  {Rho}}]{park1998}%
  \BibitemOpen
  \bibfield  {author} {\bibinfo {author} {\bibfnamefont {Tae-Sun}\ \bibnamefont
  {Park}}, \bibinfo {author} {\bibfnamefont {Kuniharu}\ \bibnamefont
  {Kubodera}}, \bibinfo {author} {\bibfnamefont {Dong-Pil}\ \bibnamefont
  {Min}}, \ and\ \bibinfo {author} {\bibfnamefont {Mannque}\ \bibnamefont
  {Rho}},\ }\bibfield  {title} {\enquote {\bibinfo {title} {Effective field
  theory for low-energy two-nucleon systems},}\ }\href {\doibase
  10.1103/PhysRevC.58.R637} {\bibfield  {journal} {\bibinfo  {journal} {Phys.
  Rev. C}\ }\textbf {\bibinfo {volume} {58}},\ \bibinfo {pages} {R637--R640}
  (\bibinfo {year} {1998})}\BibitemShut {NoStop}%
\bibitem [{\citenamefont {Chen}\ \emph {et~al.}(1999)\citenamefont {Chen},
  \citenamefont {Rupak},\ and\ \citenamefont {Savage}}]{chen1999}%
  \BibitemOpen
  \bibfield  {author} {\bibinfo {author} {\bibfnamefont {Jiunn-Wei}\
  \bibnamefont {Chen}}, \bibinfo {author} {\bibfnamefont {Gautam}\ \bibnamefont
  {Rupak}}, \ and\ \bibinfo {author} {\bibfnamefont {Martin~J.}\ \bibnamefont
  {Savage}},\ }\bibfield  {title} {\enquote {\bibinfo {title} {Nucleon-nucleon
  effective field theory without pions},}\ }\href {\doibase
  10.1016/S0375-9474(99)00298-5} {\bibfield  {journal} {\bibinfo  {journal}
  {Nuclear Physics A}\ }\textbf {\bibinfo {volume} {653}},\ \bibinfo {pages}
  {386 -- 412} (\bibinfo {year} {1999})}\BibitemShut {NoStop}%
\bibitem [{\citenamefont {Chen}\ and\ \citenamefont
  {Savage}(1999)}]{chen1999b}%
  \BibitemOpen
  \bibfield  {author} {\bibinfo {author} {\bibfnamefont {Jiunn-Wei}\
  \bibnamefont {Chen}}\ and\ \bibinfo {author} {\bibfnamefont {Martin~J.}\
  \bibnamefont {Savage}},\ }\bibfield  {title} {\enquote {\bibinfo {title}
  {$n\stackrel{\ensuremath{\rightarrow}}{p}d\ensuremath{\gamma}$ for big-bang
  nucleosynthesis},}\ }\href {\doibase 10.1103/PhysRevC.60.065205} {\bibfield
  {journal} {\bibinfo  {journal} {Phys. Rev. C}\ }\textbf {\bibinfo {volume}
  {60}},\ \bibinfo {pages} {065205} (\bibinfo {year} {1999})}\BibitemShut
  {NoStop}%
\bibitem [{\citenamefont {Bravyi}\ and\ \citenamefont
  {Kitaev}(2002)}]{bravyi2002}%
  \BibitemOpen
  \bibfield  {author} {\bibinfo {author} {\bibfnamefont {Sergey~B.}\
  \bibnamefont {Bravyi}}\ and\ \bibinfo {author} {\bibfnamefont {Alexei~Yu.}\
  \bibnamefont {Kitaev}},\ }\bibfield  {title} {\enquote {\bibinfo {title}
  {Fermionic quantum computation},}\ }\href {\doibase 10.1006/aphy.2002.6254}
  {\bibfield  {journal} {\bibinfo  {journal} {Annals of Physics}\ }\textbf
  {\bibinfo {volume} {298}},\ \bibinfo {pages} {210 -- 226} (\bibinfo {year}
  {2002})}\BibitemShut {NoStop}%
\bibitem [{\citenamefont {Whitfield}\ \emph {et~al.}(2016)\citenamefont
  {Whitfield}, \citenamefont {Havl\'{\i}\ifmmode~\check{c}\else \v{c}\fi{}ek},\
  and\ \citenamefont {Troyer}}]{whitfield2016}%
  \BibitemOpen
  \bibfield  {author} {\bibinfo {author} {\bibfnamefont {James~D.}\
  \bibnamefont {Whitfield}}, \bibinfo {author} {\bibfnamefont {Vojt\ifmmode
  \check{e}\else~\v{e}\fi{}ch}\ \bibnamefont {Havl\'{\i}\ifmmode~\check{c}\else
  \v{c}\fi{}ek}}, \ and\ \bibinfo {author} {\bibfnamefont {Matthias}\
  \bibnamefont {Troyer}},\ }\bibfield  {title} {\enquote {\bibinfo {title}
  {Local spin operators for fermion simulations},}\ }\href {\doibase
  10.1103/PhysRevA.94.030301} {\bibfield  {journal} {\bibinfo  {journal} {Phys.
  Rev. A}\ }\textbf {\bibinfo {volume} {94}},\ \bibinfo {pages} {030301}
  (\bibinfo {year} {2016})}\BibitemShut {NoStop}%
\bibitem [{\citenamefont {Setia}\ \emph {et~al.}(2019)\citenamefont {Setia},
  \citenamefont {Bravyi}, \citenamefont {Mezzacapo},\ and\ \citenamefont
  {Whitfield}}]{setia2019}%
  \BibitemOpen
  \bibfield  {author} {\bibinfo {author} {\bibfnamefont {Kanav}\ \bibnamefont
  {Setia}}, \bibinfo {author} {\bibfnamefont {Sergey}\ \bibnamefont {Bravyi}},
  \bibinfo {author} {\bibfnamefont {Antonio}\ \bibnamefont {Mezzacapo}}, \ and\
  \bibinfo {author} {\bibfnamefont {James~D.}\ \bibnamefont {Whitfield}},\
  }\bibfield  {title} {\enquote {\bibinfo {title} {Superfast encodings for
  fermionic quantum simulation},}\ }\href {\doibase
  10.1103/PhysRevResearch.1.033033} {\bibfield  {journal} {\bibinfo  {journal}
  {Phys. Rev. Research}\ }\textbf {\bibinfo {volume} {1}},\ \bibinfo {pages}
  {033033} (\bibinfo {year} {2019})}\BibitemShut {NoStop}%
\bibitem [{\citenamefont {Steudtner}\ and\ \citenamefont
  {Wehner}(2019)}]{steudtner2019}%
  \BibitemOpen
  \bibfield  {author} {\bibinfo {author} {\bibfnamefont {Mark}\ \bibnamefont
  {Steudtner}}\ and\ \bibinfo {author} {\bibfnamefont {Stephanie}\ \bibnamefont
  {Wehner}},\ }\bibfield  {title} {\enquote {\bibinfo {title} {Quantum codes
  for quantum simulation of fermions on a square lattice of qubits},}\ }\href
  {\doibase 10.1103/PhysRevA.99.022308} {\bibfield  {journal} {\bibinfo
  {journal} {Phys. Rev. A}\ }\textbf {\bibinfo {volume} {99}},\ \bibinfo
  {pages} {022308} (\bibinfo {year} {2019})}\BibitemShut {NoStop}%
\bibitem [{\citenamefont {{Derby}}\ and\ \citenamefont
  {{Klassen}}(2020)}]{derby2020}%
  \BibitemOpen
  \bibfield  {author} {\bibinfo {author} {\bibfnamefont {Charles}\ \bibnamefont
  {{Derby}}}\ and\ \bibinfo {author} {\bibfnamefont {Joel}\ \bibnamefont
  {{Klassen}}},\ }\bibfield  {title} {\enquote {\bibinfo {title} {{Low Weight
  Fermionic Encodings for Lattice Models}},}\ }\href@noop {} {\bibfield
  {journal} {\bibinfo  {journal} {arXiv e-prints}\ ,\ \bibinfo {eid}
  {arXiv:2003.06939}} (\bibinfo {year} {2020})},\ \Eprint
  {http://arxiv.org/abs/2003.06939} {arXiv:2003.06939 [quant-ph]} \BibitemShut
  {NoStop}%
\bibitem [{\citenamefont {Nielsen}\ and\ \citenamefont
  {Chuang}(2010)}]{nielsen2010}%
  \BibitemOpen
  \bibfield  {author} {\bibinfo {author} {\bibfnamefont {M.A.}\ \bibnamefont
  {Nielsen}}\ and\ \bibinfo {author} {\bibfnamefont {I.L.}\ \bibnamefont
  {Chuang}},\ }\href@noop {} {\emph {\bibinfo {title} {Quantum Computation and
  Quantum Information}}}\ (\bibinfo  {publisher} {Cambridge University Press},\
  \bibinfo {year} {2010})\BibitemShut {NoStop}%
\bibitem [{\citenamefont {Lloyd}(1996)}]{lloyd1996}%
  \BibitemOpen
  \bibfield  {author} {\bibinfo {author} {\bibfnamefont {Seth}\ \bibnamefont
  {Lloyd}},\ }\bibfield  {title} {\enquote {\bibinfo {title} {Universal quantum
  simulators},}\ }\href {\doibase 10.1126/science.273.5278.1073} {\bibfield
  {journal} {\bibinfo  {journal} {Science}\ }\textbf {\bibinfo {volume}
  {273}},\ \bibinfo {pages} {1073--1078} (\bibinfo {year} {1996})}\BibitemShut
  {NoStop}%
\bibitem [{\citenamefont {Mazzola}\ \emph {et~al.}(2019)\citenamefont
  {Mazzola}, \citenamefont {Ollitrault}, \citenamefont {Barkoutsos},\ and\
  \citenamefont {Tavernelli}}]{mazzola2019}%
  \BibitemOpen
  \bibfield  {author} {\bibinfo {author} {\bibfnamefont {Guglielmo}\
  \bibnamefont {Mazzola}}, \bibinfo {author} {\bibfnamefont {Pauline~J.}\
  \bibnamefont {Ollitrault}}, \bibinfo {author} {\bibfnamefont
  {Panagiotis~Kl.}\ \bibnamefont {Barkoutsos}}, \ and\ \bibinfo {author}
  {\bibfnamefont {Ivano}\ \bibnamefont {Tavernelli}},\ }\bibfield  {title}
  {\enquote {\bibinfo {title} {Nonunitary operations for ground-state
  calculations in near-term quantum computers},}\ }\href {\doibase
  10.1103/PhysRevLett.123.130501} {\bibfield  {journal} {\bibinfo  {journal}
  {Phys. Rev. Lett.}\ }\textbf {\bibinfo {volume} {123}},\ \bibinfo {pages}
  {130501} (\bibinfo {year} {2019})}\BibitemShut {NoStop}%
\bibitem [{\citenamefont {Terashima}\ and\ \citenamefont
  {Ueda}(2005)}]{ueda2003}%
  \BibitemOpen
  \bibfield  {author} {\bibinfo {author} {\bibfnamefont {Hiroaki}\ \bibnamefont
  {Terashima}}\ and\ \bibinfo {author} {\bibfnamefont {Masahito}\ \bibnamefont
  {Ueda}},\ }\href {\doibase 10.1142/S0219749905001456} {\bibfield  {journal}
  {\bibinfo  {journal} {Int. J. Quantum Inform.}\ }\textbf {\bibinfo {volume}
  {3}},\ \bibinfo {pages} {633--647} (\bibinfo {year} {2005})}\BibitemShut
  {NoStop}%
\bibitem [{\citenamefont {Childs}\ and\ \citenamefont
  {Wiebe}(2012)}]{childs2012}%
  \BibitemOpen
  \bibfield  {author} {\bibinfo {author} {\bibfnamefont {Andrew~M.}\
  \bibnamefont {Childs}}\ and\ \bibinfo {author} {\bibfnamefont {Nathan}\
  \bibnamefont {Wiebe}},\ }\bibfield  {title} {\enquote {\bibinfo {title}
  {Hamiltonian simulation using linear combinations of unitary operations},}\
  }\href {\doibase 10.26421/QIC12.11-12} {\bibfield  {journal} {\bibinfo
  {journal} {Quantum Information and Computation}\ }\textbf {\bibinfo {volume}
  {12}},\ \bibinfo {pages} {0901--0924} (\bibinfo {year} {2012})}\BibitemShut
  {NoStop}%
\bibitem [{\citenamefont {Berry}\ \emph {et~al.}(2015)\citenamefont {Berry},
  \citenamefont {Childs}, \citenamefont {Cleve}, \citenamefont {Kothari},\ and\
  \citenamefont {Somma}}]{berry2015}%
  \BibitemOpen
  \bibfield  {author} {\bibinfo {author} {\bibfnamefont {Dominic~W.}\
  \bibnamefont {Berry}}, \bibinfo {author} {\bibfnamefont {Andrew~M.}\
  \bibnamefont {Childs}}, \bibinfo {author} {\bibfnamefont {Richard}\
  \bibnamefont {Cleve}}, \bibinfo {author} {\bibfnamefont {Robin}\ \bibnamefont
  {Kothari}}, \ and\ \bibinfo {author} {\bibfnamefont {Rolando~D.}\
  \bibnamefont {Somma}},\ }\bibfield  {title} {\enquote {\bibinfo {title}
  {Simulating hamiltonian dynamics with a truncated taylor series},}\ }\href
  {\doibase 10.1103/PhysRevLett.114.090502} {\bibfield  {journal} {\bibinfo
  {journal} {Phys. Rev. Lett.}\ }\textbf {\bibinfo {volume} {114}},\ \bibinfo
  {pages} {090502} (\bibinfo {year} {2015})}\BibitemShut {NoStop}%
\bibitem [{\citenamefont {Low}\ and\ \citenamefont {Chuang}(2017)}]{low2017}%
  \BibitemOpen
  \bibfield  {author} {\bibinfo {author} {\bibfnamefont {Guang~Hao}\
  \bibnamefont {Low}}\ and\ \bibinfo {author} {\bibfnamefont {Isaac~L.}\
  \bibnamefont {Chuang}},\ }\bibfield  {title} {\enquote {\bibinfo {title}
  {Optimal hamiltonian simulation by quantum signal processing},}\ }\href
  {\doibase 10.1103/PhysRevLett.118.010501} {\bibfield  {journal} {\bibinfo
  {journal} {Phys. Rev. Lett.}\ }\textbf {\bibinfo {volume} {118}},\ \bibinfo
  {pages} {010501} (\bibinfo {year} {2017})}\BibitemShut {NoStop}%
\bibitem [{\citenamefont {Childs}\ \emph {et~al.}(2017)\citenamefont {Childs},
  \citenamefont {Kothari},\ and\ \citenamefont {Somma}}]{childs2017}%
  \BibitemOpen
  \bibfield  {author} {\bibinfo {author} {\bibfnamefont {Andrew~M.}\
  \bibnamefont {Childs}}, \bibinfo {author} {\bibfnamefont {Robin}\
  \bibnamefont {Kothari}}, \ and\ \bibinfo {author} {\bibfnamefont
  {Rolando~D.}\ \bibnamefont {Somma}},\ }\bibfield  {title} {\enquote {\bibinfo
  {title} {Quantum algorithm for systems of linear equations with exponentially
  improved dependence on precision},}\ }\href {\doibase 10.1137/16M1087072}
  {\bibfield  {journal} {\bibinfo  {journal} {SIAM Journal on Computing}\
  }\textbf {\bibinfo {volume} {46}},\ \bibinfo {pages} {1920--1950} (\bibinfo
  {year} {2017})}\BibitemShut {NoStop}%
\bibitem [{\citenamefont {Vidal}\ and\ \citenamefont
  {Dawson}(2004)}]{vidal2004}%
  \BibitemOpen
  \bibfield  {author} {\bibinfo {author} {\bibfnamefont {G.}~\bibnamefont
  {Vidal}}\ and\ \bibinfo {author} {\bibfnamefont {C.~M.}\ \bibnamefont
  {Dawson}},\ }\bibfield  {title} {\enquote {\bibinfo {title} {Universal
  quantum circuit for two-qubit transformations with three controlled-not
  gates},}\ }\href {\doibase 10.1103/PhysRevA.69.010301} {\bibfield  {journal}
  {\bibinfo  {journal} {Phys. Rev. A}\ }\textbf {\bibinfo {volume} {69}},\
  \bibinfo {pages} {010301} (\bibinfo {year} {2004})}\BibitemShut {NoStop}%
\bibitem [{\citenamefont {Vatan}\ and\ \citenamefont
  {Williams}(2004)}]{vatan2004}%
  \BibitemOpen
  \bibfield  {author} {\bibinfo {author} {\bibfnamefont {Farrokh}\ \bibnamefont
  {Vatan}}\ and\ \bibinfo {author} {\bibfnamefont {Colin}\ \bibnamefont
  {Williams}},\ }\bibfield  {title} {\enquote {\bibinfo {title} {Optimal
  quantum circuits for general two-qubit gates},}\ }\href {\doibase
  10.1103/PhysRevA.69.032315} {\bibfield  {journal} {\bibinfo  {journal} {Phys.
  Rev. A}\ }\textbf {\bibinfo {volume} {69}},\ \bibinfo {pages} {032315}
  (\bibinfo {year} {2004})}\BibitemShut {NoStop}%
\bibitem [{\citenamefont {et. al}(2019)}]{qiskit}%
  \BibitemOpen
  \bibfield  {author} {\bibinfo {author} {\bibfnamefont {H{\'e}ctor~Abraham}\
  \bibnamefont {et. al}},\ }\href {\doibase 10.5281/zenodo.2562110} {\enquote
  {\bibinfo {title} {Qiskit: An open-source framework for quantum computing},}\
  } (\bibinfo {year} {2019})\BibitemShut {NoStop}%
\bibitem [{\citenamefont {5qubit~backed: IBM Q~team}(2020)}]{IBMQ_Vigo}%
  \BibitemOpen
  \bibfield  {author} {\bibinfo {author} {\bibnamefont {5qubit~backed: IBM
  Q~team}},\ }\href {https://quantum-computing.ibm.com} {\enquote {\bibinfo
  {title} {{I}{B}{M} {V}igo backend specification v1.0.2},}\ } (\bibinfo {year}
  {2020}),\ \bibinfo {note} {retrieved from
  https://quantum-computing.ibm.com}\BibitemShut {NoStop}%
\bibitem [{\citenamefont {Li}\ and\ \citenamefont {Benjamin}(2017)}]{li2017}%
  \BibitemOpen
  \bibfield  {author} {\bibinfo {author} {\bibfnamefont {Ying}\ \bibnamefont
  {Li}}\ and\ \bibinfo {author} {\bibfnamefont {Simon~C.}\ \bibnamefont
  {Benjamin}},\ }\bibfield  {title} {\enquote {\bibinfo {title} {Efficient
  variational quantum simulator incorporating active error minimization},}\
  }\href {\doibase 10.1103/PhysRevX.7.021050} {\bibfield  {journal} {\bibinfo
  {journal} {Phys. Rev. X}\ }\textbf {\bibinfo {volume} {7}},\ \bibinfo {pages}
  {021050} (\bibinfo {year} {2017})}\BibitemShut {NoStop}%
\bibitem [{\citenamefont {Endo}\ \emph {et~al.}(2018)\citenamefont {Endo},
  \citenamefont {Benjamin},\ and\ \citenamefont {Li}}]{Endo2018}%
  \BibitemOpen
  \bibfield  {author} {\bibinfo {author} {\bibfnamefont {Suguru}\ \bibnamefont
  {Endo}}, \bibinfo {author} {\bibfnamefont {Simon~C.}\ \bibnamefont
  {Benjamin}}, \ and\ \bibinfo {author} {\bibfnamefont {Ying}\ \bibnamefont
  {Li}},\ }\bibfield  {title} {\enquote {\bibinfo {title} {Practical quantum
  error mitigation for near-future applications},}\ }\href {\doibase
  10.1103/PhysRevX.8.031027} {\bibfield  {journal} {\bibinfo  {journal} {Phys.
  Rev. X}\ }\textbf {\bibinfo {volume} {8}},\ \bibinfo {pages} {031027}
  (\bibinfo {year} {2018})}\BibitemShut {NoStop}%
\bibitem [{\citenamefont {Shende}\ and\ \citenamefont
  {Markov}(2009)}]{shende2009}%
  \BibitemOpen
  \bibfield  {author} {\bibinfo {author} {\bibfnamefont {Vivek~V.}\
  \bibnamefont {Shende}}\ and\ \bibinfo {author} {\bibfnamefont {Igor~L.}\
  \bibnamefont {Markov}},\ }\bibfield  {title} {\enquote {\bibinfo {title} {On
  the cnot-cost of toffoli gates},}\ }\href@noop {} {\bibfield  {journal}
  {\bibinfo  {journal} {Quantum Info. Comput.}\ }\textbf {\bibinfo {volume}
  {9}},\ \bibinfo {pages} {461–486} (\bibinfo {year} {2009})}\BibitemShut
  {NoStop}%
\bibitem [{\citenamefont {Low}\ and\ \citenamefont {Chuang}(2019)}]{low2019}%
  \BibitemOpen
  \bibfield  {author} {\bibinfo {author} {\bibfnamefont {Guang~Hao}\
  \bibnamefont {Low}}\ and\ \bibinfo {author} {\bibfnamefont {Isaac~L.}\
  \bibnamefont {Chuang}},\ }\bibfield  {title} {\enquote {\bibinfo {title}
  {Hamiltonian {S}imulation by {Q}ubitization},}\ }\href {\doibase
  10.22331/q-2019-07-12-163} {\bibfield  {journal} {\bibinfo  {journal}
  {{Quantum}}\ }\textbf {\bibinfo {volume} {3}},\ \bibinfo {pages} {163}
  (\bibinfo {year} {2019})}\BibitemShut {NoStop}%
\bibitem [{\citenamefont {Gily\'{e}n}\ \emph {et~al.}(2019)\citenamefont
  {Gily\'{e}n}, \citenamefont {Su}, \citenamefont {Low},\ and\ \citenamefont
  {Wiebe}}]{qsvt2019}%
  \BibitemOpen
  \bibfield  {author} {\bibinfo {author} {\bibfnamefont {Andr\'{a}s}\
  \bibnamefont {Gily\'{e}n}}, \bibinfo {author} {\bibfnamefont {Yuan}\
  \bibnamefont {Su}}, \bibinfo {author} {\bibfnamefont {Guang~Hao}\
  \bibnamefont {Low}}, \ and\ \bibinfo {author} {\bibfnamefont {Nathan}\
  \bibnamefont {Wiebe}},\ }\bibfield  {title} {\enquote {\bibinfo {title}
  {Quantum singular value transformation and beyond: Exponential improvements
  for quantum matrix arithmetics},}\ }in\ \href {\doibase
  10.1145/3313276.3316366} {\emph {\bibinfo {booktitle} {STOC 2019:
  Proceedings}}},\ \bibinfo {series and number} {STOC 2019}\ (\bibinfo
  {publisher} {Association for Computing Machinery},\ \bibinfo {address} {New
  York, NY, USA},\ \bibinfo {year} {2019})\ p.\ \bibinfo {pages}
  {193–204}\BibitemShut {NoStop}%
\bibitem [{\citenamefont {Roggero}(2020)}]{roggero2020B}%
  \BibitemOpen
  \bibfield  {author} {\bibinfo {author} {\bibfnamefont {A.}~\bibnamefont
  {Roggero}},\ }\bibfield  {title} {\enquote {\bibinfo {title}
  {Spectral-density estimation with the gaussian integral transform},}\ }\href
  {\doibase 10.1103/PhysRevA.102.022409} {\bibfield  {journal} {\bibinfo
  {journal} {Phys. Rev. A}\ }\textbf {\bibinfo {volume} {102}},\ \bibinfo
  {pages} {022409} (\bibinfo {year} {2020})}\BibitemShut {NoStop}%
\bibitem [{\citenamefont {Barenco}\ \emph {et~al.}(1995)\citenamefont
  {Barenco}, \citenamefont {Bennett}, \citenamefont {Cleve}, \citenamefont
  {DiVincenzo}, \citenamefont {Margolus}, \citenamefont {Shor}, \citenamefont
  {Sleator}, \citenamefont {Smolin},\ and\ \citenamefont
  {Weinfurter}}]{barenco1995}%
  \BibitemOpen
  \bibfield  {author} {\bibinfo {author} {\bibfnamefont {Adriano}\ \bibnamefont
  {Barenco}}, \bibinfo {author} {\bibfnamefont {Charles~H.}\ \bibnamefont
  {Bennett}}, \bibinfo {author} {\bibfnamefont {Richard}\ \bibnamefont
  {Cleve}}, \bibinfo {author} {\bibfnamefont {David~P.}\ \bibnamefont
  {DiVincenzo}}, \bibinfo {author} {\bibfnamefont {Norman}\ \bibnamefont
  {Margolus}}, \bibinfo {author} {\bibfnamefont {Peter}\ \bibnamefont {Shor}},
  \bibinfo {author} {\bibfnamefont {Tycho}\ \bibnamefont {Sleator}}, \bibinfo
  {author} {\bibfnamefont {John~A.}\ \bibnamefont {Smolin}}, \ and\ \bibinfo
  {author} {\bibfnamefont {Harald}\ \bibnamefont {Weinfurter}},\ }\bibfield
  {title} {\enquote {\bibinfo {title} {Elementary gates for quantum
  computation},}\ }\href {\doibase 10.1103/PhysRevA.52.3457} {\bibfield
  {journal} {\bibinfo  {journal} {Phys. Rev. A}\ }\textbf {\bibinfo {volume}
  {52}},\ \bibinfo {pages} {3457--3467} (\bibinfo {year} {1995})}\BibitemShut
  {NoStop}%
\bibitem [{\citenamefont {Babbush}\ \emph {et~al.}(2019)\citenamefont
  {Babbush}, \citenamefont {Berry},\ and\ \citenamefont {Neven}}]{babbush2019}%
  \BibitemOpen
  \bibfield  {author} {\bibinfo {author} {\bibfnamefont {Ryan}\ \bibnamefont
  {Babbush}}, \bibinfo {author} {\bibfnamefont {Dominic~W.}\ \bibnamefont
  {Berry}}, \ and\ \bibinfo {author} {\bibfnamefont {Hartmut}\ \bibnamefont
  {Neven}},\ }\bibfield  {title} {\enquote {\bibinfo {title} {Quantum
  simulation of the sachdev-ye-kitaev model by asymmetric qubitization},}\
  }\href {\doibase 10.1103/PhysRevA.99.040301} {\bibfield  {journal} {\bibinfo
  {journal} {Phys. Rev. A}\ }\textbf {\bibinfo {volume} {99}},\ \bibinfo
  {pages} {040301} (\bibinfo {year} {2019})}\BibitemShut {NoStop}%
\bibitem [{\citenamefont {He}\ \emph {et~al.}(2020)\citenamefont {He},
  \citenamefont {Nachman}, \citenamefont {de~Jong},\ and\ \citenamefont
  {Bauer}}]{he2020}%
  \BibitemOpen
  \bibfield  {author} {\bibinfo {author} {\bibfnamefont {Andre}\ \bibnamefont
  {He}}, \bibinfo {author} {\bibfnamefont {Benjamin}\ \bibnamefont {Nachman}},
  \bibinfo {author} {\bibfnamefont {Wibe~A.}\ \bibnamefont {de~Jong}}, \ and\
  \bibinfo {author} {\bibfnamefont {Christian~W.}\ \bibnamefont {Bauer}},\
  }\bibfield  {title} {\enquote {\bibinfo {title} {Zero-noise extrapolation for
  quantum-gate error mitigation with identity insertions},}\ }\href {\doibase
  10.1103/PhysRevA.102.012426} {\bibfield  {journal} {\bibinfo  {journal}
  {Phys. Rev. A}\ }\textbf {\bibinfo {volume} {102}},\ \bibinfo {pages}
  {012426} (\bibinfo {year} {2020})}\BibitemShut {NoStop}%
\bibitem [{\citenamefont {Kandala}\ \emph {et~al.}(2017)\citenamefont
  {Kandala}, \citenamefont {Mezzacapo}, \citenamefont {Temme}, \citenamefont
  {Takita}, \citenamefont {Brink}, \citenamefont {Chow},\ and\ \citenamefont
  {Gambetta}}]{kandala2017}%
  \BibitemOpen
  \bibfield  {author} {\bibinfo {author} {\bibfnamefont {Abhinav}\ \bibnamefont
  {Kandala}}, \bibinfo {author} {\bibfnamefont {Antonio}\ \bibnamefont
  {Mezzacapo}}, \bibinfo {author} {\bibfnamefont {Kristan}\ \bibnamefont
  {Temme}}, \bibinfo {author} {\bibfnamefont {Maika}\ \bibnamefont {Takita}},
  \bibinfo {author} {\bibfnamefont {Markus}\ \bibnamefont {Brink}}, \bibinfo
  {author} {\bibfnamefont {Jerry~M}\ \bibnamefont {Chow}}, \ and\ \bibinfo
  {author} {\bibfnamefont {Jay~M}\ \bibnamefont {Gambetta}},\ }\bibfield
  {title} {\enquote {\bibinfo {title} {Hardware-efficient variational quantum
  eigensolver for small molecules and quantum magnets},}\ }\href@noop {}
  {\bibfield  {journal} {\bibinfo  {journal} {Nature}\ }\textbf {\bibinfo
  {volume} {549}},\ \bibinfo {pages} {242--246} (\bibinfo {year}
  {2017})}\BibitemShut {NoStop}%
\bibitem [{\citenamefont {Endo}\ \emph {et~al.}(2019)\citenamefont {Endo},
  \citenamefont {Zhao}, \citenamefont {Li}, \citenamefont {Benjamin},\ and\
  \citenamefont {Yuan}}]{endo2019}%
  \BibitemOpen
  \bibfield  {author} {\bibinfo {author} {\bibfnamefont {Suguru}\ \bibnamefont
  {Endo}}, \bibinfo {author} {\bibfnamefont {Qi}~\bibnamefont {Zhao}}, \bibinfo
  {author} {\bibfnamefont {Ying}\ \bibnamefont {Li}}, \bibinfo {author}
  {\bibfnamefont {Simon}\ \bibnamefont {Benjamin}}, \ and\ \bibinfo {author}
  {\bibfnamefont {Xiao}\ \bibnamefont {Yuan}},\ }\bibfield  {title} {\enquote
  {\bibinfo {title} {Mitigating algorithmic errors in a hamiltonian
  simulation},}\ }\href {\doibase 10.1103/PhysRevA.99.012334} {\bibfield
  {journal} {\bibinfo  {journal} {Phys. Rev. A}\ }\textbf {\bibinfo {volume}
  {99}},\ \bibinfo {pages} {012334} (\bibinfo {year} {2019})}\BibitemShut
  {NoStop}%
\end{thebibliography}
 \end{document}